\documentclass[aps,pra,onecolumn,notitlepage]{revtex4-1}

\usepackage{paralist}
\usepackage{amsmath,amssymb,amsfonts,amsthm}
\usepackage{multirow}
\usepackage{xcolor}
\usepackage{slashbox}
\usepackage{graphicx}
\usepackage{hyperref}
\definecolor{darkred}{rgb}{0.5,0,0}
\definecolor{darkgreen}{rgb}{0,0.5,0}
\definecolor{darkblue}{rgb}{0,0,0.5}
\hypersetup{colorlinks,breaklinks,linkcolor=darkred,citecolor=darkgreen,urlcolor=darkblue}

\usepackage[T1]{fontenc}
\usepackage{dsfont}
\newcommand{\id}{\mathds{1}}

\def\ket#1{|#1\rangle}
\def\braket#1#2{\langle#1|#2\rangle}
\def\ketbra#1{|#1\rangle\langle#1|}
\def\bra#1{\langle#1|}
\def\mub#1#2{\ket{\varphi_{#2}^{#1}}}

\def\bmub#1#2{\bra{\varphi_{#2}^{#1}}}

\def\pmub#1#2{\ketbra{\varphi_{#2}^{#1}}}
\newcommand\hb[2]{\genfrac{}{}{0pt}{}{#1}{#2}}
\newcommand{\tA}{\hat{A}}
\newcommand{\tB}{\hat{B}}

\newcommand{\mi}{\mathrm{i}}
\newcommand{\me}{\mathrm{e}}
\newcommand{\md}{\mathrm{d}}
\newcommand{\mr}{\mathrm{r}}
\newcommand{\pp}{\mathrm{p}}
\newcommand{\mjm}{\mathrm{jm}}
\newcommand{\mg}{\mathrm{g}}
\newcommand{\ff}{\mathbb{F}_d}

\newcommand{\vj}{{\vec{j}}}

\newcommand{\ird}{\eta^\md}
\newcommand{\irr}{\eta^\mr}
\newcommand{\irp}{\eta^\pp}
\newcommand{\irjm}{\eta^\mjm}
\newcommand{\irg}{\eta^\mg}
\renewcommand{\leq}{\leqslant}
\renewcommand{\le}{\leqslant}
\renewcommand{\geq}{\geqslant}
\renewcommand{\ge}{\geqslant}
\renewcommand{\arraystretch}{1.4}
\DeclareMathOperator{\JM}{\bf{JM}}
\DeclareMathOperator{\POVM}{\bf{POVM}}
\DeclareMathOperator{\tr}{tr}
\DeclareMathOperator{\Tr}{Tr}
\DeclareMathOperator{\Sp}{Sp}

\theoremstyle{definition}
\newtheorem{defn}{Definition}
\theoremstyle{plain}

\newtheorem{ctrex}{Counterexample}

\usepackage[toc,page]{appendix}

\newcommand{\nocontentsline}[3]{}
\newcommand{\tocless}[2]{\bgroup\let\addcontentsline=\nocontentsline#1{#2}\egroup}

\makeatletter
\def\l@subsubsection#1#2{}
\renewcommand*{\@fnsymbol}[1]{\ensuremath{\ifcase#1\or \dagger\or \ddagger\or 
    \mathsection \or ~\or \|\or **\or \dagger\dagger
\or \ddagger\ddagger \else\@ctrerr\fi}}

\makeatother

\begin{document}
\title{Incompatibility robustness of quantum measurements: a unified framework}
\author{S{\'e}bastien Designolle$^{\ast,}$}
\email{sebastien.designolle@unige.ch}
\affiliation{Group of Applied Physics, University of Geneva, 1211 Geneva, Switzerland}
\author{M{\'a}t{\'e} Farkas$^{\ast,}$}
\email{mate.frks@gmail.com}
\affiliation{Institute of Theoretical Physics and Astrophysics, National Quantum Information Centre, Faculty of Mathematics, Physics and Informatics, University of Gdansk, 80-952 Gdansk, Poland}
\author{J\k{e}drzej Kaniewski}
\email{jkaniewski@fuw.edu.pl}
\affiliation{Center for Theoretical Physics, Polish Academy of Sciences, Al.~Lotnik{\'o}w 32/46, 02-668 Warsaw, Poland}
\affiliation{Faculty of Physics, University of Warsaw, Pasteura 5, 02-093 Warsaw, Poland}
\thanks{$\!\!\!\!\!^\ast$ These authors contributed equally to this work.}
\date{\today}

\begin{abstract}
  In quantum mechanics performing a measurement is an invasive process which generally disturbs the system.
  Due to this phenomenon, there exist incompatible quantum measurements, i.e., measurements that cannot be simultaneously performed on a single copy of the system.
  It is then natural to ask what the most incompatible quantum measurements are.
  To answer this question, several measures have been proposed to quantify how incompatible a set of measurements is, however their properties are not well-understood.
  In this work, we develop a general framework that encompasses all the commonly used measures of incompatibility based on robustness to noise.
  Moreover, we propose several conditions that a measure of incompatibility should satisfy, and investigate whether the existing measures comply with them.
  We find that some of the widely used measures do not fulfil these basic requirements.
  We also show that when looking for the most incompatible pairs of measurements, we obtain different answers depending on the exact measure.
  For one of the measures, we analytically prove that projective measurements onto two mutually unbiased bases are among the most incompatible pairs in every dimension.
  However, for some of the remaining measures we find that some peculiar measurements turn out to be even more incompatible.
\end{abstract}

\maketitle

\tableofcontents

\section{Introduction}

It is well-known that the concept of a measurement in quantum physics challenges our everyday intuition.
In a classical theory objects have properties, whether we look at them or not, and a measurement simply reveals to us their pre-existing values.
In quantum mechanics, on the other hand, performing a measurement is an invasive process, which necessarily disturbs the state (except for some special cases).
Moreover, even if we have complete knowledge about the system, we can only predict the probabilities of different outcomes, which can be computed using the Born rule.
An intriguing consequence of the quantum formalism is the existence of measurements that are \emph{incompatible}, i.e., that cannot be measured simultaneously given only one copy of the system.
The best known example consists of the position and momentum of a quantum mechanical particle, which cannot be measured simultaneously with arbitrary precision.

In this work we study the incompatibility of measurements with a finite number of outcomes.
These measurements assign to each physical state $\rho$ a discrete probability distribution $\{p_a(\rho)\}_a$, whose elements we interpret as the probability of outcome $a$ on the state $\rho$.
We say that two measurements are \emph{compatible} (or \emph{jointly measurable}) if there exists a single measurement, referred to as the \emph{parent measurement}, that is able to universally replace the two \cite{Lud54,BLPY16}.
More specifically, on any state the outcome probabilities of both measurements can be recovered from the outcome probabilities of the parent measurement.
Therefore, the two measurements can be performed simultaneously by performing the parent measurement.
If such a parent measurement does not exist, we say that the measurements are \emph{incompatible} (or \emph{not jointly measurable}).
We remark here that other notions of compatibility, such as commutativity, non-disturbance and coexistence, are also used in the literature \cite{Lud54,HW10}; let us for completeness briefly explain how they are related.
Commutativity of a measurement pair implies non-disturbance, which in turn implies joint measurability, which then implies coexistence.
Moreover, it is known that none of the converse implications hold in general, therefore these notions are strictly distinct \cite{RRW13}.
In this work we focus solely on the notion of joint measurability, because the existence (or not) of a parent measurement has a clear operational meaning.
Therefore,
throughout the present paper we use the terms {``(in)compatibility''} and {``(non-)joint measurability''} interchangeably.
It is important to notice that whenever two measurements are compatible, they cannot be used to produce quantum advantage in tasks like Bell nonlocality \cite{WPF09} or Einstein--Podolsky--Rosen steering \cite{QVB14,UBGP15}.
Moreover, it was recently shown that joint measurability is equivalent to a specific notion of classicality, namely, preparation non-contextuality \cite{TU19,GQA19}.
Hence, one may think of compatible measurements as ``classical'', and incompatible measurements as a resource for the above tasks. Therefore, it is of fundamental importance to characterise and understand the structure of incompatible measurements.

What is particularly important is to go beyond the dichotomy of compatible and incompatible measurements, and quantify \emph{to what extent} a pair of measurements is incompatible.
A natural framework for this quantification, often used in the literature, is to define measures based on robustness to noise.
Briefly speaking, robustness-based measures of incompatibility quantify the minimal amount of noise that needs to be added to a pair of measurements to make them compatible.
The more noise is required, the more incompatible the measurements are.
Note that measures of this type are directly relevant to experiments, because in real-world implementations measurements are always noisy, due to inevitable experimental imperfections.

Robustness-based measures are also natural measures of incompatibility in the context of resource theories \cite{CFS16,Fri17}.
Here one considers a set of ``free'' objects (compatible measurements) and quantify the usefulness of ``resource'' objects (incompatible measurements) by so-called resource monotones.
While in this work we do not develop a full resource theory of incompatibility, we note that robustness-based measures are good candidates for resource monotones if they satisfy certain natural properties~\cite{HKR15,SSC19,CG19,OB19}.
In resource theories one defines ``free operations'' that do not create resource (that is, do not map compatible measurements to incompatible ones).
Properly defined resource monotones should then be monotonic under such free operations.
Once measures with the desired properties are found, the question ``what are the most incompatible pairs of measurements?'' is well-defined with respect to each of these measures.

Several robustness-based measures have been proposed in the literature (see Ref.~\cite{HMZ16} for an introduction),
the essential difference between them being the assumed noise model.
Nevertheless, some basic properties of these measures have not been determined and little effort has been dedicated to understanding the similarities and differences among them.
In this work we make the following contributions to fill this gap.
\begin{itemize}
  \item We develop a framework in which a robustness-based measure can be defined with respect to an arbitrary noise model.
    We identify the minimum assumptions on the noise model that ensure that the resulting measure satisfies some basic requirements, i.e., we provide an explicit connection between the properties of the noise model and the desired properties of the measure.
  \item We apply our framework to study five measures already introduced in the literature in a unified fashion.
    By giving explicit counterexamples we show that some widely used measures do not satisfy certain natural properties motivated by resource theories.
  \item We show that when looking for the most incompatible pairs, we obtain different answers depending on the specific measure of incompatibility.
    For one of the measures we analytically prove that mutually unbiased bases are among the most incompatible pairs of measurements in every dimension.
    For three other measures we can explicitly show that, for dimensions larger than two, mutually unbiased bases are \emph{not} among the most incompatible pairs.
    Our study for the last measure is inconclusive.
\end{itemize}

In Section~\ref{sec:prelim} we define incompatibility robustness in a fashion that is independent of the specific noise model, introduce the natural properties that the measures should desirably satisfy and relate them to the properties of the noise model, formulate the notion of most incompatible measurement pairs, and discuss the measures' semidefinite programming formulation and how to use this formulation to derive bounds on them.
Then in Section~\ref{sec:measures} we introduce the five measures already used in the literature, illustrate them on a simple example, analyse their relevant properties, and derive new bounds on each of them.
At the end of this section we discuss the relations between the measures, apply our results to compute all the different measures for mutually unbiased bases, then summarise the main results in a compact form.
In Section~\ref{sec:most} we address the question of the most incompatible pairs of measurements under the five measures.
Finally, in Section~\ref{sec:conclusion} we summarise the new findings and pose some important open questions arising from our work.

We note here that the notion of incompatibility naturally generalises to more than two measurements, but for simplicity in the main text we restrict ourselves to pairs of measurements.
For a formal treatment of larger sets of measurements, and results regarding them, we refer the interested reader to Appendix~\ref{app:more}.

\section{Definitions and basic properties}
\label{sec:prelim}

In this section we formalise the main definitions and concepts outlined in the introduction.
We give a mathematically precise definition of (in)compatibility and of robustness-based measures for an arbitrary noise model.
Then we specify a few natural properties the measures should satisfy, and give concrete conditions on the noise model under which these are automatically fulfilled.
We also rigorously formulate the notion of ``most incompatible measurements'', and discuss how to efficiently search for them.
Finally, we introduce the notion of semidefinite programming, and how to use it to derive bounds on robustness-based measures.

\subsection{Incompatible measurements}
\label{subsec:jm}

Throughout this paper we analyse the most general model of quantum measurements, positive operator valued measures (POVMs).
For this model, we establish that the physical system lives on a $d$-dimensional Hilbert space, $\mathcal{H}\simeq\mathbb{C}^d$.
The relevant objects are all elements of the set of linear operators on this space, $\mathcal{B}(\mathbb{C}^d)$.
The state of the system is described by a positive semidefinite operator with unit trace, denoted by $\rho$.
A POVM with $n$ outcomes is a set of $n$ positive semidefinite operators, $\{A_a\}_{a=1}^n$, such that $\sum_{a=1}^nA_a = \id$, where $\id$ is the identity operator.
The probability of observing outcome $a$ is given by the Born rule, $p_a(\rho) = \tr(A_a\rho)$.
In the following, we will use the terms ``measurement'' and ``POVM'' interchangeably.

We will often refer to the following three important classes of POVMs.
\emph{Rank-one} POVMs are measurements whose elements are rank-one operators, $A_a \propto \ketbra{\varphi_a}$, where $\ketbra{\varphi_a}$ is the projector onto $\ket{\varphi_a}\in\mathbb{C}^d$.
Note that such measurements cannot have fewer elements than the dimension of the Hilbert space, that is, $n\geq d$ with the above notation.
\emph{Projective} measurements are POVMs whose elements are projectors.
Note that such measurements cannot have more non-zero elements than the dimension of the Hilbert space.
Since the set of measurements with $n$ outcomes acting on dimension $d$ is a convex set, we will talk about \emph{extremal} POVMs (in the convex geometry sense).
Recall that every POVM can be written as a convex combination of extremal POVMs
and these have been extensively studied in Ref.~\cite{DAr04}.

The ability to recover the outcome probabilities of two POVMs on any state from the statistics of a single measurement is referred to as \emph{joint measurability} and can be formulated in the following way.

\begin{defn}
  \label{def:jm}
  Given two POVMs, $\{A_a\}_{a=1}^{n_A}$ and $\{B_b\}_{b=1}^{n_B}$, we say that they are \emph{jointly measurable} (or \emph{compatible}) if there exists a POVM $\{G_{ab}\}_{a=1,b=1}^{n_A,n_B}$ such that $\sum_{b=1}^{n_B} G_{ab} = A_a$ for all $a$, and $\sum_{a=1}^{n_A} G_{ab} = B_b$ for all $b$.
  We call such a POVM a \emph{parent} measurement of $\{A_a\}_{a=1}^{n_A}$ and $\{B_b\}_{b=1}^{n_B}$.
\end{defn}
This definition captures the idea that the parent measurement provides a joint outcome distribution of the two initial measurements on every state.
It is worth pointing out that the notion of joint measurability in which the parent POVM is allowed an arbitrary (finite) outcome set and arbitrary classical post-processing turns out to be equivalent to the one above (see e.g., Ref.~\cite[Section~3.1]{HMZ16}).

We note that a parent POVM is not necessarily unique for a fixed pair of measurements~\cite{HRS08,GC18}.
It is clear that in order to recover the outcome probabilities of $A$ and $B$, one only needs to measure $G$ and add up the relevant probabilities (in the following we sometimes drop the outcome indices to refer to the POVMs, when it does not lead to confusion; this notation is to be understood as $A = \{A_a\}_{a=1}^{n_A}$).
A simple example of a jointly measurable pair is the trivial measurement pair, $\{\frac{\id}{n_A}\}_{a=1}^{n_A}$ and $\{\frac{\id}{n_B}\}_{b=1}^{n_B}$ with the parent POVM $\{\frac{\id}{n_An_B}\} _{a=1,b=1}^{n_A,n_B}$.
In fact any POVM pair with pairwise commuting measurement operators, $[A_a,B_b] = 0$ for all $a$ and $b$, is jointly measurable.
This can be seen by employing the parent POVM $G$ with elements $G_{ab} = A_aB_b$, which is guaranteed to be positive in this case.
Note that commutativity becomes necessary and sufficient if one of the two measurements is projective, see Ref.~\cite[Proposition 8]{HRS08} for a proof.

If a parent POVM does not exist, we say that $A$ and $B$ are \emph{not jointly measurable} (or \emph{incompatible}).
A standard example of incompatible $d$-outcome measurement pairs in dimension $d\ge 2$ is a pair of projective measurements onto two \emph{mutually unbiased bases} (MUBs) \cite{DEBZ10}.
These consist of rank-one projectors $A^\mathrm{MUB}=\{\ketbra{\varphi_a}\}_{a=1}^d$ and $B^\mathrm{MUB}=\{\ketbra{\psi_b}\}_{b=1}^d$ onto the orthonormal bases $\{\ket{\varphi_a}\}_{a=1}^d$ and $\{\ket{\psi_b}\}_{b=1}^d$, such that all the pairwise overlaps (moduli of inner products) are uniform: $|\braket{\varphi_a}{\psi_b}| = 1/\sqrt{d}$ for all $a,b$.
As these measurements are projective and non-commuting, they are incompatible.

In the following we will denote the set of POVM pairs with outcome numbers $n_A$ and $n_B$ in dimension $d$ by $\POVM_d^{n_A,n_B}$, and its elements by $(A,B)$.
Note that POVM pairs inherit the convex structure of POVMs (denoted by $\POVM_d^n$), therefore convex combinations of them are well-defined.
For the subset corresponding to jointly measurable pairs, we will use the notation $\JM_d^{n_A,n_B}$, but drop the indices whenever it does not lead to confusion.
Note that the set $\JM^{n_A,n_B}_d$ is a convex subset of $\POVM^{n_A,n_B}_d$: it is straightforward to verify that if $(A^0,B^0)\in\JM^{n_A,n_B}_d$ with parent POVM $G^0$, and $(A^1,B^1)\in \JM^{n_A,n_B}_d$ with parent POVM $G^1$, then $(1-p)(A^0,B^0) + p(A^1,B^1)\in \JM^{n_A,n_B}_d$ with parent POVM $(1-p)G^0 + pG^1$ for all $p\in[0,1]$.
That is, taking convex combinations preserves joint measurability.

\subsection{Incompatibility robustness}
\label{subsec:ir}

In order to talk about noisy measurements, we define what we mean by a \emph{noise model}.
\begin{defn}\label{def:noise}
  A \emph{noise model} ${\bf N}$ is a map ${\bf N}:\POVM_d^n \to \mathbb{P}(\POVM_d^n)$, where $\mathbb{P}$ is the  set of all subsets, that maps every POVM $A\in\POVM_d^n$ to a subset of all $n$-outcome POVMs in dimension $d$, that is, ${\bf N}:A\mapsto{\bf N}_A \subseteq \POVM_d^n$. We will refer to ${\bf N}_A$ as the \emph{noise set} of $A$ under this noise model.
\end{defn}
Given a noise model, we can define \emph{noisy versions} of POVMs as convex combinations of POVMs with elements of their corresponding noise sets. Specifically, if $M\in {\bf N}_A$ and $\eta \in [0,1]$, then a noisy version of $A$ with \emph{visibility} $\eta$ is the POVM
\begin{equation}\label{eqn:noisy}
  \eta A + (1-\eta)M \in \POVM_d^n.
\end{equation} 

Noise models will be crucial for our analysis, as different noise models give rise to different measures of incompatibility.
Initially, for a unified treatment of robustness based measures, we will discuss properties that do not depend on the precise choice of the noise model, and only introduce explicit choices in Section \ref{sec:measures}, where we analyse the five specific measures.

In order to apply it to incompatibility, we extend the concept of a noise model to pairs of measurements: in this case, the noise model ${\bf N}$ is a map ${\bf N}:\POVM_d^{n_A,n_B}\to\mathbb{P}(\POVM_d^{n_A,n_B})$ that maps every pair $(A,B)\in\POVM_d^{n_A,n_B}$ to its corresponding noise set, ${\bf N}:(A,B)\mapsto {\bf N}_{A,B}\subseteq \POVM_d^{n_A,n_B}$.
Note that the set ${\bf N}_{A,B}$ may actually depend on the measurements $A$ and $B$, and not simply on their dimension or number of outcomes (whenever the map ${\bf N}$ is not constant).
The simplest example of a noise model is ${\bf N}_{A,B} = \{(\{\frac{\id}{n_A}\}_{a=1}^{n_A},\{\frac{\id}{n_B}\}_{b=1}^{n_B})\}$, that maps every POVM pair to the one-element set containing only the trivial measurement pair.
On the other end of the spectrum, the largest possible choice of the noise model is ${\bf N}_{A,B} = \POVM^{n_A,n_B}_d$, mapping every POVM pair to the set of all POVM pairs.

We will now define a measure of incompatibility corresponding to an arbitrary noise model. To ensure that the measure is well-defined, we require that the map ${\bf N}$ is such that for every pair $(A,B)$ the noise set ${\bf N}_{A,B}$ contains at least one jointly measurable pair.
For any such noise model, one can define an incompatibility robustness measure for pairs of POVMs, i.e.,~the maximal visibility at which the noisy pair is still compatible.

\begin{defn}\label{def:robustness}
  Given two POVMs, $\{A_a\}_{a=1}^{n_A}$ and $\{B_b\}_{b=1}^{n_B}$ on $\mathbb{C}^d$, and a noise model ${\bf N}$, we say that the \emph{incompatibility robustness} $\eta^\ast_{A,B}$ of the pair $(A,B)$ with respect to this noise model is
  \begin{equation}\label{eq:robustness}
    \eta^\ast_{A,B} = \sup_{\hb{\eta\in[0,1]}{(M,N)\in {\bf N}_{A,B}}}\Big\{\eta~\Big|~\eta\cdot(A,B) + (1-\eta)\cdot(M,N)\in\JM_d^{n_A,n_B}\Big\}.
  \end{equation}
\end{defn}

\begin{figure}[h!]
  \centering
  \includegraphics[width=14cm]{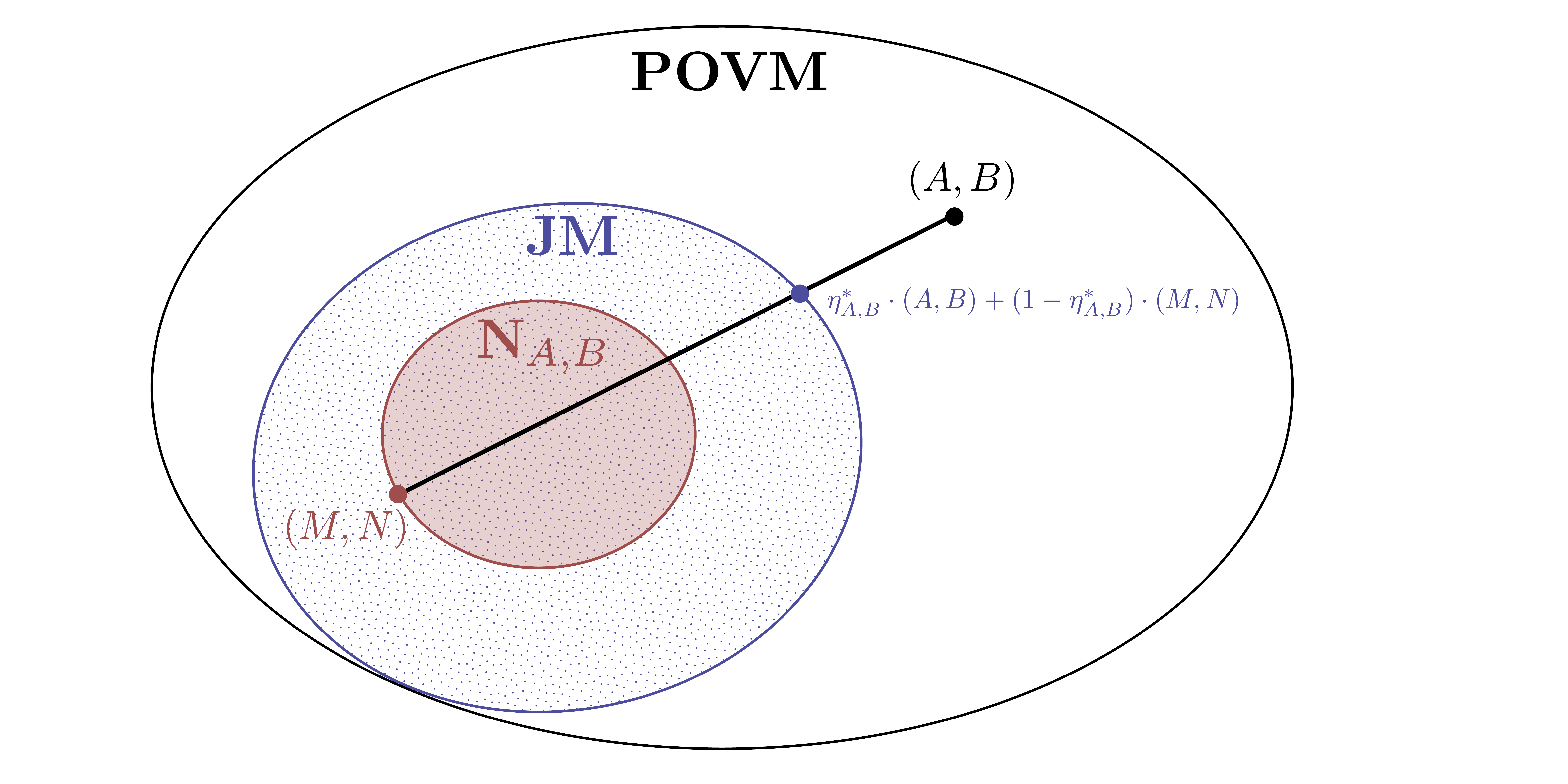}
  \caption{Schematic representation of a generic incompatibility robustness measure for a noise model which maps to closed and convex sets.
    Note that in general the noise set ${\bf N}_{A,B}$ need not be contained in the jointly measurable set $\JM$.
  One can also easily infer that the optimal noise pair $(M,N)$ must lie on the boundary of ${\bf N}_{A,B}$ and that the optimal noisy pair $\eta^\ast_{A,B}\cdot(A,B) + (1-\eta^\ast_{A,B})\cdot(M,N)$ must lie on the boundary of $\JM$.}
  \label{fig:robustness}
\end{figure}

This definition has a clear geometric interpretation, see Fig.~\ref{fig:robustness}.
Note that regardless of the choice of the noise model, $\eta^\ast_{A,B}=1$ if and only if $A$ and $B$ are jointly measurable, and that under this definition the lower $\eta^\ast_{A,B}$ is, the more incompatible the measurements are.

There are several other requirements one might impose on the noise model.
Let us briefly discuss some of these and explain what their consequences are.

\begin{itemize}
  \item If we assume that for every pair $(A, B)$, the noise set ${\bf N}_{A,B}$ is \emph{closed}, we are guaranteed that the supremum is achieved, i.e., there exists an optimal noise pair.
    In this case the supremum in Eq.~\eqref{eq:robustness} can be replaced by a maximum. Note that since we are dealing with finite-dimensional objects, it is irrelevant which topology we choose to define the notion of closedness.

  \item If we assume that for every pair $(A, B)$, the noise set ${\bf N}_{A,B}$ is \emph{convex}, we are guaranteed to find a decomposition of the form given in Eq.~\eqref{eq:robustness} for any $\eta \in [0, \eta^{\ast}_{A, B})$.
    It suffices to find a noise pair $(M', N')$ and a visibility $\eta' \ge \eta$ such that
    \begin{equation}\label{eqn:convex1}
      \eta'\cdot(A,B) + (1-\eta')\cdot(M',N') \in \JM.
    \end{equation}
    Such $(M',N')$ and $\eta'$ are guaranteed to exist, since $\eta < \eta^\ast_{A,B}$.
    Then pick $(M^{\JM{}},N^{\JM{}})\in{\bf N}_{A,B}$ such that
    \begin{equation}\label{eqn:convex2}
      (M^{\JM{}},N^{\JM{}})\in\JM,
    \end{equation}
    which is again guaranteed to exist by our fundamental assumption on the noise model.
    From the convexity of $\JM$ it follows that taking the convex combination of Eq.~\eqref{eqn:convex1} with weight $\eta/\eta'$ and Eq.~\eqref{eqn:convex2} with weight $(1-\eta/\eta')$ leads to $\eta\cdot(A,B) + (1-\eta)\cdot(M,N) \in \JM$, where
    \begin{equation}
      (M, N) = \frac{\eta}{ 1 - \eta } \cdot \frac{ 1 - \eta' }{\eta'}\cdot( M', N' ) + \left( 1 - \frac{\eta}{ 1 - \eta } \cdot \frac{ 1 - \eta' }{\eta'} \right)\cdot( M^{\JM{}}, N^{\JM{}} ),
    \end{equation}
    and the convexity of ${\bf N}_{A,B}$ ensures that $(M, N) \in {\bf N}_{A,B}$.
    Note that a looser constraint, namely that ${\bf N}_{A,B}$ is a radial set at $(M^{\JM{}},N^{\JM{}})$ (the line segments between $(M^{\JM{}},N^{\JM{}})$ and all other elements of ${\bf N}_{A,B}$ are contained in ${\bf N}_{A,B}$) is sufficient for this property.

  \item Another property one might require from the noise set is \emph{covariance with respect to unitaries}.
    Intuitively, this means that if two pairs of measurements are related by a unitary, then so should be their respective noise sets.
    More specifically, if $(A, B)$ and $(A', B')$ satisfy
    \begin{equation}
      A'_{a} = U A_{a} U^{\dagger} \quad \textnormal{and} \quad B'_{b} = U B_{b} U^{\dagger}
    \end{equation}
    for all outcomes $a$ and $b$ and for some fixed unitary $U$, then
    \begin{equation}
      (M, N) \in {\bf N}_{A,B} \iff (U M U^{\dagger}, U N U^{\dagger}) \in {\bf N}_{A',B'}.
    \end{equation}
    This property is sufficient to ensure that the resulting incompatibility robustness measure is unitarily invariant, i.e.~$\eta^{\ast}_{A, B} = \eta^{\ast}_{A', B'}$.

  \item Finally, one might require that for every choice of $(A, B)$ the corresponding noise set ${\bf N}_{A, B}$ is \emph{invariant under unitaries}, i.e.,
    \begin{equation}
      (M, N) \in {\bf N}_{A,B} \implies (U M U^{\dagger}, U N U^{\dagger}) \in {\bf N}_{A, B}
    \end{equation}
    for every unitary $U$.
    An advantage of this property is that if we assume that the noise set is convex, then we can average over the Haar measure on unitary matrices, which leads to a noise pair whose every element is proportional to the identity operator.
    We will use this property in Section~\ref{sec:mostincomp} to derive non-trivial lower bounds on the resulting incompatibility measure.
\end{itemize}

The last two properties are clearly related.
Indeed, if the noise set does not depend on the pair $(A, B)$ beyond the dimension and the outcome numbers (the map ${\bf N}$ is constant), they turn out to be equivalent.
However, in full generality these two properties are independent, i.e., we can have one without the other.
To conclude let us simply state that \emph{all the measures considered in this work satisfy all the requirements stated above}.

In Section~\ref{sec:measures}, we will replace the star in $\eta^\ast_{A,B}$ with a reference to the specific noise model in order to make clear which measure we use.
In general we are looking for noise models that give rise to measures of incompatibility that satisfy certain natural properties motivated by resource theories.

\subsection{Monotonicity}
\label{sec:monotonicity}

The natural properties we consider capture the intuition that measures of incompatibility should not decrease under operations that do not create incompatibility.
In other words, measurements should not become more incompatible under such operations.
This is well-motivated from the resource theoretic point of view, allowing for a partial order of measurement pairs based on their incompatibility robustness.

Consider an operation $\Phi:(A,B) \mapsto \Phi(A,B)$, that maps every POVM pair to another POVM pair, not necessarily preserving the dimension or the outcome numbers.
We say that this operation is \emph{joint measurability-preserving} if for all $(A,B)\in\JM$ we have that $\Phi(A,B) \in\JM$.
It is desirable that our measures are non-decreasing under such operations, that is, $\eta^\ast_{\Phi(A,B)} \ge \eta^\ast_{A,B}$ for every joint measurability-preserving operation $\Phi$.
If this inequality holds for every $(A,B)$ we say that $\eta^\ast$ is \emph{monotonic under} $\Phi$.

Whenever the joint measurability-preserving operation $\Phi$ is linear, a simple property of the noise model ${\bf N}$ implies monotonicity, namely, $\Phi( {\bf N}_{A,B}) \subseteq {\bf N}_{\Phi(A,B)}$ for all $(A,B)$.
To see this, consider a measurement pair $(A,B)$ and its corresponding noise set ${\bf N}_{A,B}$.
Following from Definition~\ref{def:robustness}, we have that
\begin{equation}
  \eta^\ast_{A,B}\cdot (A,B) + (1-\eta^\ast_{A,B})\cdot(M,N) \in \JM
\end{equation}
for some $(M,N)\in {\bf N}_{A,B}$.
Applying $\Phi$ to the left-hand side, we obtain
\begin{equation}\label{eq:etastarnoise}
  \eta^\ast_{A,B}\cdot \Phi(A,B) + (1-\eta^\ast_{A,B})\cdot\Phi(M,N) \in \JM,
\end{equation}
as $\Phi$ is linear and joint measurability-preserving.
Whenever $\Phi( {\bf N}_{A,B}) \subseteq {\bf N}_{\Phi(A,B)}$, the left-hand side of Eq.~\eqref{eq:etastarnoise} is a noisy version of $\Phi(A,B)$ with visibility $\eta^\ast_{A,B}$, which implies that $\eta^\ast_{\Phi(A,B)} \ge \eta^\ast_{A,B}$.
Therefore, if the image of the noise set under $\Phi$ is contained in the noise set of the image for every measurement pair, then $\eta^\ast$ based on this noise model is monotonic under $\Phi$.
In many cases, the stronger property $\Phi( {\bf N}_{A,B}) = {\bf N}_{\Phi(A,B)}$ holds for all $(A,B)$, and then we say that the noise model is \emph{invariant} under $\Phi$.

\begin{figure}[h!]
  \centering
  \includegraphics[width=10cm]{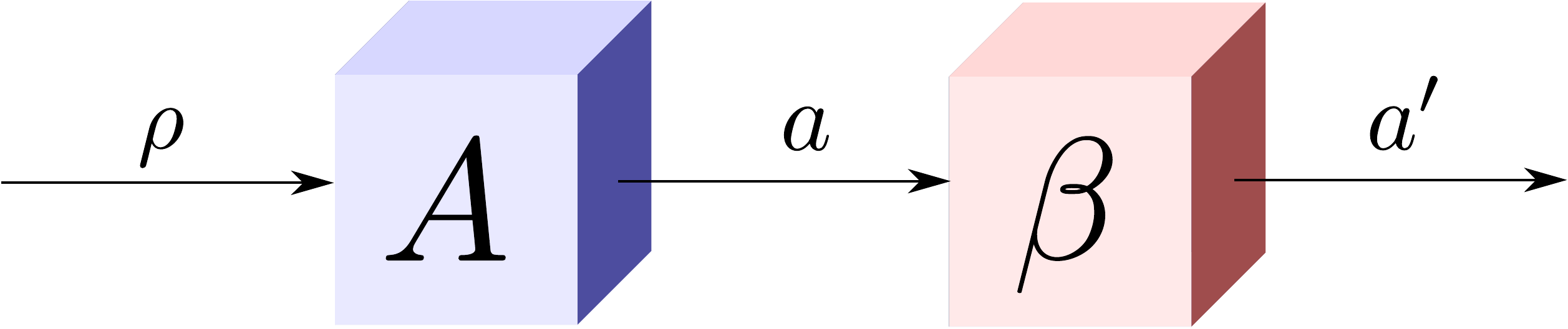}
  \caption{Schematic representation of a post-processing of a measurement.}
  \label{fig:postproc}
\end{figure}

\begin{figure}[h!]
  \centering
  \includegraphics[width=10cm]{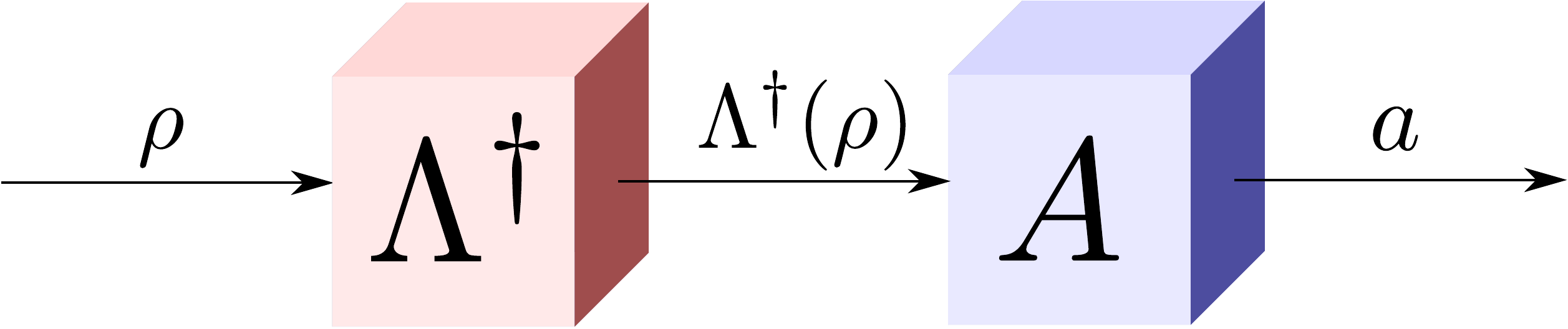}
  \caption{Schematic representation of a pre-processing of a measurement.}
  \label{fig:preproc}
\end{figure}

In this paper we will consider two natural classes of joint measurability-preserving operations, which are transformations of the measurement outputs and inputs.
The first class acts on the outputs of the measurements and is therefore called \emph{post-processings}.
The second class, on the other hand, acts on the inputs (quantum states) of the measurements, and is accordingly called \emph{pre-processings} (see Figs~\ref{fig:postproc} and \ref{fig:preproc}, respectively).
Post-processings amount to recording the outcome of the measurement and then applying a response function to it.
It can therefore be formulated in the following way.

\begin{defn}\label{def:postproc}
  A post-processing $\beta$ maps $\{A_a\}_{a=1}^{n_A}$ to $\{A^\beta_{a'}\}_{a'=1}^{n'_A}$, where
  \begin{equation}\label{eq:postproc}
    A^\beta_{a'} = \sum_{a=1}^{n_A}\beta(a'|a)A_a,
  \end{equation}
  and $\{\beta(a'|a)\}_{a'}$ is a probability distribution for every $a \in \{1,2,\ldots,n_A\}$.
\end{defn}

A post-processing is called deterministic if the probability distribution $\{\beta(a'|a)\}_{a'}$ is deterministic for all $a \in \{1,2,\ldots,n_A\}$, that is, $\beta(a'|a)\in\{0,1\}$.
If such a post-processing decreases the number of outcomes, it is referred to as \emph{coarse-graining} or \emph{binning}, e.g., the operation mapping the POVM $\{A_1,A_2,A_3\}$ to $\{A_1,A_2+A_3\}$.
What is important is that every POVM can be obtained by coarse-graining a rank-one POVM with potentially more outcomes.

Note that post-processings preserve the dimension but might change the outcome number.
For pairs $(A,B)$ the operation $\Phi^\beta: (A,B) \mapsto (A^{\beta_A}, B^{\beta_B})$ is joint measurability-preserving (note that the post-processings applied to $A$ and $B$ are independent): assume that $(A,B)\in\JM$ with parent POVM $G$.
Then it is straightforward to verify that $(A^{\beta_A}, B^{\beta_B})\in\JM$ with parent POVM $G^\beta$, where $G^\beta_{a'b'} = \sum_{ab}\beta_A(a'|a)\beta_B(b'|b)G_{ab}$.

The second class, pre-processings, amounts to first applying a quantum channel to the measured state and then performing the measurement.
Denoting the channel acting on the state by $\Lambda^\dagger$ (the dual of the map $\Lambda$), we arrive at the following definition.

\begin{defn}\label{def:preproc}
  A pre-processing $\Lambda$ maps $\{A_a\}_{a=1}^{n_A}$ to $\{A^\Lambda_a\}_{a=1}^{n_A}$, where
  \begin{equation}\label{eq:preproc}
    A^\Lambda_a = \Lambda(A_a),
  \end{equation}
  and $\Lambda:\mathcal{B}(\mathbb{C}^d)\mapsto\mathcal{B}(\mathbb{C}^{d'})$ is a completely positive unital map.
\end{defn}

Note that, for our formal treatment the unital map $\Lambda$ does only need to be positive (and not necessarily completely positive), although all the positive unital maps appearing in this work are also completely positive.

A well-known example of pre-processings is the one in Naimark's dilation theorem.
This states that for every POVM $A$ on $\mathbb{C}^d$, there exists $d'\in \mathbb{N}$, an isometry $V:\mathbb{C}^d\to\mathbb{C}^{d'}$, and a projective measurement $P$ on $\mathbb{C}^{d'}$ such that $A_a = V^\dagger P_a V$ for all $a$, that is, $ A = P^\Lambda$, where $\Lambda(.) = V^\dagger (.) V$ is a (completely) positive unital map.
That is, every POVM can be obtained by pre-processing a projective measurement acting on a potentially higher dimensional Hilbert space.

Note that pre-processings preserve the outcome number but might change the dimension.
For pairs $(A,B)$ the operation $\Phi^\Lambda: (A,B) \mapsto (A^{\Lambda}, B^{\Lambda})$ is joint measurability-preserving
(in contrast to the case of post-processing, here there is just a single pre-processing applied to both $A$ and $B$):
assume that $(A,B)\in\JM$ with parent POVM $G$.
Then it is straightforward to verify that $(A^{\Lambda}, B^{\Lambda})\in\JM$ with parent POVM $G^\Lambda$.
Note also that an incompatibility measure that is monotonic under pre-processings necessarily satisfies unitary invariance, as already mentioned in Ref.~\cite[Section~C]{HKR15}.

Finally, let us consider another natural operation that preserves joint-measurability, although it is of a different flavour than pre- and post-processings.
Namely, recall that taking convex combinations preserves joint measurability, that is, for any $(A^0,B^0)\in\JM$ and $(A^1,B^1)\in\JM$ we have that $(A^p,B^p) = (1-p)(A^0,B^0) + p(A^1,B^1) \in \JM$ for all $p\in[0,1]$ (see Section \ref{subsec:jm}).
For this reason, it is desirable that our measures do not decrease under taking convex combinations, that is, $\eta^\ast_{A^p,B^p} \ge \min\{\eta^\ast_{A^0,B^0},\eta^\ast_{A^1,B^1}\}$ for all $p \in [0, 1]$, a property sometimes referred to as \emph{quasi-concavity}.

It is easy to see that this condition holds whenever the noise model satisfies the simple property that, using the above notation, for any $(M^0,N^0)\in {\bf N}_{A^0,B^0}$ and $(M^1,N^1)\in {\bf N}_{A^1,B^1}$, we have $(M^p,N^p) = (1-p)(M^0,N^0) + p(M^1,N^1) \in {\bf N}_{A^p,B^p}$.
To see this, let us define $\eta^\ast_{\min} = \min\{\eta^\ast_{A^0,B^0},\eta^\ast_{A^1,B^1}\}$.
From the convexity of the noise set, there exist $(M^0,N^0)\in {\bf N}_{A^0,B^0}$ and $(M^1,N^1)\in {\bf N}_{A^1,B^1}$ such that $\eta^\ast_{\min}\cdot(A^0,B^0) + (1-\eta^\ast_{\min})\cdot(M^0,N^0)\in \JM$ and $\eta^\ast_{\min}\cdot(A^1,B^1) + (1-\eta^\ast_{\min})\cdot(M^1,N^1)\in \JM$ (see Section \ref{subsec:ir}).
Taking a convex combination of these two relations with coefficients $1-p$ and $p$, respectively, results in $\eta^\ast_{\min}\cdot(A^p,B^p) + (1-\eta^\ast_{\min})\cdot(M^p,N^p)\in \JM$, that is, $\eta^\ast_{A^p,B^p} \ge \min\{\eta^\ast_{A^0,B^0},\eta^\ast_{A^1,B^1}\}$.
All the noise models considered in this paper satisfy the requirement stated above and therefore the corresponding measures are non-decreasing under convex combinations.

A stronger property that is often desired is \emph{joint concavity}, which using the above notation reads $\eta^\ast_{A^p,B^p} \ge p\eta^\ast_{A^0,B^0} + (1-p)\eta^\ast_{A^1,B^1}$ (note that throughout this paper we will write ``concavity'' and ``convexity'' instead of ``joint concavity'' and ``joint convexity'', for simplicity).
However, what one naturally deduces by looking at the noise model turns out to be slightly different.
More specifically, if the noise set is convex for every pair and the noise model is a constant map we may conclude that the inverse of the measure is convex, i.e., $1/\eta^\ast_{A^p,B^p} \leq (1-p)/\eta^\ast_{A^0,B^0} + p/\eta^\ast_{A^1,B^1}$, similarly to the proof in Ref.~\cite[Proposition 2]{Haa15}.
It is easy to see that the concavity of $\eta^{\ast}$ implies that $1/\eta^{\ast}$ is convex \cite[Eq.~(3.11)]{BV04}, but the converse does not hold in general.
In fact, in Appendix~\ref{app:ctrex}, using an explicit counterexample, we show that none of the measures studied in this paper are concave.
It is common to use the measure $t^\ast=1/\eta^\ast-1$ instead of $\eta^\ast$ because it is easy to prove its convexity, and it also has the appealing property that it vanishes for every $(A,B) \in \JM$ (a property referred to as \emph{faithfulness} in Ref.~\cite{SL19} --- also note that in \cite{SOCH+19}, faithfulness, post-processing monotonicity and convexity were postulated as natural properties of any measure of incompatibility).
Moreover, whenever $\eta^\ast$ is monotonic under pre- or post-processings, then so is $t^\ast$ (with opposite relation in the inequality defining monotonicity).
Nevertheless, in the following we will study $\eta^\ast$ since it suits our purposes better and it is easily interconvertible with~$t^\ast$.

In Section~\ref{sec:measures}, we will investigate the properties introduced above for each specific measure.
As all these measures are quasi-concave and none of them are concave, we will only explicitly address pre- and post-processing monotonicity of $\eta^\ast$, and convexity of the corresponding inverse measure, $t^\ast$.

\subsection{Most incompatible measurements}
\label{sec:mostincomp}

For any given measure of incompatibility, one can ask what the most incompatible pairs of POVMs are.
To make this question well-defined, we introduce the following quantity.
\begin{defn}
  \label{def:chi}
  Given a measure of incompatibility, $\eta^\ast$, we define $\chi^\ast(d;n_A,n_B)$ to be its lowest possible value for dimension $d$ and outcome numbers $n_A$ and $n_B$.
  \begin{equation}\label{eq:xi}
    \chi^\ast(d;n_A,n_B) = \min\left\{\eta^\ast_{A,B}~|~(A,B)\in\POVM^{n_A,n_B}_d\right\}.
  \end{equation}
\end{defn}
The minimum in this definition is justified, as the set $\POVM^{n_A,n_B}_{d}$ is closed and bounded.
For a fixed measure this definition yields a real number from the range $[0, 1]$ for all positive integers $d, n_{A}, n_{B}$.
Sometimes, however, we might be interested in less detailed information.
We might just ask the question ``what are the most incompatible measurement pairs in dimension $d$?'', regardless of the outcome numbers, leading to the quantity
\begin{equation}
  \chi^\ast(d) = \inf_{n_{A}, n_{B}} \chi^\ast(d;n_A,n_B),
\end{equation}
where the infimum is taken over positive integers and it is not clear whether $\chi^\ast(d)$ is achieved for any finite $n_{A}$ and $n_{B}$.
Alternatively, we might only fix the outcome numbers, leading to $\chi^\ast(n_A,n_B)$, or fix neither the dimension nor the outcome numbers, leading to $\chi^\ast$.

One might wonder whether a non-trivial lower bound on $\chi^\ast$ can be derived based only on the previously assumed property of the noise model, namely, that for every POVM pair the corresponding noise set contains at least one jointly measurable pair, but this turns out not to be the case.
For every pair of incompatible measurements $(A, B)$ we can choose the noise set to contain a single jointly measurable pair with the property that the interior of the line segment connecting $(A,B)$ and the noise pair lies outside the jointly measurable set.
Clearly, in this case $\eta^{\ast}_{A, B} = 0$ for all incompatible pairs $(A,B)$, and $\eta^\ast$ defined through this construction is just the indicator function of joint measurability.

However, a mild additional assumption on the noise model allows us to get a non-trivial lower bound on $\chi^\ast$.
Suppose that for every incompatible pair $(A, B)$ there exists a valid noise pair $(M, N)$ such that the measurement operators of $A$ commute with those of $N$ and similarly the measurement operators of $B$ commute with those of $M$.
Then, the POVM given by
\begin{equation}
  G_{ab} = \frac{1}{2} ( A_{a} N_{b} + M_{a} B_{b} )
\end{equation}
is a valid parent POVM for $\frac12(A+M)$ and $\frac12(B+N)$, therefore it ensures that $\eta^{\ast}_{A, B} \geq \frac{1}{2}$, and we conclude that $\chi^\ast\ge\frac12$.
Clearly, the above condition is fulfilled whenever we are guaranteed to find a noise pair where all the elements are proportional to the identity (a direct consequence of the unitary invariance property discussed in Section~\ref{subsec:ir}).
This is the case for all the measures that we study.

To make the search for the most incompatible pairs of measurements efficient, it is crucial to identify operations under which the measure is monotonic, as it significantly shrinks the set over which we need to optimise.
Specifically, if we want to compute $\chi^\ast(d;n_A,n_B)$ and we deal with a measure that is non-decreasing under convex combinations, we only need to consider pairs of extremal measurements.
If our goal is to compute $\chi^\ast(d)$, i.e., we do not care about the number of outcomes, and our measure is monotonic under post-processings, we do not need to consider measurement pairs that are post-processings of another pair.
Since every POVM can be written as a post-processing (coarse-graining) of some rank-one POVM with possibly more outcomes, for post-processing monotonic measures the value $\chi^\ast(d)$ can be found by searching only over rank-one measurements.
Similarly, if we aim to compute $\chi^\ast(n_A,n_B)$, i.e., we do not care about the dimension, and our measure is monotonic under pre-processings, we do not need to consider measurement pairs that are pre-processings of another pair.
Due to Naimark's dilation theorem, every POVM can be obtained by pre-processing a projective measurement that possibly acts on a higher dimensional space, therefore projective measurements achieve $\chi^\ast(n_A,n_B)$ for pre-processing monotonic measures.

\subsection{Semidefinite programming}
\label{sec:pre_sdp}

It is clear from Eq.~\eqref{eq:robustness} that incompatibility robustness measures are defined through an optimisation problem.
The class of optimisation problems that arises in our case is called semidefinite programming and can be seen as a generalisation of linear programming \cite{BV04}.
A semidefinite program (SDP) is an optimisation problem whose optimisation variables are matrices, and whose objective function and constraints are linear functions of these variables.
The constraints can be either matrix equalities or matrix inequalities (recall that for matrices the inequality $A \geq B$ is equivalent to $A - B$ being a positive semidefinite matrix).
For every SDP, later referred to as the \emph{primal}, another SDP, called the \emph{dual}, can be defined such that its solution bounds the primal one.
In this paper the primal SDP is a maximisation problem and the dual SDP is a minimisation problem whose solution upper bounds the primal solution.
In all the examples that we study in this work, the solutions of these two SDPs in fact coincide, as we will see in Section~\ref{sec:ird_def}.
Thanks to this feature, it is possible to efficiently solve such SDPs on a computer, which gives us a tool to study incompatibility robustness measures \emph{numerically}.
This tool we often employed using the MATLAB computing environment together with the YALMIP~\cite{Lof04}, SDPT3~\cite{TTT99} and MOSEK~\cite{mosek} optimisation toolboxes.
However, the main objective of our work is to study these measures \emph{analytically}.
In order to do so, we find \emph{feasible points} for the SDPs, that is, assignments of variables that satisfy all the constraints, but that are not necessarily optimal.
By finding feasible points for the primal and dual problems, we obtain lower and upper bounds, respectively, on the value of the optimisation problem.
In the next two sections we introduce objects that will come in useful for finding such feasible points.

\subsubsection{Lower bounds}
\label{sec:pre_low}

Feasible points for the primal SDP lead to lower bounds on the incompatibility robustness.
For a fixed pair $(A,B)$ feasible points correspond to a noise pair $(M,N)$, a visibility $\eta$, and a parent POVM $G$ for ${\eta\cdot(A,B) + (1-\eta)\cdot(M,N)}$, all of these satisfying the constraints of the SDP.
That is, the noise pair should satisfy $(M,N)\in {\bf N}_{A,B}$, and the visibility must be in the range $\eta\in[0,1]$.
Crucially, the parent POVM $G$ should give $\eta A + (1-\eta)M$ and $\eta B + (1-\eta)N$ as marginals (which also guarantees its proper normalisation), and all its measurement operators should be positive semidefinite.
In order to find feasible parent POVMs satisfying these properties, we introduce an ansatz solution.
This ansatz encompasses all possible choices of the parent POVM elements that are linear combinations of the elements of $A$ and $B$, their square-roots, and products thereof, such that the normalisation of the parent POVM is ensured.
Namely, let
\begin{equation}
  \label{eqn:ansatz_low}
  G_{ab}\propto\,\{A_a,B_b\} + (\alpha_{b}A_a + \beta_{a}B_b) + \gamma_{ab}\id + \delta(A_a^\frac12 B_b A_a^\frac12 + B_b^\frac12 A_a B_b^\frac12)
\end{equation}
for some real parameters $\alpha_{b}, \beta_{b}$, $\gamma_{ab}$ and $\delta$.
It is clear then that $\sum_{ab}G_{ab}\propto\id$.

In this construction the anticommutator term plays a crucial role.
When the measurement operators of the two POVMs commute, i.e., we have $A_{a} B_{b} = B_{b} A_{a}$ for all $a$ and $b$, the anticommutator is guaranteed to be positive semidefinite.
We can therefore set $G_{ab} = \frac12\{ A_{a}, B_{b} \}$, which is a valid parent POVM for $A$ and $B$.
For non-commuting measurement operators, however, the anticommutator might have some negative eigenvalues for which the remaining terms are supposed to compensate.
Note that the same construction for parent POVMs has recently been used in Ref.~\cite{CCT19}.

For a pair of rank-one POVMs checking the positivity of Eq.~\eqref{eqn:ansatz_low} becomes analytically tractable: in this case
we can write the operator as a direct sum of an operator acting on the two-dimensional subspace spanned by the eigenvectors of $A_{a}$ and $B_{b}$, and a multiple of the identity on the orthogonal subspace (which is non-trivial for $d\ge3$).
This allows us to explicitly compute the eigenvalues and check positivity.
For this reason, for our methods to work efficiently and provide tight bounds, it is extremely important that the measure we study is monotonic under post-processings.
This is because in this case it is enough to look at rank-one POVMs in order to find the most incompatible pairs, and the robustness of any POVM pair can be bounded by the robustness of their rank-one decompositions.

Note that computing the marginals of the POVM in Eq.~\eqref{eqn:ansatz_low} is also easy in general, except for the terms multiplying the parameter $\delta$.
However, for most constructions we will choose $\delta=0$, and only include this term in a special (albeit very important) case.

As an example, let us present a known result initially presented for pairs in Ref.~\cite{HSTZ14} and then generalised to arbitrary number of measurements \cite{HKRS15,HMZ16,CHT18}.
The idea is to try to perform two measurements simultaneously by duplicating the input state and then feeding each measurement with one of the copies.
By virtue of the no-cloning theorem, the duplication process cannot be perfect.
Thanks to a duality between noiseless measurements acting on noisy states and noisy measurements acting on noiseless states, one can obtain a parent POVM from this procedure
\begin{equation}
  \label{eqn:cloning_povm}
  G_{ab}=\frac{1}{2(d+1)}\Big[\{A_a,B_b\} + \tr(B_b)A_a + \tr(A_a)B_b\Big],
\end{equation}
which is indeed of the form \eqref{eqn:ansatz_low}.
The positivity of $G_{ab}$ defined in this way follows straightforwardly from the fact that $[A_a/\tr(A_a)+B_b/\tr(B_b)]^2 \geq 0$ (we assume that $\tr A_{a} \tr B_{b} > 0$; the other cases are trivial).
This parent POVM gives rise to a universal lower bound on some measures, see Eq.~\eqref{eqn:cloning}.

\subsubsection{Upper bounds}
\label{sec:pre_up}

In order to derive upper bounds on incompatibility robustness measures, we need to find feasible points for the dual SDPs.
These SDPs have a similar structure for all the different measures that we study in this work, and therefore some quantities will often appear in the upper bounds.
For this reason, we define them here:
\begin{equation}
  \label{eqn:f_lambda}
  f=\sum\limits_a\frac{\tr A_a^2}{d}+\sum\limits_b\frac{\tr B_b^2}{d}\quad\text{and}\quad\lambda=\max\limits_{a,b}\Big\{\max \Sp\big(A_a+B_b\big)\Big\},
\end{equation}
where $\Sp(M)$ is the spectrum of the operator $M$ (note that $A_a+B_b$ is always positive semidefinite).
It is easy to see that $f \leq 2$ and the inequality is saturated if and only if both measurements are projective.
We will also need the following four quantities:
\begin{equation}
  \label{eqn:g}
  \begin{gathered}
    g^\md=\sum\limits_a\left(\frac{\tr A_a}{d}\right)^2+\sum\limits_b\left(\frac{\tr B_b}{d}\right)^2,\quad g^\mr=\frac{1}{n_A}+\frac{1}{n_B},\\ g^\pp=\min\limits_a\frac{\tr A_a}{d}+\min_b\frac{\tr B_b}{d},\quad\text{and}\quad g^\mjm=\min\limits_{a,b}\Big\{\min \Sp\big(A_a+B_b\big)\Big\}.
  \end{gathered}
\end{equation}
Note that $g^\md=g^\mr=g^\pp=2/d$ whenever both measurements are rank-one projective.

\subsection{Example}
\label{sec:pre_ill}

We will compute all the studied incompatibility robustness measures for a pair of rank-one projective qubit measurements parametrised as
\begin{equation}
  \label{eqn:runex}
  A_{a}(\theta) = \frac{1}{2} \big[ \id + (-1)^{a} ( \cos \theta \, \sigma_{z} + \sin \theta \, \sigma_{x} ) \big]\quad\text{and}\quad B_{b}(\theta) = \frac{1}{2} \big[ \id + (-1)^{b} ( \cos \theta \, \sigma_{z} - \sin \theta \, \sigma_{x} ) \big],
\end{equation}
where $\sigma_z$ and $\sigma_x$ are the Pauli $Z$ and $X$ matrices, $\theta \in [0, \pi/4]$ and $a,b=1,2$.
Note that we choose the angle $\theta$ to be half of the angle between the Bloch vectors of the two measurements.
For this pair of rank-one projective measurements, we can compute the different parameters defined in Eqs~\eqref{eqn:f_lambda} and \eqref{eqn:g}, namely, $f=2$, $\lambda=1+\cos\theta$, $g^\md=g^\mr=g^\pp=1$, and $ g^\mjm=1-\cos\theta$.
In the following, when discussing any measure of incompatibility for this pair, we will use $\eta^\ast_\theta$ as a shorthand for $\eta^\ast_{A(\theta),B(\theta)}$.
We will also make use of the following compact notation to write down the primal and dual variables:
\begin{equation}
  G=\begin{bmatrix}G_{11} & G_{12}\\ G_{21} & G_{22}\end{bmatrix}
  \quad\text{and}\quad
  (X,Y)=\left(\begin{bmatrix} X_1\\ X_2 \end{bmatrix}, \begin{bmatrix} Y_1\\ Y_2 \end{bmatrix}\right),
  \label{eqn:notation}
\end{equation}
where the elements $G_{ab}$, $X_a$ and $Y_b$ are $2\times2$ Hermitian matrices.

\section{Five relevant measures}
\label{sec:measures}

In this section we introduce five different explicit noise models, which give rise to five different robustness-based measures of incompatibility that are commonly used in the literature.
For each measure we write down both the primal and the dual SDPs, analyse their desired properties, illustrate their computation on a pair of rank-one projective qubit measurements, and derive explicit lower and upper bounds on them.
A compact summary of the main results can be found at the end of this section in Table~\ref{tab:magic}.

\subsection{Incompatibility depolarising robustness}
\label{sec:ird}

\subsubsection{Definition and properties}
\label{sec:ird_def}

In this case the noise model is defined by the map
\begin{equation}
  \label{eqn:noised}
  {\bf N}^\md_{A,B} = \left\{\left(\Big\{\tr (A_a) \frac{\id_d}{d}\Big\}_{a=1}^{n_A},\Big\{\tr (B_b)\frac{\id_d}{d}\Big\}_{b=1}^{n_B}\right)\right\}.
\end{equation}
The noise set depends on the specific measurements, which makes this measure different than all the other measures considered in this work.
It has been investigated in many works \cite{CHT12,HKR15,HKRS15,BQG+17,BN18,BN182,DSFB19,CCT19}, often in relation with Einstein--Podolsky--Rosen steering.
This specific type of noise has also been considered in scenarios different from incompatibility~\cite{OGWA17}.
The physical motivation is as follows: take a depolarising quantum channel $\Lambda_\eta^\dagger(.)$, which acts on states as $\Lambda_\eta^\dagger(\rho) = \eta\rho + (1-\eta)\tr(\rho)\id/d$, that is, by mixing them with white noise.
If we measure a system that has undergone such an evolution, we obtain the outcome probabilities $p(a) = \tr[A_{a}\Lambda_\eta^\dagger(\rho)] = \tr[\Lambda_\eta(A_{a})\rho]$, where $\Lambda_\eta(A_{a}) = \eta A_{a} + (1-\eta)\tr(A_{a})\id/d$ is the dual of the depolarising channel, which leads precisely to the type of noise set defined in Eq.~\eqref{eqn:noised}.

The corresponding incompatibility robustness, as introduced in Definition~\ref{def:robustness}, can be computed via the SDPs
\begin{equation}
  \ird_{A,B}=\left\{
    \begin{array}{cl}
      \max\limits_{\eta,\{G_{ab}\}_{ab}} &\eta \\
      \text{s.t.}
      &G_{ab} \geq 0, \quad \eta \leq 1 \\
      &\sum\limits_b G_{ab} = \eta A_a+(1-\eta)\tr A_a\frac{\id}{d}\\
      &\sum\limits_a G_{ab} = \eta B_b+(1-\eta)\tr B_b\frac{\id}{d} \\
  \end{array}\right.\quad=\left\{
    \begin{array}{cl}
      \min\limits_{\hb{\{X_a\}_a}{\{Y_b\}_b}} &1 +\sum\limits_a\tr(X_aA_a)+\sum\limits_b\tr(Y_bB_b)\\
      \text{s.t.}
      &X_a=X_a^\dagger,\quad Y_b=Y_b^\dagger,\quad X_a+Y_b \geq 0 \vphantom{\sum\limits_a}\\
      &1 +\sum\limits_a\tr (X_aA_a)+\sum\limits_b\tr (Y_bB_b) \\
      &\qquad\geq\sum\limits_a\frac{\tr A_a}{d}\tr X_a+\sum\limits_b\frac{\tr B_b}{d}\tr Y_b
  \end{array}\right.,
  \label{eqn:ird_sdp}
\end{equation}
where in the following the first formulation will be referred to as the primal, and the second as the dual.
The primal variables $G_{ab}$ and $\eta$ are simply the measurement operators of the parent POVM and the visibility, respectively.
The dual variables $X_{a}$ and $Y_{b}$ are Lagrange multipliers corresponding to the primal equality constraints.
Note that the normalisation of $G$ is not enforced as it follows from the other constraints.
For an explicit derivation of the dual problem, see Ref.~\cite[Appendix~A]{DSFB19}.
Slater's theorem states that whenever a strictly feasible point (a point satisfying all the constraints strictly) exists for either the primal or the dual, the duality gap is zero, thus the primal and dual solutions coincide \cite{BV04}.
In this case, we can take $X_a=Y_b=\delta\,\id$, which is a strictly feasible point of the dual for sufficiently large $\delta$.
Thus, the theorem applies and justifies the equality between the two problems in Eq.~\eqref{eqn:ird_sdp}.
Similar arguments apply to all pairs of primal-dual SDPs that we discuss in this work.

As the noise set ${\bf N}^\md_{A,B}$ defined in Eq.~\eqref{eqn:noised} is invariant under post-processings by linearity of the trace, it follows from Section~\ref{sec:monotonicity} that $\ird$ is monotonic under post-processings.
It turns out, however, that $\ird$ does not satisfy the other two natural properties introduced in Section~\ref{sec:monotonicity}, namely monotonicity under non trace-preserving pre-processings and convexity of the inverse;
see Appendix~\ref{app:ctrex} for counterexamples.
Note that the monotonicity under pre-processings was incorrectly claimed in Ref.~\cite[Proposition 2]{HKR15}.

\subsubsection{Example}
\label{sec:ird_ill}

From a result by Busch \cite[Theorem 4.5]{Bus86} on the joint measurability of pairs of two-outcome qubit measurements, also rephrased by Uola et al.~more recently \cite[Section~III C]{ULMH16}, we get
\begin{equation}
  \ird_\theta=\frac{1}{\cos\theta+\sin\theta}.
  \label{eqn:ird_ill}
\end{equation}
This value is plotted in Fig.~\ref{fig:runex} together with the other measures.
For completeness and later reference, we give optimal solutions to both the primal and the dual stated in Eq.~\eqref{eqn:ird_sdp}
\begin{equation}
  G=\frac{1}{\cos\theta+\sin\theta}
  \begin{bmatrix}
    \cos\theta\,\frac{\id-\sigma_z}{2}&\sin\theta\,\frac{\id-\sigma_x}{2}\\
    \sin\theta\,\frac{\id+\sigma_x}{2}&\cos\theta\,\frac{\id+\sigma_z}{2}
  \end{bmatrix}
  \quad\text{and}\quad
  (X,Y)=\frac{1}{4(\cos\theta+\sin\theta)}
  \left(\begin{bmatrix}
      \id+(\sigma_z+\sigma_x)\\
      \id-(\sigma_z+\sigma_x)
    \end{bmatrix},
    \begin{bmatrix}
      \id+(\sigma_z-\sigma_x)\\
      \id-(\sigma_z-\sigma_x)
  \end{bmatrix}\right),
\end{equation}
where we have used the notation introduced in Eq.~\eqref{eqn:notation}.

\subsubsection{Lower bound}
\label{sec:ird_low}

As mentioned before, a lower bound on $\ird$ is already known \cite{HSTZ14,HKRS15,HMZ16,CHT18}.
The parent POVM given in Eq.~\eqref{eqn:cloning_povm} is indeed a feasible point for the primal in Eq.~\eqref{eqn:ird_sdp} together with
\begin{equation}
  \label{eqn:cloning}
  \eta=\frac12\left(1+\frac{1}{d+1}\right).
\end{equation}

For a pair $(A,B)$ of rank-one measurements in dimension $d \geq 2$, this bound can be improved.
Let us introduce a feasible point for the primal in Eq.~\eqref{eqn:ird_sdp} with $G$ of the form \eqref{eqn:ansatz_low}, where
\begin{equation}
  \label{eqn:ird_low_ansatz}
  \begin{pmatrix}\alpha_{b}\\\beta_{a}\end{pmatrix}=\frac{-2+\sqrt{d^2+4d-4}}{d}\begin{pmatrix}\tr B_b\\\tr A_a\end{pmatrix},\quad\gamma_{ab}=\left(\frac{d+2-\sqrt{d^2+4d-4}}{2d}\right)^2\tr A_a\tr B_b,\quad\text{and}\quad \delta=0.
\end{equation}
For a proof that this leads to valid measurement operators $G_{ab}$ and for a measurement-dependent refinement we refer the reader to Appendix~\ref{app:ird_low}.
This construction gives a lower bound on $\ird$ for all pairs of rank-one measurements.
However, since the measure is monotonic under post-processings, the bound is actually universal, i.e., for an arbitrary pair $(A, B)$ of measurements in dimension $d$ we have
\begin{equation}
  \label{eqn:ird_low}
  \ird_{A,B}\geq \frac{d-2+\sqrt{d^2+4d-4}}{4(d-1)}.
\end{equation}
Importantly, this bound turns out to be strictly better than Eq.~\eqref{eqn:cloning}, which was the best lower bound known so far.

\subsubsection{Upper bound}
\label{sec:ird_up}

Following the idea used in Ref.~\cite{DSFB19}, we provide a valid assignment of the dual variables $X_{a}$ and $Y_{b}$ for the dual problem given in Eq.~\eqref{eqn:ird_sdp} to get an upper bound on $\ird$, namely,
\begin{equation}
  \label{eqn:ird_up_ansatz}
  X_a=\frac{\frac{\lambda}{2}\,\id-A_a}{(f-g^\md)d}\quad\text{and}\quad
  Y_b=\frac{\frac{\lambda}{2}\,\id-B_b}{(f-g^\md)d}
\end{equation}
where $f$ and $\lambda$ are defined in Eq.~\eqref{eqn:f_lambda} and $g^\md$ in Eq.~\eqref{eqn:g}.
Here we implicitly assume that $f \neq g^{\md}$, but one can show that the equality $f = g^{\md}$ holds if and only if all POVM elements of $A$ and $B$ are proportional to $\id$, in which case the pair is trivially compatible (see Appendix~\ref{app:footnote13AappendixT2}).
The resulting upper bound is given by
\begin{equation}
  \ird_{A,B}\leq\frac{\lambda-g^\md}{f-g^\md}=1-\frac{f-\lambda}{f-g^\md},
  \label{eqn:ird_up}
\end{equation}
where the last equality makes clear that this upper bound is non-trivial whenever $f>\lambda$ (since $f>g^\md$ from Appendix~\ref{app:footnote13AappendixT2}).
In the following we always implicitly assume that this condition is satisfied when we discuss the various upper bounds.

\subsection{Incompatibility random robustness}
\label{sec:irr}

\subsubsection{Definition and properties}
\label{sec:irr_def}

In this case the noise model is defined by the map
\begin{equation}
  \label{eqn:noiser}
  {\bf N}^\mr_{A,B} = \left\{\left(\Big\{\frac{\id_d}{n_A}\Big\}_{a=1}^{n_A},\Big\{\frac{\id_d}{n_B}\Big\}_{b=1}^{n_B}\right)\right\},
\end{equation}
a single element containing the trivial measurement, i.e., the measurement generating a uniform distribution of outcomes regardless of the state.
It has been investigated in many works \cite{UBGP15,CS16,CHT18,BN18,BN182,CCT19}, and also in the framework of general probabilistic theories \cite{BRGK13,JP17}.

The corresponding incompatibility robustness, as introduced in Definition~\ref{def:robustness}, can be computed via the SDPs~\cite{CS16}
\begin{equation}
  \irr_{A,B}=\left\{
    \begin{array}{cl}
      \max\limits_{\eta,\{G_{ab}\}_{ab}} &\eta\\
      \text{s.t.}
      &G_{ab} \geq 0, \quad \eta \leq 1 \\
      &\sum\limits_b G_{ab} = \eta A_a+(1-\eta)\frac{\id}{n_A} \\
      &\sum\limits_a G_{ab} = \eta B_b+(1-\eta)\frac{\id}{n_B} \\
  \end{array}\right.\quad=\left\{
    \begin{array}{cl}
      \min\limits_{\hb{\{X_a\}_a}{\{Y_b\}_b}} &1 +\sum\limits_a\tr(X_aA_a)+\sum\limits_b\tr(Y_bB_b)\\
      \text{s.t.}
      &X_a=X_a^\dagger,\quad Y_b=Y_b^\dagger,\quad X_a+Y_b \geq 0 \vphantom{\sum\limits_a}\\
      &1 +\sum\limits_a\tr (X_aA_a)+\sum\limits_b\tr (Y_bB_b) \\
      &\qquad\geq\sum\limits_a\frac{1}{n_A}\tr X_a+\sum\limits_b\frac{1}{n_B}\tr Y_b
  \end{array}\right..
  \label{eqn:irr_sdp}
\end{equation}
Note that the normalisation of $G$ is not enforced as it follows from the other constraints.

As the noise set ${\bf N}^\mr_{A,B}$ defined in Eq.~\eqref{eqn:noiser} is invariant under pre-processings (recall that pre-processings are unital), it follows from Section~\ref{sec:monotonicity} that $\irr$ is monotonic under pre-processings.
Moreover, as this set is also convex and independent of the specific form of $A$ and $B$ (the map ${\bf N}^\mathrm{r}$ is constant), we know from Section~\ref{sec:monotonicity} that $1/\irr$ is convex.
However, this measure is not monotonic under non outcome number-preserving post-processings, see Appendix~\ref{app:ctrex} for a counterexample.

\subsubsection{Example}
\label{sec:irr_ill}

For rank-one projective measurements $\ird$ and $\irr$ coincide, therefore
\begin{equation}
  \label{eqn:irr_ill}
  \irr_\theta=\ird_\theta=\frac{1}{\cos\theta+\sin\theta}.
\end{equation}

\subsubsection{Lower bound}
\label{sec:irr_low}

As $\irr$ is not monotonic under post-processings, we cannot use a solution for rank-one measurements as in Section~\ref{sec:ird_low} to deduce a general lower bound.
Thus, we consider an arbitrary pair $(A,B)$ of measurements in dimension $d$ and we introduce a feasible point for the primal in Eq.~\eqref{eqn:irr_sdp} with $G$ of the form \eqref{eqn:ansatz_low}, where
\begin{equation}
  \label{eqn:irr_low_ansatz}
  \alpha_{b}=\sqrt{\frac{n_A}{n_B}},\quad\beta_{a}=\sqrt{\frac{n_B}{n_A}},\quad\gamma_{ab}=0,\quad\text{and}\quad\delta=0
\end{equation}
from which we obtain the bound
\begin{equation}
  \label{eqn:irr_low}
  \irr_{A,B}\geq\frac12\left(1+\frac{1}{\sqrt{n_An_B}+1}\right).
\end{equation}
The positivity of this parent POVM follows from
\begin{equation}
  0\leq\sqrt{n_An_B}\left(\frac{A_a}{\sqrt{n_B}}+\frac{B_b}{\sqrt{n_A}}\right)^2=\{A_a,B_b\}+\sqrt{\frac{n_A}{n_B}}A_a^2+\sqrt{\frac{n_B}{n_A}}B_b^2\leq\{A_a,B_b\}+\sqrt{\frac{n_A}{n_B}}A_a+\sqrt{\frac{n_B}{n_A}}B_b,
\end{equation}
where the last inequality is due to $A_a^2\leq A_a$ and $B_b^2\leq B_b$.

\subsubsection{Upper bound}
\label{sec:irr_up}

In the case of $\irr$ we choose the dual variables as
\begin{equation}
  \label{eqn:irr_up_ansatz}
  X_a=\frac{\frac{\lambda}{2}\,\id-A_a}{(f-g^\mr)d}\quad\text{and}\quad 
  Y_b=\frac{\frac{\lambda}{2}\,\id-B_b}{(f-g^\mr)d}
\end{equation}
where $f$ and $\lambda$ are defined in Eq.~\eqref{eqn:f_lambda} and $g^\mr$ in Eq.~\eqref{eqn:g}.
Here we implicitly assume that $f \neq g^{\mr}$, but one can show that the equality $f = g^{\mr}$ holds if and only if all POVM elements of $A$ and $B$ are proportional to $\id$, in which case the pair is trivially compatible (see Appendix~\ref{app:footnote13AappendixT2}).
The resulting upper bound is given by
\begin{equation}
  \irr_{A,B}\leq\frac{\lambda-g^\mr}{f-g^\mr}.
  \label{eqn:irr_up}
\end{equation}

\subsection{Incompatibility probabilistic robustness}
\label{sec:irp}

\subsubsection{Definition and properties}
\label{sec:irp_def}

In this case the noise model is defined by the map
\begin{equation}
  \label{eqn:noisep}
  {\bf N}^\pp_{A,B} = \left\{\bigg(\left\{p_a\,\id_d\right\}_{a=1}^{n_A},\left\{q_b\,\id_d\right\}_{b=1}^{n_B}\bigg)\Bigg|\,p_a\geq0,q_b\geq0,\sum\limits_ap_a=1=\sum\limits_bq_b\right\},
\end{equation}
where $\{p_a\}_a$ and $\{q_b\}_b$ are probability distributions.
This measure has been investigated in many works \cite{HSTZ14,HKR15,AHK+16,HMZ16,Hei16,JP17,CHT18,Jen18,BN18,BN182}, and also in the framework of general probabilistic theories \cite{BHSS12,Pla16}.

The corresponding incompatibility robustness, as introduced in Definition~\ref{def:robustness}, can be computed via the SDPs
\begin{equation}
  \irp_{A,B}=\left\{
    \begin{array}{cl}
      \max\limits_{\hb{\eta,\{G_{ab}\}_{ab}}{\{\tilde{p}_a\}_a,\{\tilde{q}_b\}_b}} &\eta \\
      \text{s.t.}
      &G_{ab} \geq 0, \quad \tilde{p}_a\geq0, \quad \tilde{q}_b\geq0\\
      &\sum\limits_a\tilde{p}_a=1-\eta=\sum\limits_b\tilde{q}_b \\
      &\sum\limits_b G_{ab} = \eta A_a+\tilde{p}_a\id\\
      &\sum\limits_a G_{ab} = \eta B_b+\tilde{q}_b\id \\
  \end{array}\right.\quad=\left\{
    \begin{array}{cl}
      \min\limits_{\hb{\{X_a\}_a,\xi}{\{Y_b\}_b,\upsilon}} &1 +\sum\limits_a\tr(X_aA_a)+\sum\limits_b\tr(Y_bB_b)\\[3pt]
      \text{s.t.}
      &X_a=X_a^\dagger,\quad Y_b=Y_b^\dagger,\quad X_a+Y_b \geq 0 \vphantom{\sum\limits_a}\\
      &1 +\sum\limits_a\tr (X_a A_a) +\sum\limits_b\tr (Y_b B_b) \geq \xi+\upsilon\\
      &\xi\geq\tr X_a,\quad\upsilon\geq\tr Y_b\vphantom{\sum\limits_a}\\
  \end{array}\right..
  \label{eqn:irp_sdp}
\end{equation}
Note that, in order to make the problem linear in its variables, we have introduced sub-normalised probability distributions $\tilde{p}_a=(1-\eta)p_a$ and $\tilde{q}_b=(1-\eta)q_b$.
Note also that the normalisation of $G$ and the constraint $\eta\leq1$ are not enforced as they follow from the other constraints.
As the noise set ${\bf N}^\pp_{A,B}$ defined in Eq.~\eqref{eqn:noisep} contains both ${\bf N}^\md_{A,B}$ of Eq.~\eqref{eqn:noised} and ${\bf N}^\mr_{A,B}$ of Eq.~\eqref{eqn:noiser}, the constraints of the primal in Eq.~\eqref{eqn:irp_sdp} are looser than the ones in Eq.~\eqref{eqn:ird_sdp} and \eqref{eqn:irr_sdp}.
By duality, the constraints of the dual in Eq.~\eqref{eqn:irp_sdp} are then tighter than the ones in Eq.~\eqref{eqn:ird_sdp} and \eqref{eqn:irr_sdp}, which can indeed be seen by plugging suitable convex combinations of the constraints $\xi\geq\tr X_a$ and $\upsilon\geq\tr Y_b$ into ${1 +\sum_a\tr (X_a A_a) +\sum_b\tr (Y_b B_b) \geq \xi+\upsilon}$.

As the noise set ${\bf N}^\pp_{A,B}$ defined in Eq.~\eqref{eqn:noisep} is invariant under pre- and post-processings (by unitality and linearity, respectively), it follows from Section~\ref{sec:monotonicity} that $\irp$ is monotonic under pre- and post-processings.
Moreover, as this set is also convex and independent of the specific form of $A$ and $B$ (the map ${\bf N}^\mathrm{p}$ is constant), we know from Section~\ref{sec:monotonicity} that $1/\irp$ is convex.
Thus, $\irp$ is the first measure that satisfies all the properties introduced in Section~\ref{sec:prelim} except for concavity.

\subsubsection{Example}
\label{sec:irp_ill}

The dual feasible points from Section~\ref{sec:ird_ill} satisfy the additional trace constraints of the dual given in Eq.~\eqref{eqn:irp_sdp}.
Thus, the measures $\ird$ and $\irp$ coincide on this family of measurements:
\begin{equation}
  \irp_\theta=\ird_\theta=\frac{1}{\cos\theta+\sin\theta}.
\end{equation}
Note, however, that $\ird$ and $\irp$ differ in general, even for rank-one projective measurement pairs (see Section~\ref{sec:higher_dim} for an explicit example).

\subsubsection{Lower bound}
\label{sec:irp_low}

Since the noise set ${\bf N}^\pp_{A,B}$ contains both ${\bf N}^\md_{A,B}$ and ${\bf N}^\mr_{A,B}$ for all $(A,B)$, lower bounds on $\ird$ and $\irr$ immediately apply to $\irp$.

\subsubsection{Upper bound}
\label{sec:irp_up}

In the case of $\irp$ we choose the dual variables as
\begin{equation}
  \label{eqn:irp_up_ansatz}
  X_a=\frac{\frac{\lambda}{2}\,\id-A_a}{(f-g^\pp)d},\quad
  Y_b=\frac{\frac{\lambda}{2}\,\id-B_b}{(f-g^\pp)d},\quad
  \xi=\max\limits_a\tr X_a,\quad\text{and}\quad\upsilon=\max\limits_b\tr Y_b,
\end{equation}
where $f$ and $\lambda$ are defined in Eq.~\eqref{eqn:f_lambda} and $g^\pp$ in Eq.~\eqref{eqn:g}.
Here we implicitly assume that $f \neq g^{\pp}$, but one can show that the equality $f = g^{\pp}$ holds if and only if all POVM elements of $A$ and $B$ are proportional to $\id$, in which case the pair is trivially compatible (see Appendix~\ref{app:footnote13AappendixT2}).
The resulting upper bound is given by
\begin{equation}
  \irp_{A,B}\leq\frac{\lambda-g^\pp}{f-g^\pp}.
  \label{eqn:irp_up}
\end{equation}

\subsection{Incompatibility jointly measurable robustness}
\label{sec:irjm}

\subsubsection{Definition and properties}
\label{sec:irjm_def}

In this case the noise model is defined by the map
\begin{equation}
  \label{eqn:noisejm}
  {\bf N}^\mjm_{A,B} = \JM_d^{n_A,n_B},
\end{equation}
the set of jointly measurable pairs of POVMs with $n_A$ and $n_B$ outcomes in dimension $d$.
To the best of our knowledge, this measure has only been considered in Ref.~\cite[Section~II C]{CS16}.

The corresponding incompatibility robustness, as introduced in Definition~\ref{def:robustness}, can be computed via the SDPs
\begin{equation}
  \irjm_{A,B}=\left\{
    \begin{array}{cl}
      \max\limits_{\eta,\hb{\{G_{ab}\}_{ab}}{\{\tilde{H}_{ab}\}_{ab}}} &\eta \\
      \text{s.t.}
      &G_{ab}\geq0,\quad\sum\limits_{ab} G_{ab} = \id,\quad\tilde{H}_{ab} \geq 0 \\
      &\sum\limits_b (G_{ab}-\tilde{H}_{ab}) = \eta A_a \\
      &\sum\limits_a (G_{ab}-\tilde{H}_{ab}) = \eta B_b \\
  \end{array}\right.\quad=\left\{
    \begin{array}{cl}
      \min\limits_{N,\hb{\{X_a\}_a}{\{Y_b\}_b}} &\tr N\\
      \text{s.t.}
      &N=N^\dagger,\quad X_a=X_a^\dagger,\quad Y_b=Y_b^\dagger \vphantom{\sum\limits_a}\\
      &N\geq X_a+Y_b \geq 0\vphantom{\sum\limits_a}\\
      &\sum\limits_a\tr(X_aA_a)+\sum\limits_b\tr(Y_bB_b)\geq1\\
  \end{array}\right..
  \label{eqn:irjm_sdp}
\end{equation}
Note that the noise POVMs do not explicitly appear in the primal problem, since optimising over jointly measurable pairs is equivalent to optimising over the parent measurement, here denoted by $H$.
To make the problem linear in its variables, we have introduced a sub-normalised parent POVM of the noise, $\tilde{H}=(1-\eta)H$.
Note also that the constraint $\eta\leq1$ is not enforced as it follows from summing up one of the marginal constraints.

In analogy with $\irp$, the measure $\irjm$ also satisfies the properties introduced in Section~\ref{sec:prelim}, namely monotonicity under pre- and post-processings, and convexity of the inverse.

\subsubsection{Example}
\label{sec:irjm_ill}

The value of this measure for a pair of rank-one projective qubit measurements is strictly higher than for the previous measures, whenever the pair is incompatible.
Specifically,
\begin{equation}
  \irjm_\theta=\frac{2}{1+\cos\theta+\sin\theta}.
  \label{eqn:irjm_ill}
\end{equation}
This value is plotted in Fig.~\ref{fig:runex} together with the other measures.
Interestingly, even for such a simple example the primal problem given in Eq.~\eqref{eqn:irjm_sdp} admits multiple optimal solutions.
More specifically, we obtain a continuous one-parameter family, which reads
\begin{equation}
  \label{eqn:jm_qubit}
  G=
  \begin{bmatrix}
    r\,\frac{\id-\sigma_z}{2}&(1-r)\,\frac{\id-\sigma_x}{2}\\
    (1-r)\,\frac{\id+\sigma_x}{2}&r\,\frac{\id+\sigma_z}{2}
  \end{bmatrix}
  ,\quad
  \tilde{H}=(1-\irjm_\theta)
  \begin{bmatrix}
    s\,\frac{\id+\sigma_z}{2}&(1-s)\,\frac{\id+\sigma_x}{2}\\
    (1-s)\,\frac{\id-\sigma_x}{2}&s\,\frac{\id-\sigma_z}{2}
  \end{bmatrix},
  \quad\text{where}\quad
  r = \irjm_\theta(s+\cos\theta)-s
\end{equation}
and $s$ is a free parameter taken from the interval $[0,1]$ to ensure the positivity of the elements of $H$.
Different values of $s$ correspond to applying noise along different axes: for $s=0$ the noise only affects the $X$ direction, while for $s = 1$ it only affects the $Z$ direction.
A feasible optimal point for the dual given in Eq.~\eqref{eqn:irjm_sdp} reads
\begin{equation}
  (X,Y)=\frac{1}{4(1+\cos\theta+\sin\theta)}
  \left(\begin{bmatrix}
      \id-(\sigma_z+\sigma_x)\\
      \id+(\sigma_z+\sigma_x)
    \end{bmatrix},
    \begin{bmatrix}
      \id-(\sigma_z-\sigma_x)\\
      \id+(\sigma_z-\sigma_x)
  \end{bmatrix}\right),
  \quad\text{and}\quad
  N=\frac{1}{1+\cos\theta+\sin\theta}\,\id.
\end{equation}

\subsubsection{Lower bound}
\label{sec:irjm_low}

Let us consider a pair $(A,B)$ of rank-one measurements in dimension $d$.
Finding a feasible point for the primal in Eq.~\eqref{eqn:irjm_sdp} is not an easy task, as we have to find two parent POVMs at once.
For $G_{ab}$, we make the same choice as for $\ird$, i.e., Eq.~\eqref{eqn:ird_low_ansatz} in Section~\ref{sec:ird_low}.
We choose the subnormalised noise POVM $\tilde{H}$ to be of the form \eqref{eqn:ansatz_low} with
\begin{equation}
  \label{eqn:irjm_low_ansatz}
  \begin{pmatrix}\alpha_{b}\\\beta_{a}\end{pmatrix}=\frac{-2-\sqrt{d^2+4d-4}}{d}\begin{pmatrix}\tr B_b\\\tr A_a\end{pmatrix},\quad\gamma_{ab}=\left(\frac{d+2+\sqrt{d^2+4d-4}}{2d}\right)^2\tr A_a\tr B_b,\quad\text{and}\quad\delta=0,
\end{equation}
which leads to
\begin{equation}
  \label{eqn:irjm_low}
  \eta=\frac{2\sqrt{d^2+4d-4}}{3d-2+\sqrt{d^2+4d-4}}\leq\irjm_{A,B}.
\end{equation}
Details about this specific point can be found in Appendix~\ref{app:irjm_low} together with a measurement-dependent refinement.
As $\irjm$ is monotonic under post-processings, this bound on pairs of rank-one measurements extends to all pairs of measurements in dimension $d$.

\subsubsection{Upper bound}
\label{sec:irjm_up}
Consider the following feasible point for the dual given in Eq.~\eqref{eqn:irjm_sdp}:
\begin{equation}
  \label{eqn:irjm_up_ansatz}
  X_a=\frac{A_a-\frac{g^\mjm}{2}\,\id}{(f- g^\mjm)d}, \quad Y_b=\frac{B_b-\frac{g^\mjm}{2}\,\id}{(f- g^\mjm)d},\quad\text{and}\quad N=\frac{\lambda- g^\mjm}{f- g^\mjm}\cdot\frac{\id}{d}
\end{equation}
where $f$ and $\lambda$ are defined in Eq.~\eqref{eqn:f_lambda} and $g^\mjm$ in Eq.~\eqref{eqn:g}.
Here we implicitly assume that $f \neq g^{\mjm}$, but one can show that the equality $f = g^{\mjm}$ holds if and only if all POVM elements of $A$ and $B$ are proportional to $\id$, in which case the pair is trivially compatible (see Appendix~\ref{app:footnote13AappendixT2}).
The above feasible point immediately implies that
\begin{equation}
  \label{eqn:irjm_up}
  \irjm_{A,B} \leq \frac{\lambda- g^\mjm}{f- g^\mjm}.
\end{equation}

\subsection{Incompatibility generalised robustness}
\label{sec:irg}

\subsubsection{Definition and properties}
\label{sec:irg_def}

In this case the noise model is defined by the map
\begin{equation}
  \label{eqn:noiseg}
  {\bf N}^\mg_{A,B} = \POVM_d^{n_A,n_B},
\end{equation}
the set of all POVM pairs with $n_A$ and $n_B$ outcomes, respectively, in dimension $d$.
To the best of our knowledge, this measure was first introduced in Ref.~\cite{Haa15} and studied further in Refs~\cite{UBGP15,CS16,KBUP17,BQG+17}.
Recently, it was given an operational meaning through state discrimination tasks \cite{CHT19,UKS+19,SSC19}.

The corresponding incompatibility robustness, as introduced in Definition~\ref{def:robustness}, can be computed via the SDPs
\begin{equation}
  \irg_{A,B}=\left\{
    \begin{array}{cl}
      \max\limits_{\eta,\{G_{ab}\}_{ab}} &\eta \\
      \text{s.t.}
      &G_{ab} \geq 0,\quad\sum\limits_{ab} G_{ab} = \id \\
      &\sum\limits_b G_{ab} \geq \eta A_a\\
      &\sum\limits_a G_{ab} \geq \eta B_b\\
  \end{array}\right.\quad=\left\{
    \begin{array}{cl}
      \min\limits_{N,\{X_a\}_a} &\tr N\\
      \text{s.t.}
      &N=N^\dagger,\quad N\geq X_a+Y_b\\
      &X_a\geq0, \quad Y_b\geq0\vphantom{\sum\limits_a}\\
      &\sum\limits_a\tr(X_aA_a)+\sum\limits_b\tr(Y_bB_b)\geq1\\
  \end{array}\right..
  \label{eqn:irg_sdp}
\end{equation}
Note that in the primal, the noise POVMs do not appear, because we can explicitly solve for these variables, which gives rise to matrix inequalities instead of equalities for the marginals.
These looser constraints give us additional freedom and allow us to employ operator inequalities.
Note also that the constraint $\eta\leq1$ is not enforced as it follows from summing up one of the marginal constraints.
The constraints in the primal in Eq.~\eqref{eqn:irg_sdp} are looser than in the primal in Eq.~\eqref{eqn:irjm_sdp}, because the noise set is larger for all measurement pairs.
In turn, the feasible set of the dual problem shrinks, as the dual constraints $X_a\geq0$ and $Y_b\geq0$ are tighter than $X_a+Y_b\geq0$.

In analogy with $\irp$ and $\irjm$, the measure $\irg$ also satisfies the properties we introduced in Section~\ref{sec:prelim}, namely monotonicity under pre- and post-processings, and convexity of the inverse.

\subsubsection{Example}
\label{sec:irg_ill}

The value of this measure for the running example is even higher than for the previous measures, specifically
\begin{equation}
  \irg_\theta=\frac{\sqrt2+1}{\sqrt2+\cos\theta+\sin\theta}.
  \label{eqn:irg_ill}
\end{equation}
This value is plotted in Fig.~\ref{fig:runex} together with the other measures.
A feasible point for the primal in Eq.~\eqref{eqn:irg_sdp} reads
\begin{equation}
  G=
  \begin{bmatrix}
    r\,\frac{\id-\sigma_z}{2}&(1-r)\,\frac{\id-\sigma_x}{2}\\
    (1-r)\,\frac{\id+\sigma_x}{2}&r\,\frac{\id+\sigma_z}{2}
  \end{bmatrix},
  \quad\text{where}\quad
  r = \frac{1-\sin\theta+(\sqrt2+1)\cos\theta}{\sqrt2(\sqrt2+\cos\theta+\sin\theta)},
\end{equation}
and for the dual,
\begin{equation}
  (X,Y)=\frac{\sqrt2}{4(\sqrt2+\cos\theta+\sin\theta)}
  \left(\begin{bmatrix}
      \id-\frac{\sigma_z+\sigma_x}{\sqrt2}\\
      \id+\frac{\sigma_z+\sigma_x}{\sqrt2}
    \end{bmatrix},
    \begin{bmatrix}
      \id-\frac{\sigma_z-\sigma_x}{\sqrt2}\\
      \id+\frac{\sigma_z-\sigma_x}{\sqrt2}
  \end{bmatrix}\right),
  \quad\text{and}\quad
  N=\frac{\sqrt2+1}{2(\sqrt2+\cos\theta+\sin\theta)}\,\id.
\end{equation}

\subsubsection{Lower bound}
\label{sec:irg_low}

For a pair $(A,B)$ of rank-one measurements in dimension $d$, let us introduce a feasible point for the primal in Eq.~\eqref{eqn:irg_sdp} with $G$ of the form \eqref{eqn:ansatz_low}, where
\begin{equation}
  \label{eqn:irg_low_ansatz}
  \begin{pmatrix}\alpha_{b}\\\beta_{a}\end{pmatrix}=\frac{1}{2\sqrt{d}}\begin{pmatrix}\tr B_b\\\tr A_a\end{pmatrix},\quad\gamma_{ab}=0,\quad\text{and}\quad\delta=\frac{\sqrt{d}}{2},
\end{equation}
so that we obtain the bound
\begin{equation}
  \label{eqn:irg_low}
  \eta=\frac12\left(1+\frac{1}{\sqrt{d}}\right)\leq\irg_{A,B}.
\end{equation}
A proof of feasibility of this specific point is given below.
For more details, see Appendix~\ref{app:irg_low} which also contains a measurement-dependent refinement.
As $\irg$ is monotonic under post-processings, this bound on pairs of rank-one measurements extends to all pairs of measurements in dimension $d$.

The novelty in Eq.~\eqref{eqn:irg_low_ansatz}, as compared to the parent POVMs used for the other measures, is the fact that $\delta$ is non-zero.
What enables us to introduce this term is the extra freedom in the primal in Eq.~\eqref{eqn:irg_sdp}, namely, the inequalities in the marginal constraints instead of equalities, which allows us to analyse the marginals for non-zero~$\delta$.

For the proof of feasibility, we write the parent POVM defined by the coefficients in Eq.~\eqref{eqn:irg_low_ansatz} as
\begin{equation}
  \label{eqn:irg_low2}
  G_{ab} = \frac{1}{4(d+\sqrt{d})}\left[\tr(B_b) A_a + \tr(A_a) B_b + 2\sqrt{d} \{A_a, B_b\} + d\left(A_a^{\frac12} B_b A_a^{\frac12} + B_b^{\frac12} A_a B_b^{\frac12}\right)\right].
\end{equation}
Since $A_{a}$ and $B_{b}$ are rank-one, we can write $A_a = \tr(A_a)P_a$ and $B_b = \tr(B_b)Q_b$ for some $P_a=\ketbra{\varphi_a}$ and $Q_b=\ketbra{\psi_b}$.
Therefore, we can rewrite \eqref{eqn:irg_low2} as
\begin{equation}
  G_{ab} = \frac{\tr(A_a)\tr(B_b)}{4(d+\sqrt{d})}\left[(P_a + \sqrt{d} P_aQ_b)^\dagger(P_a + \sqrt{d} P_aQ_b)+(Q_b + \sqrt{d}Q_bP_a)^\dagger(Q_b + \sqrt{d}Q_bP_a)\right] \geq 0,
\end{equation}
which shows that $G$ is a valid POVM.

Next we should compute its marginals.
The first one reads
\begin{equation}
  \label{eqn:irg_low3}
  \sum_b G_{ab} = \frac{1}{4(d+\sqrt{d})}\left[d A_a + \tr(A_a)\id + 4 \sqrt{d} A_a + d\left(A_a + \sum_bB_b^{\frac12} A_a B_b^{\frac12}\right)\right],
\end{equation}
where the terms are ordered as in Eq.~\eqref{eqn:irg_low2} for clarity.
Moreover, we have that for every $\ket{\xi}$,
\begin{equation}
  \begin{split}
    d\,\bra{\xi}\sum_bB_b^{\frac12} A_a B_b^{\frac12}\ket{\xi} & \left.= \sum_{b'}\tr(B_{b'})\sum_b\tr(B_b)\tr(A_a) \braket{\xi}{\psi_b} \braket{\psi_b} {\varphi_a} \braket{\varphi_a}{\psi_b} \braket{\psi_b}{\xi} \right.\\
    & \left.= \tr(A_a)\sum_{b'}\left|\sqrt{\tr(B_{b'})}\right|^2\sum_b\left|\sqrt{\tr(B_b)}
    \braket{\xi}{\psi_b}\braket{\psi_b} {\varphi_a}\right|^2 \right.\\
    & \left.\geq \tr(A_a)\left|\sum_b \tr(B_b)\braket{\xi}{\psi_b}
      \braket{\psi_b} {\varphi_a}\right|^2 = \tr(A_a)|\braket{\xi}
    {\varphi_a}|^2 = \bra{\xi}A_a \ket{\xi}, \right.
  \end{split}
\end{equation}
where we used the Cauchy--Schwarz inequality.
Therefore, $d\sum_bB_b^{1/2} A_a B_b^{1/2} \ge A_a$, which
together with $\tr(A_a)\id \ge A_a$ enables us to lower bound the marginal \eqref{eqn:irg_low3}, namely,
\begin{equation}
  \sum_b G_{ab} \ge \frac{2(d + 2\sqrt{d} + 1)}{4(d + \sqrt{d})}A_a = \frac12 \left(1+\frac{1}{\sqrt{d}}\right)A_a.
\end{equation}
By symmetry of Eq.~\eqref{eqn:irg_low2} the same conclusion holds for the second marginal, which shows that the point defined in Eqs~\eqref{eqn:irg_low_ansatz} and~\eqref{eqn:irg_low} is indeed feasible.

\subsubsection{Upper bound}
\label{sec:irg_up}

Consider the following feasible point for the dual given in Eq.~\eqref{eqn:irg_sdp}:
\begin{equation}
  \label{eqn:irg_up_ansatz}
  X_a= \frac{A_a}{fd}, \quad Y_b= \frac{B_b}{fd}, \quad\text{and}\quad N=\frac{\lambda}{f}\cdot\frac{\id}{d}
\end{equation}
where $f$ and $\lambda$ are defined in Eq.~\eqref{eqn:f_lambda}.
This immediately implies that
\begin{equation}
  \label{eqn:irg_up}
  \irg_{A,B}\leq\frac{\lambda}{f}.
\end{equation}

\subsection{Relations between the measures}

Certain inclusions between the noise sets defined in Eqs~\eqref{eqn:noised}, \eqref{eqn:noiser}, \eqref{eqn:noisep}, \eqref{eqn:noisejm}, and \eqref{eqn:noiseg}, imply an ordering of the measures.
More specifically, from
\begin{equation}
  \label{eqn:order}
  ({\bf N}_{A,B}^\md\cup{\bf N}_{A,B}^\mr) \subseteq{\bf N}_{A,B}^\pp\subseteq{\bf N}_{A,B}^\mjm\subseteq{\bf N}_{A,B}^\mg,
\end{equation}
we conclude that
\begin{equation}
  \label{eqn:order_measures}
  \max \{ \ird_{A,B},\irr_{A,B} \} \leq \irp_{A,B} \leq \irjm_{A,B} \leq \irg_{A,B}
\end{equation}
for every pair $(A,B)$.
It turns out that $\ird$ and $\irr$ are incomparable (see Appendix~\ref{app:ctrex} for an example).
A more detailed analysis allows us to prove that some of the inequalities given in Eq.~\eqref{eqn:order_measures} are in fact strict.
Specifically, in Appendix~\ref{app:relations} we derive improved relations between $\ird$ and $\irjm$, $\ird$ and $\irg$, and $\irr$ and $\irg$, which imply that for a pair of incompatible measurements $(A,B)$ the separations between these measures are strict, i.e., ${\ird_{A,B} < \irjm_{A,B}}$, ${\ird_{A,B} < \irg_{A,B}}$, and ${\irr_{A,B} < \irg_{A,B}}$.
Moreover, the examples given in Section~\ref{sec:MUB} show that in some cases $\ird$ coincides with $\irp$, as well as $\irr$ with $\irp$ and $\irjm$ with $\irg$.
The question whether the separation between $\irp$ and $\irjm$ is strict or not is left open.

\subsection{Mutually unbiased bases}
\label{sec:MUB}

We have mentioned earlier that mutually unbiased bases constitute a standard example of a pair of incompatible measurements on a $d$-dimensional system.
Indeed, they might seem like natural candidates for the most incompatible pair of measurements in dimension $d$.
In this section we show that for a pair of MUBs all the previously introduced measures can be computed analytically.
The specific values we obtain will be compared against the findings of Section~\ref{sec:most}, in which we look for the most incompatible pairs of measurements.

For a pair $(A^\mathrm{MUB},B^\mathrm{MUB})$ of projective measurements onto two MUBs in dimension $d$ (see Section~\ref{subsec:jm}), we will use $\eta^\ast_{\mathrm{MUB}}(d)$ as a shorthand for $\eta^\ast_{A^\mathrm{MUB},B^\mathrm{MUB}}$.
Note that although in higher dimensions not all pairs of MUBs are unitarily equivalent, they nevertheless give the same value for all the measures studied in this work.
Hence, for these measures the quantity $\eta^\ast_{\mathrm{MUB}}(d)$ turns out to be well-defined.

In dimension $d=2$ a pair of MUB measurements is a special case of the example introduced in Section~\ref{sec:pre_ill}, corresponding to $\theta=\pi/4$.
Therefore Eqs~\eqref{eqn:ird_ill}, \eqref{eqn:irjm_ill}, and \eqref{eqn:irg_ill} imply that
\begin{equation}
  \label{eqn:ir_mub2}
  \ird_\mathrm{MUB}(2)=\irr_\mathrm{MUB}(2)=\irp_\mathrm{MUB}(2)=\frac{1}{\sqrt2},\quad\irjm_\mathrm{MUB}(2)=2(\sqrt2-1),\quad\text{and}\quad\irg_\mathrm{MUB}(2)=\frac12\left(1+\frac{1}{\sqrt2}\right).
\end{equation}

For a pair of projective measurements onto two MUBs in dimension $d\geq3$, the parameters given in Eqs~\eqref{eqn:f_lambda} and \eqref{eqn:g} equal $f=2$, $\lambda=1+1/\sqrt{d}$, $g^\md=g^\mr=g^\pp=2/d$, and $ g^\mjm=0$.
It turns out that for MUBs the upper bounds given in Eqs~\eqref{eqn:ird_up}, \eqref{eqn:irjm_up}, and \eqref{eqn:irg_up} are actually tight.
Therefore, the only missing component is a feasible point for the primal.

For $\ird$ and $\irr$ our feasible solution consists of
\begin{equation}
  \eta = \frac12\left(1+\frac{1}{\sqrt{d} + 1}\right)
\end{equation}
and
\begin{equation}
  \label{eqn:ir_mub}
  G_{ab}=\frac{1}{2(\sqrt{d}+1)}\left(\{A_a,B_b\}+\frac{1}{\sqrt{d}}A_a+\frac{1}{\sqrt{d}}B_b\right).
\end{equation}
This parent POVM, inspired by Ref.~\cite[Section~IV]{ULMH16}, is of the form of Eq.~\eqref{eqn:ansatz_low}.
The positivity of these operators can be confirmed using the techniques presented in Appendix \ref{app:low} and let us stress that the proof crucially relies on the fact that the bases are mutually unbiased.
For $\irp$ we must explicitly include the weights and we choose them to be uniform $p_a = q_b = 1/d$ for all $a,b$.
This assignment saturates the upper bound given in Eq.~\eqref{eqn:ird_up}, which implies that
\begin{equation}
  \label{eqn:ir_mub3}
  \ird_\mathrm{MUB}(d)=\irr_\mathrm{MUB}(d)=\irp_\mathrm{MUB}(d)=\frac12\left(1+\frac{1}{\sqrt{d}+1}\right).
\end{equation}
For $\irg$ we use the same parent POVM, but the more flexible form of noise allows for higher visibility:
\begin{equation}
  \label{eqn:visibility_g_jm}
  \eta = \frac12\left(1+\frac{1}{\sqrt{d}}\right).
\end{equation}
For $\irjm$ we must supplement our solution with a sub-normalised parent POVM of the noise pair
\begin{equation}
  \label{eqn:ir_mubH}
  \tilde{H}_{ab} = \frac{1-\irjm_\mathrm{MUB}(d)}{d(d-2)}\left[\id+\frac{d}{d-1}\Big(\{A_a,B_b\}-A_a-B_b\Big)\right],
\end{equation}
which has already been used in Ref.~\cite{CHT19}, and is of the form of Eq.~\eqref{eqn:ansatz_low}.
This construction is only valid for $d \geq 3$, because for $d = 2$ the corresponding noise pair $\{(\id-A_a)/(d-1)\}_a$ and $\{(\id-B_b)/(d-1)\}_b$ is not jointly measurable (see Eq.~\eqref{eqn:jm_qubit} for a family of optimal feasible points for the primal).
In both cases the visibility given in Eq.~\eqref{eqn:visibility_g_jm} saturates the upper bounds~\eqref{eqn:irg_ill} and \eqref{eqn:irjm_ill}, respectively, which implies that for all $d \geq 3$, we have
\begin{equation}
  \label{eqn:ir_mub4}
  \irjm_\mathrm{MUB}(d)=\irg_\mathrm{MUB}(d)=\frac12\left(1+\frac{1}{\sqrt{d}}\right).
\end{equation}
Note that the value $\irg_\mathrm{MUB}(d)$ was already derived in~\cite{Haa15}. Also notice that Eq.~\eqref{eqn:ir_mub4} together with Eq.~\eqref{eqn:irg_low} implies that MUBs are among the most incompatible measurement pairs with respect to $\irg$ in every dimension.

\subsection{Summary}

\begin{figure}[h!]
  \centering
  \includegraphics[width=12cm]{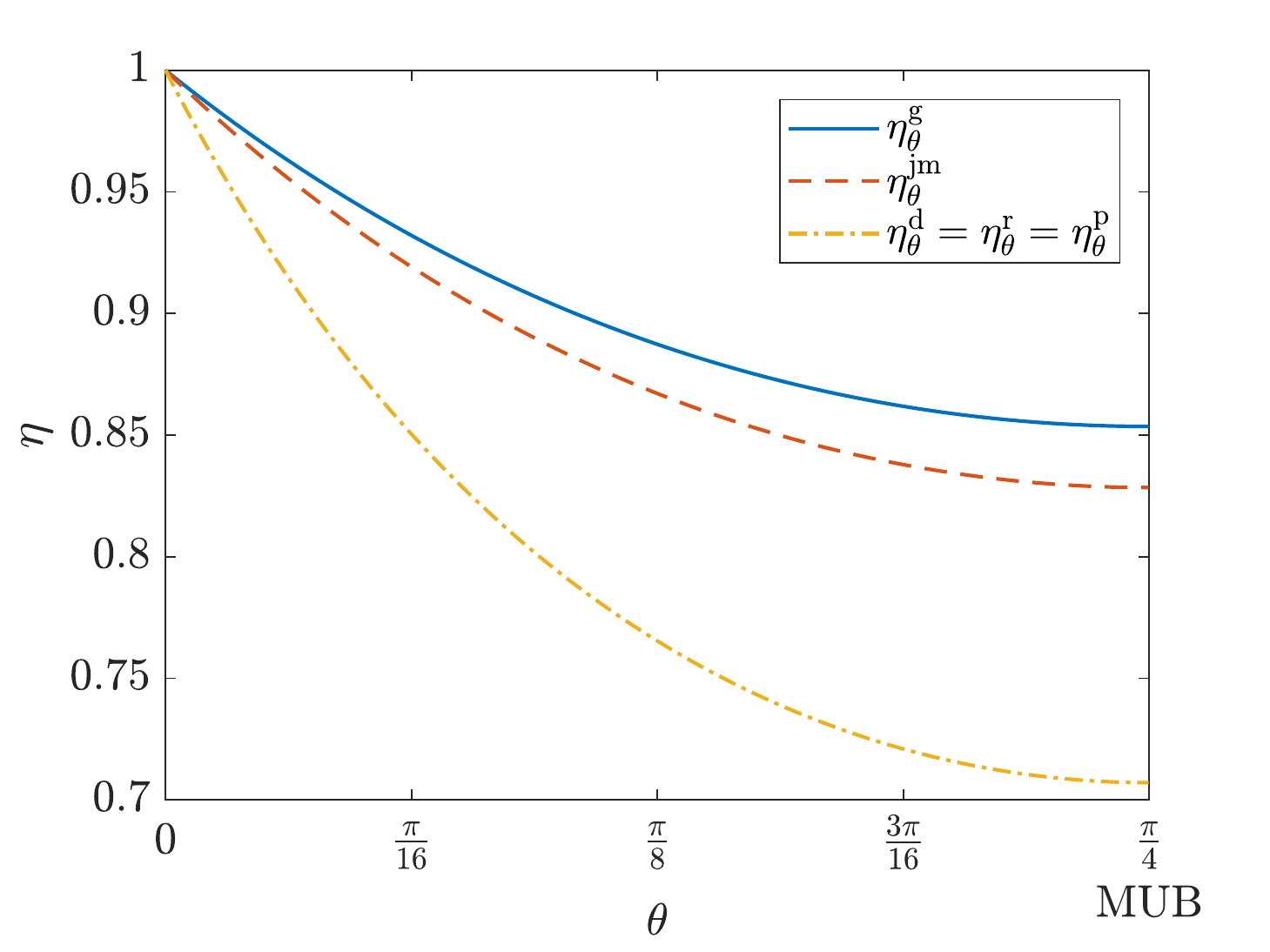}
  \caption{
    The value of all the different measures (see Table \ref{tab:magic}) for a pair of rank-one projective measurements on a qubit such that the angle between the Bloch vectors of these measurements equals $2\theta$; see Eq.~\eqref{eqn:runex}.
    Note that the rightmost point where $\theta=\pi/4$ corresponds to qubit MUBs, which demonstrates the fact that MUBs are the most incompatible rank-one projective qubit measurements under all these measures.
    From bottom to top, the curves are $\ird=\irr=\irp$ from Eq.~\eqref{eqn:ird_ill}, then $\irjm$ from Eq.~\eqref{eqn:irjm_ill}, and finally $\irg$ from Eq.~\eqref{eqn:irg_ill}.
    Although $\ird$, $\irr$, and $\irp$ coincide in this case, this is not the case in general.
  }
  \label{fig:runex}
\end{figure}

In Table~\ref{tab:magic} we give a compact summary of the results for the differents robustness-based measures of incompatibility: definition of the noise sets, properties introduced in Section~\ref{sec:monotonicity}, lower and upper bounds, and value for a specific example of two projective measurements onto MUBs (see Section~\ref{sec:MUB}).
In Fig.~\ref{fig:runex} we plot the values of $\eta^\ast_\theta$ achieved by a pair of rank-one projective measurements acting on a qubit.

\begin{table}[h!]
  \centering
  \renewcommand{\arraystretch}{1}
  \begin{tabular}{|c|c|c|c|c|c|c|c|}
    \hline
    &Form of the noise & Post & Pre & Cvx & Lower bound & MUB value & Upper bound \vphantom{$\vcenter{\rule{0pt}{20pt}}$} \\ \hline
    $\ird$ &$\left\{\Big(\left\{\tr A_a \frac{\id}{d}\right\}_a,\left\{\tr B_b\frac{\id}{d}\right\}_b\Big)\right\}$& yes & no & no & $\displaystyle \frac{d-2+\sqrt{d^2+4d-4}}{4(d-1)}$ & & $\displaystyle\frac{\lambda-g^\md}{f-g^\md}$ \vphantom{$\vcenter{\rule{0pt}{36pt}}$}\\\cline{1-6}\cline{8-8}
    $\irr$ &$\left\{\left(\left\{\frac{\id}{n_A}\right\}_a,\left\{\frac{\id}{n_B}\right\}_b\right)\right\}$& no & yes & yes & $\displaystyle \frac12\left(1+\frac{1}{\sqrt{n_An_B}+1}\right)$ & {$\displaystyle \frac12\left(1+\frac{1}{\sqrt{d}+1}\right)$} & $\displaystyle\frac{\lambda-g^\mr}{f-g^\mr}$ \vphantom{$\vcenter{\rule{0pt}{36pt}}$}\\\cline{1-6}\cline{8-8}
    $\irp$ &$\left\{\Big(\big\{p_a\,\id\big\}_a,\big\{q_b\,\id\big\}_b\Big)\right\}$& \multicolumn{3}{c|}{yes} & $\max \{ \ird,\irr \}$ & & $\displaystyle\frac{\lambda-g^\pp}{f-g^\pp}$ \vphantom{$\vcenter{\rule{0pt}{36pt}}$}\\ \hline
    $\irjm$ &$\JM_d^{n_A,n_B}$& \multicolumn{3}{c|}{yes} & $\displaystyle \frac{2\sqrt{d^2+4d-4}}{3d-2+\sqrt{d^2+4d-4}}$ & $\!\!\left\{\begin{array}{ll}2(\sqrt{2} - 1)&d = 2\\\frac12\left(1+\frac{1}{\sqrt{d}}\right)&d \geq 3\end{array}\right.\!\!$ & $\displaystyle \frac{\lambda- g^\mjm}{f- g^\mjm}$\vphantom{$\vcenter{\rule{0pt}{36pt}}$} \\\hline
    $\irg$ &$\POVM_d^{n_A,n_B}$& \multicolumn{3}{c|}{yes} & \multicolumn{2}{c|}{$\displaystyle \frac12\left(1+\frac{1}{\sqrt{d}}\right)$} & $\displaystyle \frac{\lambda}{f}$ \vphantom{$\vcenter{\rule{0pt}{36pt}}$} \\\hline
  \end{tabular}
  \caption{
    Summary of the results on the depolarising, random, probabilistic, jointly measurable, and general incompatibility robustness of pairs of POVMs.
    Recall that $d$ is the dimension, while $n_{A}$ and $n_{B}$ are the outcome numbers.
    ``Post'' and ``Pre'' stand for post-processing and pre-processing monotonicity, respectively, see Section~\ref{sec:monotonicity}.
    ``Cvx'' stands for the convexity of the inverse of the measure, see Section~\ref{sec:monotonicity}.
    For a pair of rank-one projective measurements $(A,B)$, the quantities appearing in the upper bounds are $f=2$, $\lambda=\max_{a,b}\{\max\Sp(A_a+B_b)\}$, $ g^\mjm=\min_{a,b}\{\min\Sp(A_a+B_b)\}$, and $g^\md=g^\mr=g^\pp=2/d$; see Eqs~\eqref{eqn:f_lambda} and \eqref{eqn:g} for definitions.
  }
  \label{tab:magic}
  \renewcommand{\arraystretch}{1.4}
\end{table}

\section{Most incompatible pairs of measurements}
\label{sec:most}

In this section, we address the question of the most incompatible measurement pairs in dimension $d$, for all the measures introduced in Section~\ref{sec:measures}.
This question has already been raised and partially answered in previous works: in infinite dimension for $\irp$ in Ref.~\cite{HSTZ14} and numerically for $\ird$ and $\irg$ in Ref.~\cite{BQG+17}.
Perhaps surprisingly, we find that the answer depends on which incompatibility measure we consider.
We have already seen that projective measurements onto a pair of mutually unbiased bases are among the most incompatible pairs under $\irg$ in every dimension.
On the other hand, for the measures $\ird$ and $\irp$ we give explicit constructions of pairs which are more incompatible than MUBs for any dimension $d\geq3$.
For $\irjm$, our study is inconclusive, and we do not find measurements that are more incompatible than MUBs in any dimension.
First we discuss the special case of $\irr$, then we solve the qubit case for all the measures, and finally we discuss higher dimensions.

\subsection{Incompatibility random robustness}
\label{sec:irr_special}

Recall that in order to find the most incompatible measurement pair in dimension $d$ regardless of the outcome numbers, it is enough to consider rank-one POVMs if the measure in consideration is monotonic under post-processings.
As we see from Table~\ref{tab:magic}, this is not the case for $\irr$, which, at first glance, makes this problem hard to tackle.
However, what turns out is that for this measure the answer is trivial.
Consider a pair of measurements $(A,B)$ and increase artificially the number of outcomes by adding zero POVM elements to both measurements.
Let us add these elements one-by-one, and denote the POVM pair at step $i$ by $(A^i,B^i)$.
In Appendix~\ref{app:irr_up} we show that if $\lambda<2$ and $2(\lambda-1)<f$, we have
\begin{equation}
  \lim_{i\to\infty}\irr_{A^i,B^i} \leq \frac{2-\lambda}{f-2(\lambda-1)},
\end{equation}
where $f$ and $\lambda$ are defined in Eq.~\eqref{eqn:f_lambda}.
It is then clear that whenever $f=2$ and $\lambda<2$ (e.g., any pair of rank-one projective measurements onto two bases that do not have any eigenvectors in common), this limit reaches $\frac12$.
As it coincides with the trivial lower bound mentioned in Section~\ref{sec:mostincomp}, this shows that $\chi^\mr(d)=\frac12$ for $d\geq2$.
In the rest of this section, we will not discuss this measure anymore.
However, recall that for pairs of rank-one projective measurements $\irr$ coincides with $\ird$, and therefore some of the results later in this section also apply to this measure.

\subsection{Qubit case}
\label{sec:qubit}

In Section~\ref{sec:MUB} we have shown that for a pair of mutually unbiased bases all the incompatibility measures can be computed analytically.
What is special in the case of $d = 2$ is that these values coincide with the universal lower bounds (see Table \ref{tab:magic}).
This means that pairs of projective measurements onto MUBs are among the most incompatible pairs under $\ird$, $\irp$, $\irjm$, and $\irg$ in dimension $d = 2$.
Formally, using the notation introduced in Section~\ref{sec:mostincomp}, we have that
\begin{equation}
  \chi^\md(2)=\chi^\pp(2)=\frac{1}{\sqrt2},\quad\chi^\mjm(2)=2(\sqrt2-1),\quad\text{and}\quad\chi^\mg(2)=\frac12\left(1+\frac{1}{\sqrt2}\right).
\end{equation}
For $\ird$, this was known for pairs of two-outcome POVMs \cite[Appendix~G]{DSFB19}.

It is important to point out that there exist other pairs of measurements reaching these minimal values: from the upper bounds given in Appendix~\ref{app:other_up}, it is clear that any rank-one POVM pair such that $A_a=\ketbra{a}$ and the Bloch vectors of $B$ lie in the $xy$-plane of the Bloch sphere gives rise to the same value as MUBs.
As an example, one might choose $A_a=\ketbra{a}$ and $B$ as a trine measurement in the $xy$-plane.

In Appendix~\ref{app:triplets}, we extend this result to triplets of qubit measurements.
In this case, we show that triplets of projective measurements onto MUBs are among the most incompatible measurements under $\ird$, $\irp$, $\irjm$, and $\irg$ in dimension $d = 2$.

Also note that the value of $\chi^\md(2)$ (respectively its equivalent for three measurements) has interesting consequences for Einstein--Podolsky--Rosen steering.
This is because joint measurability is intimately linked to this notion \cite{QVB14,UBGP15}, as the depolarising map in $\ird$ can be equivalently applied to the state we wish to steer, due to its self-duality.
We refer to Ref.~\cite[Appendix~F]{DSFB19} for details on this connection and only mention here that our results show that in a steering scenario with two (respectively three) measurements and an isotropic state of local dimension two, POVMs do not provide any advantage over projective measurements.

\subsection{Higher dimensions}
\label{sec:higher_dim}

\subsubsection{\texorpdfstring{Dimension $d=3$}{Dimension three}}
\label{sec:qutrit}

\begin{figure}[ht!]
  \centering
  \includegraphics[width=12cm]{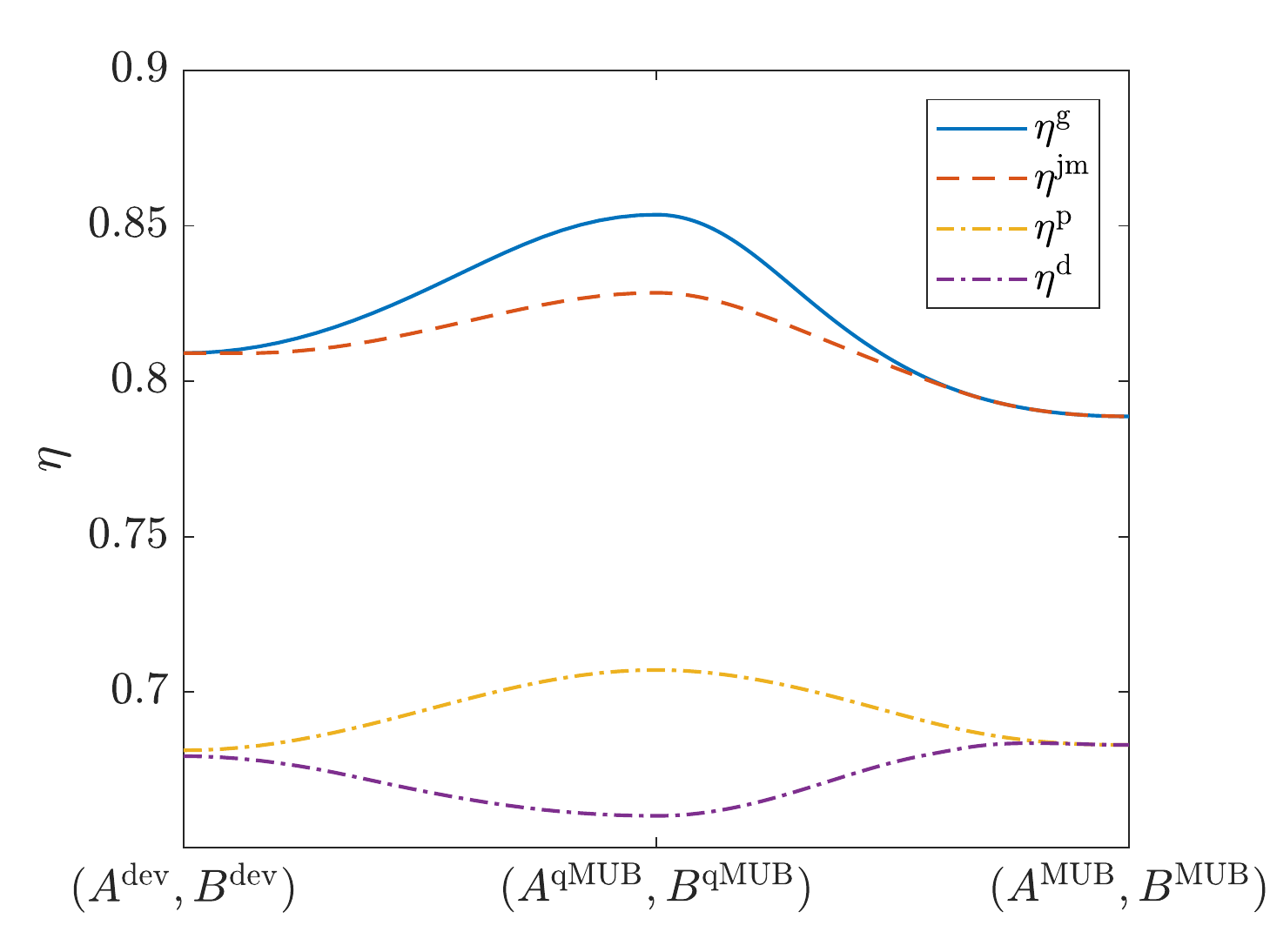}
  \caption{The (numerical) value of the four measures along a one-parameter path of rank-one projective measurements in dimension $d=3$.
    The pair $(A^\mathrm{dev},B^\mathrm{dev})$ is defined in Eq.~\eqref{eqn:abp}, $(A^\mathrm{qMUB},B^\mathrm{qMUB})$ in Eq.~\eqref{eqn:abd}, and $(A^\mathrm{MUB},B^\mathrm{MUB})$ at the beginning of this section.
    Details about the specific path used can be found in Appendix~\ref{app:path}.
  Importantly, on this path the pair $(A^\mathrm{MUB},B^\mathrm{MUB})$ achieves the minimum value with respect to $\irg$ and $\irjm$, but it is outperformed by $(A^\mathrm{dev},B^\mathrm{dev})$ with respect to $\irp$ and by $(A^\mathrm{qMUB},B^\mathrm{qMUB})$ with respect to $\ird$.}
  \label{fig:devil}
\end{figure}

In the previous section we have seen that in dimension $d=2$ pairs of projective measurements onto two MUBs are among the most incompatible pairs of measurements under $\ird$, $\irp$, $\irjm$, and $\irg$.
Starting from dimension $d=3$, the picture changes dramatically.
To show this, we plot the (numerical) value of these four measures for a particular one-parameter path of rank-one projective measurements in dimension three, see Fig.~\ref{fig:devil}.
It is evident from this plot that, contrary to the qubit case, MUBs do not achieve the lowest value of the incompatibility robustness under $\ird$ and $\irp$.
Instead, the lowest value among rank-one projective measurements is reached by other bases, which we have found through an extensive numerical search among pairs of rank-one projective measurements, using a parametrisation of unitary matrices in dimension three \cite{Bro88}.

In this section we only look at rank-one projective measurements.
Due to the unitary invariance of all the measures we assume without loss of generality that the first measurement corresponds to the computational basis $A_{a} = \ketbra{a}$, so that we only need to specify the second measurement $B$.

For $\ird$, the optimum is reached, among others, by
\begin{equation}
  \label{eqn:abd}
  B_b^\mathrm{qMUB}=U\ketbra{b}U^\dagger,\quad\text{where}\quad
  U=\begin{pmatrix}\frac{1}{\sqrt{2}}&\frac{1}{\sqrt{2}}&0\\\frac{1}{\sqrt{2}}&-\frac{1}{\sqrt{2}}&0\\0&0&1\end{pmatrix}.
\end{equation}
Note that it is simply a pair of \emph{qubit MUBs} on a two-dimensional subspace together with a trivial third outcome on the orthogonal subspace.
The incompatibility depolarising robustness of this pair, $\ird_\mathrm{qMUB}(3)\approx0.6602$ (see Eq.~\eqref{eqn:qubitMUB} below for an analytical value) outperforms substantially not only $\ird_\mathrm{MUB}(3)\approx0.6830$, but also the minimal value $0.6794$ found numerically in Ref.~\cite[Table IV]{BQG+17}.

For $\irp$, the optimum is reached, among others, by
\begin{equation}
  \label{eqn:abp}
  B_b^\mathrm{dev}=U\ketbra{b}U^\dagger,\quad\text{where}\quad
  U=\begin{pmatrix}\frac{1}{\sqrt{2}}&\frac12&\frac12\\\frac{1}{\sqrt{2}}&-\frac12&-\frac12\\0&-\frac{1}{\sqrt{2}}&\frac{1}{\sqrt{2}}\end{pmatrix},
\end{equation}
which gives $\irp_\mathrm{dev}\approx0.6813$, showing a slight \emph{deviation} from $\irp_\mathrm{MUB}(3)\approx0.6830$.

For $\irjm$, the numerical search did not yield an improvement on the MUB value, and for $\irg$ we already have an analytical proof that MUBs are among the most incompatible pairs in every dimension.

\subsubsection{\texorpdfstring{Dimension $d\ge4$}{Dimension higher than four}}
\label{sec:higher}

\begin{figure}[h!]
  \centering
  \includegraphics[width=12cm]{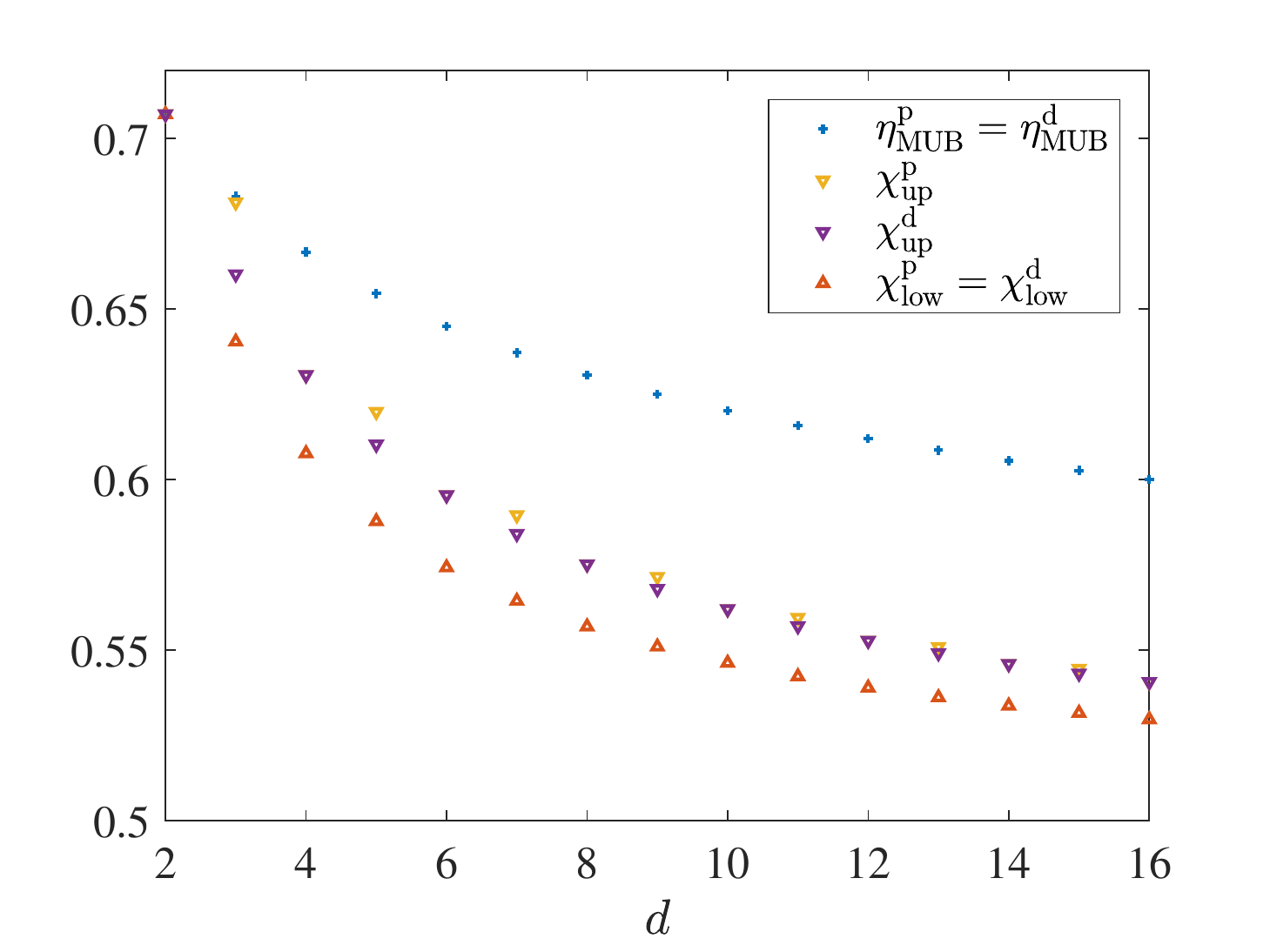}
  \caption{Illustration of the improvement over MUBs for $\ird$ and $\irp$ when the dimension $d$ ranges from 2 to 16.
    From top to bottom are depicted the MUB value (Eqs~\eqref{eqn:ir_mub3} and \eqref{eqn:ir_mub4}), the lowest value we found for $\irp$ (that is, Eq.~\eqref{eqn:qubitMUB} for even dimensions and numerical results based on an analytical construction described in the main text for odd dimensions), the lowest value we found for $\ird$ (Eq.~\eqref{eqn:qubitMUB}), and the lower bound \eqref{eqn:ird_low}.
  }
  \label{fig:chi}
\end{figure}

For $\ird$, the qubit MUB structure found in dimension $d=3$ has several natural generalisations in higher dimensions.
The general idea is to divide the Hilbert space into orthogonal subspaces of various dimensions, and define the measurements as either MUBs or trivial measurements on the different subspaces.
Among these, we found numerically that the most incompatible construction is to define a pair of qubit MUBs on a two-dimensional subspace, while on the orthogonal subspace the remaining measurement operators turn out to be irrelevant.
For simplicity, we choose trivial measurements on the orthogonal subspace, that is, $A_a = \ketbra{a}$ and $B_b = \ketbra{b}$ for $a,b \ge 3$, while $\{A_1,A_2\}$ and $\{B_1,B_2\}$ is a pair of MUBs on the qubit subspace.
For this construction, we get a lower bound in Eq.~\eqref{eqn:qubitMUB_low} and an upper bound in Eq.~\eqref{eqn:qubitMUB_up},
which give the same value and therefore the incompatibility depolarising robustness of this pair is
\begin{equation}
  \label{eqn:qubitMUB}
  \ird_\mathrm{qMUB}(d)=\frac12\left(1+\frac{\sqrt2}{d+\sqrt2}\right)<\ird_\mathrm{MUB}.
\end{equation}
In Fig.~\ref{fig:chi} we plot the improvement over MUBs that this construction achieves.
In particular, it is worth stressing that, in contrast to a pair of MUBs, this construction exhibits the same asymptotic scaling as the lower bound derived in Section~\ref{sec:ird_low}.
More specifically, expanding the right-hand side of Eq.~\eqref{eqn:ird_low} gives
\begin{equation}
  \frac{1}{2} + \frac{1}{2d} + O( d^{-2} ),
\end{equation}
whereas
\begin{align}
  \ird_\mathrm{qMUB}(d) &= \frac{1}{2} + \frac{1}{\sqrt{2} d} + O( d^{-2} ),\\
  \ird_\mathrm{MUB} &= \frac{1}{2} + \frac{1}{2 \sqrt{d}} + O(d^{-1}).
\end{align}
The reason why this pair performs so well is the fact that the two measurements are highly incompatible on the qubit subspace, while the noise is spread uniformly over the entire space.
Note that an analogous structure has been found while searching for the quantum state whose nonlocal statistics are the most robust to white noise \cite{ADGL02}.
Supported by the optimisation in dimension $d=3$ together with one billion random instances in dimensions $d=4$ and $d=5$, and the asymptotic scalings, we conjecture that this pair is among the most incompatible pairs of rank-one projective measurements under $\ird$ for all dimensions.
For general pairs of measurements we leave the question open.

For $\irp$, fixing MUBs on a qubit subspace no longer determines the incompatibility robustness any more, as the noise can now be adjusted to have different weights on the different subspaces.
In fact the construction that uses trivial measurements on the orthogonal subspace does not surpass the $d$-dimensional MUB value any more.
However, employing some other rank-one projective measurements on the orthogonal subspace gives rise to measurements that outperform MUBs.
In even dimensions, by decomposing the space into many qubit subspaces and by having MUBs on each of them, we can reach again the value of Eq.~\eqref{eqn:qubitMUB}.
For instance in dimension $d=4$ this means
\begin{equation}
  \label{eqn:qMUB4}
  B_b=U\ketbra{b}U^\dagger,\quad\text{where}\quad
  U=\begin{pmatrix}\frac{1}{\sqrt{2}}&\frac{1}{\sqrt{2}}&0&0\\\frac{1}{\sqrt{2}}&-\frac{1}{\sqrt{2}}&0&0\\0&0&\frac{1}{\sqrt{2}}&\frac{1}{\sqrt{2}}\\0&0&\frac{1}{\sqrt{2}}&-\frac{1}{\sqrt{2}}\end{pmatrix}.
\end{equation}
The parent POVM is then the same as for $\ird$ whereas the construction of the dual variables is explained in Appendix~\ref{app:ird_up}.
Our conjecture on $\ird$ then translates straightforwardly to $\irp$ in even dimensions as $\ird\leq\irp$.
In odd dimensions, this construction is not applicable.
We conjecture that in dimension $d=3$ the pair defined in Eq.~\eqref{eqn:abp} is among the most incompatible pairs of projective measurements under $\irp$.
In higher odd dimensions, taking this pair on a qutrit subspace together with MUBs on all remaining qubit subspaces always outperforms MUBs (see Fig.~\ref{fig:chi}).
As there might be some more involved construction giving a lower value, we leave the question of the lowest value of $\irp$ open for odd dimensions higher that $d=5$.
Note nonetheless that with one billion random pairs of rank-one measurements in dimension $d=5$ we were not able to surpass it.

For $\irjm$, encouraged by the optimisation in dimension $d=3$ and the one billion random sampling in dimensions $d=4$ and $d=5$, we conjecture that pairs of MUBs in any dimension cannot be outperformed by any pair of rank-one projective measurements.

Regarding $\irg$, the incompatibility generalised robustness of a pair of MUBs is precisely the universal lower bound that we derived in Eq.~\eqref{eqn:irg_low}.
This means that MUBs are among the most incompatible pairs among all pairs of measurements in dimension $d$, regardless of the number of outcomes.
Formally, using the notation introduced in Section~\ref{sec:mostincomp}, this means that
\begin{equation}
  \chi^\mg(d)=\frac12\left(1+\frac{1}{\sqrt{d}}\right).
\end{equation}

\section{Conclusions}
\label{sec:conclusion}

In this work we develop a unified framework to study various robustness-based measures of incompatibility of quantum measurements.
We find that some of the widely used measures do not satisfy some natural properties, which means that one should be cautious when dealing with them.
In particular, they are not suitable for constructing a resource theory of incompatibility.
Moreover, we find that the most incompatible measurement pair depends on the exact measure that we use, even when all the addressed natural properties are satisfied.
We are able to show that for one of the measures a pair of rank-one projective measurements onto mutually unbased bases is among the most incompatible pairs, but also that this is not the case for some other measures.
Our work shows that the different measures exhibit genuinely different properties and we conclude that despite a substantial effort dedicated to the topic, our understanding is still rather limited.

One natural future direction arising from our work would be to obtain a complete characterisation of the most incompatible measurement pairs in all scenarios for all the measures.
We expect, however, that this might be rather difficult, so one might start by restricting the task to natural scenarios, e.g., $d = n_{A} = n_{B}$ or even just searching over rank-one projective measurements.

Many results in this paper can be straightforwardly extended to the case of more than two measurements.
We refer to Appendix~\ref{app:more} for the SDP formulations of the various measures, the upper bounds and a few lower bounds.
This could serve as a good starting point for future research.

A last promising research direction arising from our work concerns the possibility of constructing a resource theory of incompatibility.
Are some of the existing measures suitable as resource monotones? Are there some additional conditions that one should require?
What is the most general class of operations that preserves joint measurability? Answering these questions will help us to understand how to quantify and classify incompatibility in a meaningful and operational manner.

\section*{Acknowledgments}

The authors thank Nicolas Brunner, Bartosz Regu{\l}a, Ren{\'e} Schwonnek, Marco Tomamichel and Roope Uola for useful discussions.
Financial support by the Swiss National Science Foundation (Starting grant DIAQ, NCCR-QSIT) is gratefully acknowledged.
M.F.~acknowledges support from the Polish NCN grant Sonata~UMO-2014/14/E/ST2/00020, and the grant Badania M\l odych Naukowc\'ow number 538-5400-B049-18, issued by the Polish Ministry of Science and Higher Education.
J.K.~acknowledges support from the National Science Centre, Poland (grant no.~2016/23/P/ST2/02122). This project is carried out under POLONEZ programme which has received funding from the European Union's Horizon 2020 research and innovation programme under the Marie Sk{\l}odowska--Curie grant agreement no.~665778.
%

\tocless\bibliography{incbib}

\bibliographystyle{ieeetr}

\appendix
\section*{APPENDIX}

\section{Counterexamples}
\label{app:ctrex}

In this Appendix, we prove some claims made in the main text through explicit examples.
Note that some of the values in this section are obtained via numerics, but as these values are solutions of SDPs, they are exact up to machine precision.

\begin{ctrex}
  \label{ctrex:ird_pre}
  The measure $\ird$ is not monotonic under pre-processings.
\end{ctrex}
Note that an incorrect proof of this statement appeared in Ref.~\cite[Proposition 2]{HKR15}.
The issue with the argument is that it implicitly assumes pre-processings to be trace-preserving.
The following counterexample exploits this loophole.
Consider a pair of qubit MUBs measurements:
\begin{equation}
  \label{eqn:pair_mub}
  A_1 = \begin{pmatrix} 1 & 0 \\ 0 & 0 \end{pmatrix},\quad
  A_2 = \begin{pmatrix} 0 & 0 \\ 0 & 1 \end{pmatrix},\quad
  B_1 = \begin{pmatrix}\frac12&\frac12\\\frac12&\frac12\end{pmatrix},\quad
  B_2 = \begin{pmatrix}\frac12&-\frac12\\-\frac12&\frac12\end{pmatrix}.
\end{equation}
For these the value $\ird_{A,B} = 1/{\sqrt{2}}$ is well-known.
See for example Ref.~\cite[Section~III~A]{ULMH16} or the example in the main text (Section~\ref{sec:ird_ill}).
Let us create new qutrit measurements $A^\Lambda$ and $B^\Lambda$ by pre-processing, specifically, by applying the map ${\Lambda(.) = K_1(.)K_1^\dagger + K_2(.)K_2^\dagger}$, where
\begin{equation}
  K_1 = \begin{pmatrix} 1 & 0 \\ 0 & 1 \\ 0 & 0 \end{pmatrix}
  \quad\text{and}\quad
  K_2 = \begin{pmatrix} 0 & 0 \\ 0 & 0 \\ 0 & 1 \end{pmatrix}\quad\text{so that}\quad\Lambda\left[\begin{pmatrix} a & b\\ c & d \end{pmatrix}\right]=\begin{pmatrix} a & b & 0 \\ c & d & 0 \\ 0 & 0 & d \end{pmatrix}.
\end{equation}
Crucially, $\tr A_2=1\neq2=\tr A^\Lambda_2$ and similarly for $B$.
From the following feasible point for the dual in \eqref{eqn:ird_sdp}:
\begin{equation}
  X_1 = \begin{pmatrix} \frac94 & 0 & 0 \\ 0 & \frac{27}{20} & 0 \\ 0 & 0 & \frac34 \end{pmatrix},\quad
  X_2 = \begin{pmatrix} \frac{27}{10} & 0 & 0 \\ 0 & \frac34 & 0 \\ 0 & 0 & \frac34 \end{pmatrix},\quad
  Y_1 = \begin{pmatrix} \frac{2\sqrt{39}-99}{40} & -\frac14 & 0 \\ -\frac14 & \frac{4 \sqrt{39}-63}{60} & 0 \\ 0 & 0 & -\frac34 \end{pmatrix},\quad
  Y_2 = \begin{pmatrix} \frac{2\sqrt{39}-99}{40} & \frac14 & 0 \\ \frac14 & \frac{4 \sqrt{39}-63}{60} & 0 \\ 0 & 0 & -\frac34 \end{pmatrix},
\end{equation}
we get the bound
\begin{equation}
  \ird_{A^\Lambda,B^\Lambda} \leq \frac{14 \sqrt{39}-3}{120} \approx 0.7036 < 0.7071 \approx \frac{1}{\sqrt{2}}=\ird_{A,B}.
\end{equation}

\begin{ctrex}
  The measure $1/\ird$ is not convex.
\end{ctrex}
Consider the following pairs $(A^0,B^0)$ and $(A^1,B^1)$ of qubit measurements
\begin{equation}
  \label{eqn:ctrex_ird_cvx}
  A_1^0= \begin{pmatrix}1&0\\0&\frac12\end{pmatrix},\quad
  A_2^0=\begin{pmatrix}0&0\\0&\frac12\end{pmatrix},\quad
  B_1^0=\begin{pmatrix}\frac12&\frac12\\\frac12&\frac12\end{pmatrix},\quad
  B_2^0=\begin{pmatrix}\frac12&-\frac12\\-\frac12&\frac12\end{pmatrix},\quad
  A_1^1=\id,\quad
  A_2^1=0,\quad
  B^1=B^0.
\end{equation}
In \cite{YLLO10}, jointly measurable pairs of two-outcome qubit measurements are fully characterised.
From this, we can compute
\begin{equation}
  \ird_{A^0,B^0}=\sqrt{\frac{5+\sqrt{5}}{10}},\quad\ird_{A^2,B^2}=1,\quad\text{and}\quad \ird_{\frac{A^0+A^1}{2},\frac{B^0+B^1}{2}}=\sqrt{\frac{25+\sqrt{13}}{34}},
\end{equation}
from which the convexity of $1/\ird$ is immediately negated as
\begin{equation}
  \frac12\left(\frac{1}{\ird_{A^0,B^0}}+\frac{1}{\ird_{A^1,B^1}}\right)\approx1.0878<1.0902\approx\frac{1}{\ird_{\frac{A^0+A^1}{2},\frac{B^0+B^1}{2}}}.
\end{equation}
Note that the non-concavity of $\ird$ follows from the non-convexity of $1/\ird$.

\begin{ctrex}
  \label{ctrex:irr_post}
  The measure $\irr$ is not monotonic under post-processings.
\end{ctrex}
Consider a pair $(A,B)$ of qubit MUBs measurements, as given in Eq.~\eqref{eqn:pair_mub}.
Let us create a new three-outcome measurement $A^\beta$ by the post-processing
\begin{equation}
  \beta(1|1)=\beta(2|1)=\frac12,\quad\beta(3|1)=\beta(1|2)=\beta(2|2)=0,\quad\text{and}\quad \beta(3|1)=1\quad\text{so that}\quad A^\beta_1 = A^\beta_2 = \frac{A_1}{2}\quad\text{and}\quad A^\beta_3 = A_2.
\end{equation}
The incompatibility random robustness of $A^\beta$ and $B$ is lower than $1/\sqrt2$, which can be seen by the feasible point for the dual in \eqref{eqn:irr_sdp}
\begin{equation}
  X_1 = X_2 = \begin{pmatrix} \frac34 & 0 \\ 0 & \frac{27}{10} \end{pmatrix},\quad
  X_3 = \begin{pmatrix} \frac{27}{20} & 0 \\ 0 & \frac94 \end{pmatrix},\quad
  Y_1 = \begin{pmatrix} \frac{4 \sqrt{39}-63}{60}& -\frac14 \\ -\frac14 & \frac{2\sqrt{39}-99}{40} \end{pmatrix},\quad
  Y_2 = \begin{pmatrix} \frac{4 \sqrt{39}-63}{60}& \frac14 \\ \frac14 & \frac{2\sqrt{39}-99}{40} \end{pmatrix},
\end{equation}
which gives rise to
\begin{equation}
  \irr_{A^\beta,B} \leq \frac{14 \sqrt{39}-3}{120} \approx 0.7036 < 0.7071 \approx \frac{1}{\sqrt{2}}=\irr_{A,B}.
\end{equation}

\begin{ctrex}
  The measures $\ird$ and $\irr$ are incomparable.
\end{ctrex}
Using Ref.~\cite{YLLO10} and the pair of measurements $(A^0,B^0)$ defined in Eq.~\eqref{eqn:ctrex_ird_cvx}, one gets
\begin{equation}
  \ird_{A^0,B^0}=\sqrt{\frac{5+\sqrt{5}}{10}}\approx0.8507<0.8660\approx\frac{\sqrt3}{2}=\irr_{A^0,B^0}.
\end{equation}

To get the other direction, we consider a pair of two-outcome measurements in dimension $d=3$, namely,
\begin{equation}
  A_1^2=\begin{pmatrix}1&0&0\\0&0&0\\0&0&0\end{pmatrix},\quad
  A_2^2=\begin{pmatrix}0&0&0\\0&1&0\\0&0&1\end{pmatrix},\quad
  B_1^2=\begin{pmatrix}\frac{1}{32}&\frac18&-\frac18\\\frac18&\frac34&-\frac18\\-\frac18&-\frac18&\frac34\end{pmatrix},\quad
  B_2^2=\begin{pmatrix}\frac{31}{32}&-\frac18&\frac18\\-\frac18&\frac14&\frac18\\\frac18&\frac18&\frac14\end{pmatrix},
\end{equation}
which gives $\irr_{A^2,B^2}\approx0.8799<0.8816\approx\ird_{A^2,B^2}$.

\begin{ctrex}
  None of the measures defined in the main text is concave.
\end{ctrex}

Consider the following pairs $(A^0,B^0)$ and $(A^1,B^1)$ of qubit measurements
\begin{equation}
  \label{eqn:ctrex_irr_ccv}
  A_a^0=\ketbra{a},\quad
  B_1^0= \begin{pmatrix}\frac{1}{20}&\frac{1}{20}\\\frac{1}{20}&\frac{19}{20}\end{pmatrix},\quad
  B_2^0=\begin{pmatrix}\frac{19}{20}&-\frac{1}{20}\\-\frac{1}{20}&\frac{1}{20}\end{pmatrix},\quad
  A_a^1=U_A\ketbra{a}U_A^\dagger,\quad B_b^1=U_B\ketbra{b}U_B^\dagger,
\end{equation}
where
\begin{equation}
  U_A=\begin{pmatrix}\sqrt{\frac{19}{20}}&\sqrt{\frac{1}{20}}\\\sqrt{\frac{1}{20}}&-\sqrt{\frac{19}{20}}\end{pmatrix}
  \quad\text{and}\quad
  U_B=\begin{pmatrix}\sqrt{\frac{1}{5}}&\sqrt{\frac{1}{5}}\\\sqrt{\frac{1}{5}}&-\sqrt{\frac{4}{5}}\end{pmatrix}.
\end{equation}
With this example, the concavity of all five measures studied in the main text is negated, that is,
\begin{equation}
  \eta^\ast_{\frac{A^0+A^1}{2},\frac{B^0+B^1}{2}}<\frac{\eta^\ast_{A^0,B^0}+\eta^\ast_{A^1,B^1}}{2},
\end{equation}
as one can confirm by solving the respective SDPs up to machine precision.

\section{Relations between the measures}
\label{app:relations}

In the main text we observed that inclusions between the different noise sets immediately imply certain inequalities between the measures.
More specifically, Eq.~\eqref{eqn:order_measures} states that
\begin{equation}
  \label{eqn:order_app}
  \max\{\ird_{A,B},\irr_{A,B}\} \leq \irp_{A,B} \leq \irjm_{A,B} \leq \irg_{A,B}.
\end{equation}
In this Appendix we show that these relations can be strengthened, which leads to strict separations between some of the measures.

In order to tighten the inequality between $\ird$ and $\irjm$, we take the optimal point for the primal for $\ird$ in Eq.~\eqref{eqn:ird_sdp}, and construct from it a feasible point for the primal for $\irjm$ in Eq.~\eqref{eqn:irjm_sdp}.
Specifically, for a pair of measurements $(A,B)$ we subtract some fraction of the original POVM element from the noise reaching the optimum in the primal for $\ird$ in Eq.~\eqref{eqn:ird_sdp}, such that the remaining noise is jointly measurable and can thus serve as a feasible point for the primal for $\irjm$ in Eq.~\eqref{eqn:irjm_sdp}:
\begin{equation}
  \label{eqn:irdvsirjm2}
  \ird_{A,B} A_a + (1-\ird_{A,B})\tr A_a\frac{\id}{d} = \left(\ird_{A,B}+\frac{1-\ird_{A,B}}{d}\epsilon\right)A_a + \left(1-\ird_{A,B}-\frac{1-\ird_{A,B}}{d}\epsilon\right)\frac{\tr(A_a)\id-\epsilon A_a}{d-\epsilon},
\end{equation}
and similarly for $B_b$.
The challenge now is to determine the largest value of $\epsilon$ for which the noise pair
\begin{equation}
  \label{eqn:irdvsirjm3}
  \left(\left\{\frac{\tr(A_a)\id-\epsilon A_a}{d-\epsilon}\right\}_a,\left\{\frac{\tr(B_b)\id-\epsilon B_b}{d-\epsilon}\right\}_b\right)
\end{equation}
is jointly measurable. This can be done by finding rank-one POVMs which can be post-processed to give $A$ and $B$, respectively. Let $\{R_r\}_{r}$ be a rank-one POVM which under post-processing $\beta_{R}$ gives $\{A_a\}_{a}$ and similarly let $\{S_s\}_{s}$ give $\{B_b\}_{b}$ under $\beta_{S}$. The parent POVM given in Eq.~\eqref{eqn:irjm_low_ansatz} implies that the noise pair
\begin{equation}
  \label{eqn:irdvsirjm4}
  \left(\left\{\frac{\tr(R_r)\id-\epsilon R_r}{d-\epsilon}\right\}_r,\left\{\frac{\tr(S_s)\id-\epsilon S_s}{d-\epsilon}\right\}_s\right)
\end{equation}
is jointly measurable for
\begin{equation}
  \epsilon=\frac{2d}{d+\sqrt{d^2+4d-4}}.
\end{equation}
Now note that $A_{a} = \sum_{r} \beta_{R}(a | r) R_{r}$ implies
\begin{equation}
  \sum_{r} \beta_{R}(a | r) \frac{\tr(R_r)\id-\epsilon R_r}{d-\epsilon} = \frac{\tr(A_a)\id-\epsilon A_a}{d-\epsilon}.
\end{equation}
Clearly, if we apply the post-processings $\beta_{R}$ and $\beta_{S}$ to the noise pair given in Eq.~\eqref{eqn:irdvsirjm4}, we will obtain the noise pair given in Eq.~\eqref{eqn:irdvsirjm3} for the same value of $\epsilon$. Since post-processings preserve joint measurability we deduce that
\begin{equation}
  \label{eqn:irdvsirjm}
  \ird_{A,B}+\frac{2(1-\ird_{A,B})}{d+\sqrt{d^2+4d-4}}\leq\irjm_{A,B}.
\end{equation}

In order to tighten the inequality \eqref{eqn:order_app} between $\ird$ and $\irg$, we take the optimal point for the primal for $\ird$ in Eq.~\eqref{eqn:ird_sdp}, and construct from it a feasible point for the primal for $\irg$ in Eq.~\eqref{eqn:irg_sdp}.
Specifically, we use $\tr(A_a)\id\geq A_a$ and $\tr(B_b)\id\geq B_b$ to obtain
\begin{equation}
  \ird_{A,B} A_a + (1-\ird_{A,B})\tr A_a\frac{\id}{d} \geq \left(\ird_{A,B}+\frac{1-\ird_{A,B}}{d}\right)A_a
\end{equation}
and a similar relation for $B_{b}$.
These together imply that
\begin{equation}
  \label{eqn:irdvsirg}
  \ird_{A,B}+\frac{1-\ird_{A,B}}{d}\leq\irg_{A,B}.
\end{equation}

In order to tighten the inequality \eqref{eqn:order_app} between $\irr$ and $\irg$, we take the optimal point for the primal for $\irr$ in Eq.~\eqref{eqn:irr_sdp}, and construct from it a feasible point for the primal for $\irg$ in Eq.~\eqref{eqn:irg_sdp}.
Specifically, we use $\id\geq A_a$ and $\id\geq B_b$ to obtain
\begin{equation}
  \irr_{A,B} A_a + (1-\irr_{A,B})\frac{\id}{n_A} \geq \left(\irr_{A,B}+\frac{1-\irr_{A,B}}{n_A}\right)A_a
\end{equation}
and a similar relation for $B_{b}$.
These together imply that
\begin{equation}
  \irr_{A,B}+\frac{1-\irr_{A,B}}{ \max\{ n_A, n_B \} }\leq\irg_{A,B}.
\end{equation}

Note that all the above improved relations are saturated by pairs of MUBs in dimension two, see Section~\ref{sec:MUB}.

\section{Bounds on the different measures}
\label{app:low}

In this Appendix we provide details about various bounds that we introduce in the main text, namely Eqs~\eqref{eqn:ird_low}, \eqref{eqn:ird_up}, \eqref{eqn:irjm_low}, and \eqref{eqn:irg_low}.
Moreover, we provide measurement-dependent refinements of the lower bounds and we generalise the upper bound on $\ird$, $\irr$, and $\irp$ for certain classes of measurements with some specific structures.

We will use the ansatz defined in Eq.~\eqref{eqn:ansatz_low}, but only for the case of rank-one measurements $A$ and $B$.
Note that in this case $A_a^{1/2} B_b A_a^{1/2} = \tr(A_aB_b)A_a/\tr A_a \propto A_a$, and similarly, $B_b^{1/2} A_a B_b^{1/2} \propto B_b$.
Therefore, we can write Eq.~\eqref{eqn:ansatz_low} as
\begin{equation}
  \label{eqn:ansatz_low_app}
  G_{ab} \propto\, \{A_a,B_b\} + (\tilde{\alpha}_{ab}A_a + \tilde{\beta}_{ab}B_b) + \gamma_{ab}\id,
\end{equation}
where the proportionality constant is fixed by the normalisation, and we introduced the new parameters $\tilde{\alpha}_{ab}$ and $\tilde{\beta}_{ab}$ that now depend on both indices.
Clearly, the operator is non-trivial only on the subspace spanned by the eigenvectors of $A_{a}$ and $B_{b}$, which allows us to compute its spectrum. The eigenvalues of \eqref{eqn:ansatz_low_app}
are then
\begin{equation}
  \label{eqn:eig}
  \frac12\bigg(\tilde{\alpha}_{ab}\tr A_a+\tilde{\beta}_{ab}\tr B_b+2\tr(A_aB_b)\pm\sqrt{(\tilde{\alpha}_{ab}\tr A_a-\tilde{\beta}_{ab} \tr B_b)^2+4\tr(A_aB_b)(\tilde{\alpha}_{ab}+\tr B_b)(\tilde{\beta}_{ab}+\tr A_a)}\bigg)+\gamma_{ab},
\end{equation}
together with $\gamma_{ab}$ when $d\ge3$.

\tocless\subsection{Incompatibility depolarising robustness}

\tocless\subsubsection{Lower bound}
\label{app:ird_low}

For a pair $(A,B)$ of rank-one measurements, an ansatz of the form \eqref{eqn:ansatz_low_app} that is easy to analyse is defined by $\tilde{\alpha}_{ab}=x\tr B_b$, $\tilde{\beta}_{ab}=x\tr A_a$, and $\gamma_{ab}=y\tr A_a\tr B_b$, so that
\begin{equation}
  \label{eqn:ird_low_app}
  G_{ab}=\frac{1}{2(1+dx)+d^2y}\Big(\{A_a,B_b\} + x(A_a\tr B_b + B_b\tr A_a) + y\tr A_a \tr B_b\id\Big).
\end{equation}
Clearly if either $A_{a} = 0$ or $B_{b} = 0$, we have $G_{ab} = 0$, so in the following we restrict ourselves to the case $\tr A_{a} \tr B_{b} > 0$.
From Eq.~\eqref{eqn:eig} we deduce that in order to have $G_{ab}\geq0$, we should have
\begin{equation}
  \label{eqn:jed_rearranged}
  y\geq0\quad\text{and}\quad x + c_{ab}^{2} \pm ( 1 + x ) c_{ab} + y \geq 0,\quad\text{where}\quad c_{ab} = \sqrt{ \frac{\tr(A_aB_b)}{\tr A_a\tr B_b} }.
\end{equation}
For $x \geq -1$ the second constraint is tighter with the minus sign which gives
\begin{equation}
  \label{eqn:ird_lines}
  y \geq -x (1-c_{ab}) + c_{ab} (1-c_{ab}).
\end{equation}
For a fixed $c_{ab}$ this defines a half-plane in the $(x,y)$ plane.
Taking the intersection of all the half-planes corresponding to $c_{ab} \in [0, 1]$ yields the region of $(x, y)$ for $x \geq -1$ which is allowed for all possible measurements.
To explicitly characterise the region we maximise the right-hand side of Eq.~\eqref{eqn:ird_lines} over $c_{ab} \in [0, 1]$ for every fixed value of $x \geq -1$.
Since the expression is a quadratic function of $c_{ab}$ the maximum is achieved at $c_{ab} = (1+x)/2$ if this value lies in the range $[0, 1]$ or at one of the endpoints $c_{ab} = 0$, $c_{ab} = 1$. A straightforward case-by-case analysis yields the allowed region for $x \geq -1$.

For $x \leq -1$ the tighter constraint reads
\begin{equation}
  y \geq -x (1+c_{ab}) - c_{ab} (1+c_{ab})
\end{equation}
and the same procedure leads to the allowed region for $x \leq -1$.
Combining the two results gives the overall allowed region:
\begin{equation}
  \label{eqn:ird_envelop}
  y\geq\left\{\begin{array}{lcl}-2(1 + x)&&\text{if }x\leq-3\\\frac{(1-x)^2}{4}&&\text{if }-3\leq x\leq 1\\0&&\text{if }1\leq x\end{array}\right.,
\end{equation}
over which we want to maximise the objective function of the primal in Eq.~\eqref{eqn:ird_sdp}, that is,
\begin{equation}
  \label{eqn:ird_low_eta}
  \eta=\frac{2+dx}{2(1+dx)+d^2y}.
\end{equation}
Since the right-hand side increases as $y$ decreases, the maximum is reached on the boundary of the allowed region.
Then we can plug $y$ with equality in Eq.~\eqref{eqn:ird_envelop} into the function \eqref{eqn:ird_low_eta} and differentiate the resulting single variable function with respect to $x$ to obtain the following optimal assignment:
\begin{equation}
  \label{eqn:ird_low_opt}
  x=\frac{-2+\sqrt{d^2+4d-4}}{d}\quad\text{and}\quad y=\left(\frac{d+2-\sqrt{d^2+4d-4}}{2d}\right)^2,
\end{equation}
which corresponds to the feasible point presented in Eq.~\eqref{eqn:ird_low} of the main text.
It is easy to check that this choice of $x$ and $y$ saturates Eq.~\eqref{eqn:ird_lines} for a particular value of $c_{ab}$, which we refer to as the \emph{critical overlap}
\begin{equation}
  c_\mathrm{crit}^\md=\frac{d-2+\sqrt{d^2+4d-4}}{2d}\geq\frac{1}{\sqrt{d}}.
\end{equation}
Note that this coincides with the MUB overlap only in dimension $d=2$.

There is an easy way to refine this bound in a measurement-dependent way: instead of requiring that Eq.~\eqref{eqn:ird_lines} holds for all values $c_{ab} \in [0, 1]$, we only require that it holds for the values that appear for the specific pair of rank-one measurements we consider.
Imposing fewer constraints means that we are optimising over a larger region, so we might hope to reach a higher value of the objective function.

\begin{figure}[h]
  \centering
  \includegraphics[width=11cm]{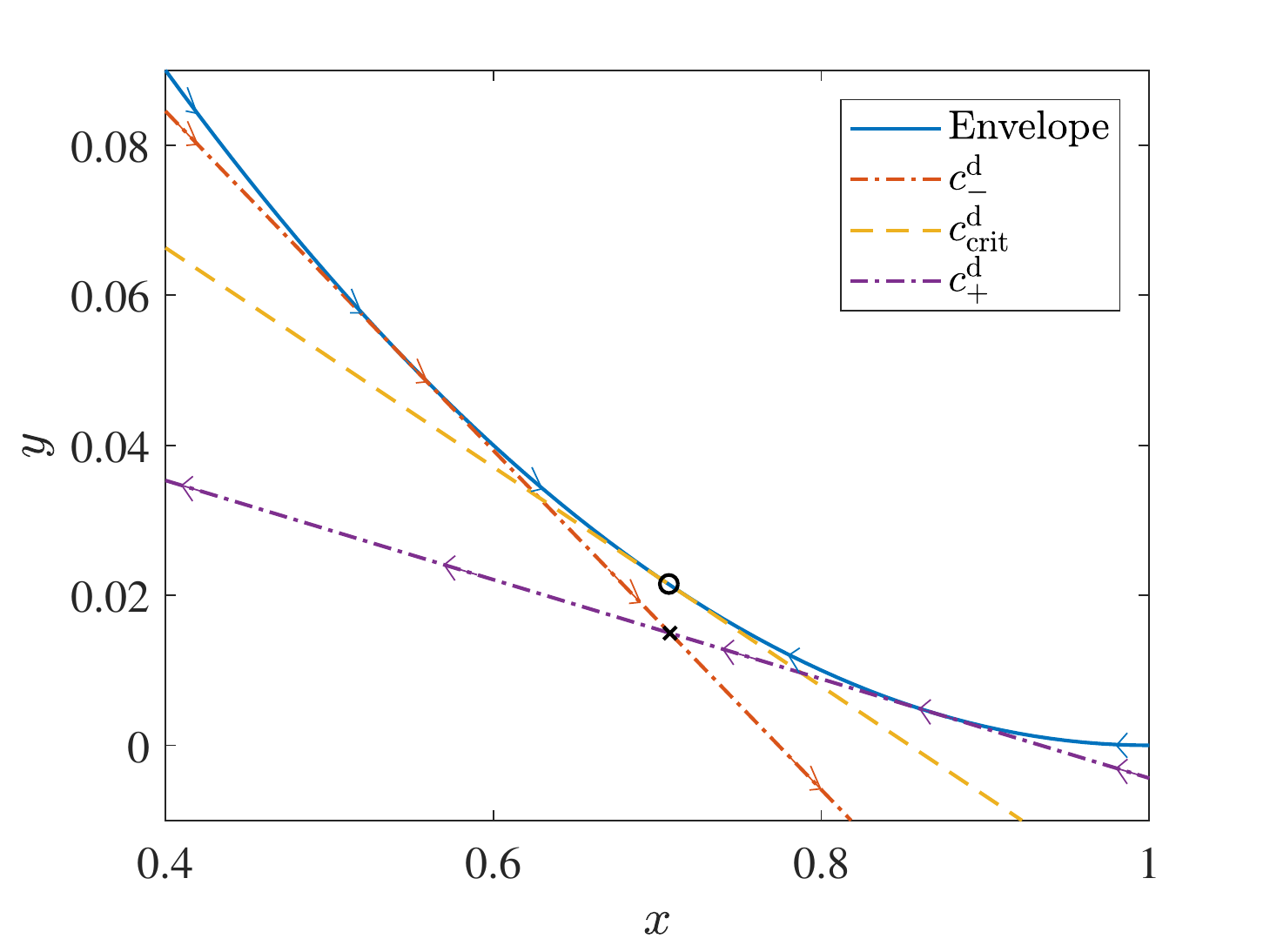}
  \caption{
    Illustration of the measurement-dependent refinement of the lower bound on $\ird$.
    The upper curve is the boundary of the allowed region defined in Eq.~\eqref{eqn:ird_envelop}, while the three lines are the boundaries of half-planes defined in Eq.~\eqref{eqn:ird_lines} for three different overlaps $c_-^\md<c_\mathrm{crit}^\md<c_+^\md$.
    Note that the arrows indicate the gradient of the objective function defined in Eq.~\eqref{eqn:ird_low_eta} along the different curves.
    In particular, the objective function is constant on the dashed line corresponding to $c_\mathrm{crit}^\md$.
    The black circle corresponds to the point given in Eq.~\eqref{eqn:ird_low_opt} while the black cross to the one given in Eq.~\eqref{eqn:cross}.
    In this figure we have used $d=3$ and $c_\pm^\md=c_\mathrm{crit}^\md\pm0.08$.
  }
  \label{fig:ird_lines}
\end{figure}

If we only care about a finite number of overlaps $c_{ab}$, the lower boundary of the relevant region is piecewise linear (see Fig.~\ref{fig:ird_lines}). If one of the overlaps equals the critical one, the bound cannot be improved, so in the following we assume that none of the overlaps equals the critical one.
It turns out that to determine the optimal assignment of $x$ and $y$ we only need to know the value of the largest overlap that is smaller than the critical one, which we denote by $c_-^\md$, and whether there are any overlaps larger than the critical one.
If there are overlaps larger than the critical one, let us denote the smallest of these by $c_+^\md$ and then the
optimal point is reached at the intersection of the two lines defined by $c_-^\md$ and $c_+^\md$ in Eq.~\eqref{eqn:ird_lines}, which gives
\begin{equation}
  \label{eqn:cross}
  x=c_-^\md+c_+^\md-1\quad\text{and}\quad y=(1-c_-^\md)(1-c_+^\md),
\end{equation}
so that the measurement-dependent refinement of Eq.~\eqref{eqn:ird_low} reads
\begin{equation}
  \label{eqn:ird_low_refinement}
  \ird_{A,B}\geq\eta^\mathrm{d,low}_{A,B}=\frac{(c_-^\md+c_+^\md-1)d+2}{2+2(c_-^\md+c_+^\md-1)d+(1-c_-^\md)(1-c_+^\md)d^2}.
\end{equation}
What is particularly interesting about this bound is that whenever $c_{-}^{\md}$ tends to 0 and $c_{+}^{\md}$ tends to 1, the bound tends to 1, i.e., these conditions are strong enough to ensure that the measurements are almost compatible.
This is clearly the case for for identical measurements, that is, for $A=B$, for which the bound equals 1.

If none of the overlaps is greater than the critical one, the optimal assignment is given by $x = c_{-}^\md$ and $y = 0$ and the resulting value corresponds to setting $c_{+}^\md = 1$ in the right-hand side of Eq.~\eqref{eqn:ird_low_refinement}.

As an example we can compute the lower bound for the embedding of qubit MUBs into higher dimensions introduced in Section~\ref{sec:higher_dim}.
In this example, $c_-^\md=1/\sqrt2$ and $c_+^\md=1$ so that we get
\begin{equation}
  \label{eqn:qubitMUB_low}
  \ird_\mathrm{qMUB}(d)\geq\frac12\left(1+\frac{\sqrt2}{d+\sqrt2}\right),
\end{equation}
which turns out to be the correct value, see Eq.~\eqref{eqn:qubitMUB_up} for a matching upper bound.

\tocless\subsubsection{Upper bound for embeddings in higher dimensions}
\label{app:ird_up}

Here we investigate how the upper bound on $\ird$ is affected by the following procedure, which we refer to as \emph{embedding}.
Consider a pair $(\tA,\tB)$ of rank-one projective measurements in dimension $d_i$ and create a new pair $(A,B)$ in dimension $d_f\ge d_i$ as follows:
\begin{equation}
  \label{eqn:ird_low_embedding}
  A_a=\left\{\begin{array}{lcl}\begin{pmatrix}\tA_a&0\\0&0\end{pmatrix}&&\text{if }1\leq a\leq d_i\\\begin{pmatrix}0&0\\0&M_{a-d_i}\end{pmatrix}&&\text{if }d_i+1\leq a\leq d_f\end{array}\right.\quad\text{and}\quad
  B_b=\left\{\begin{array}{lcl}\begin{pmatrix}\tB_b&0\\0&0\end{pmatrix}&&\text{if }1\leq b\leq d_i\\\begin{pmatrix}0&0\\0&N_{b-d_i}\end{pmatrix}&&\text{if }d_i+1\leq b\leq d_f\end{array}\right.,
\end{equation}
where $(M,N)$ is a pair of rank-one projective measurements acting on a $(d_f-d_i)$-dimensional space.

We derive an upper bound on $\ird_{A, B}$ which depends only on the quantity $\lambda$ (defined in Eq.~\eqref{eqn:f_lambda} of the main text) computed for the measurement pair $( \tA, \tB )$ and the dimensions $d_{i}$ and $d_{f}$. As long as $\lambda < 2$ the bound decreases as $d_{f}$ increases and in the limit $d_{f} \to \infty$ it converges to $\frac{1}{2}$. This can be explained by observing that as $d_{f}$ increases the noise gets spread out over the entire space and its weight on the subspace relevant for the measurements $( \tA, \tB )$ decreases. Note that the bound shows no dependence on the second pair of measurements $(M, N)$.

Let us introduce the following ansatz for the dual in Eq.~\eqref{eqn:ird_sdp}:
\begin{equation}
  \left\{\begin{array}{lcl}X_a=\begin{pmatrix}\alpha\id-\beta\tA_a&0\\0&0\end{pmatrix}&&\text{if }1\leq a\leq d_i\\X_a=\begin{pmatrix}\gamma\id&0\\0&0\end{pmatrix}&&\text{if }d_i+1\leq a\leq d_f\end{array}\right.\quad\text{and}\quad
  \left\{\begin{array}{lcl}Y_b=\begin{pmatrix}\alpha\id-\beta\tB_b&0\\0&0\end{pmatrix}&&\text{if }1\leq b\leq d_i\\Y_b=\begin{pmatrix}\gamma\id&0\\0&0\end{pmatrix}&&\text{if }d_i+1\leq b\leq d_f\end{array}\right..
\end{equation}
The scalar constraint of the dual in Eq.~\eqref{eqn:ird_sdp} reads
\begin{align}
  \label{eqn:ird_up_scalar}
  1+\sum_{a=1}^{d_i}\left(\alpha\tr \tA_a-\beta\tr \tA_a^2\right)+\sum_{b=1}^{d_i}\left(\alpha\tr \tB_b-\beta\tr \tB_b^2\right)\geq&\sum_{a=1}^{d_i}\frac{\alpha d_i-\beta\tr \tA_a}{d_f}\tr \tA_a+\sum_{a=d_i+1}^{d_f}\frac{\gamma d_i}{d_f}\tr M_a\\&+\sum_{b=1}^{d_i}\frac{\alpha d_i-\beta\tr \tB_b}{d_f}\tr \tB_b+\sum_{b=d_i+1}^{d_f}\frac{\gamma d_i}{d_f}\tr N_b\nonumber,
\end{align}
which can be further simplified using the rank-one projective assumption to
\begin{equation}
  1+2\alpha d_i\left(1-\frac{d_i}{d_f}\right)\geq2\beta d_i\left(1-\frac{1}{d_f}\right)+2\gamma d_i\left(1-\frac{d_i}{d_f}\right).
\end{equation}
It is easy to see that the constraints
\begin{equation}
  \label{eqn:three-constraints}
  \gamma\geq0,\quad\alpha+\gamma\geq\beta,\quad\text{and}\quad2\alpha\geq\beta\lambda
\end{equation}
ensure that $X_a+Y_b\geq0$.
More specifically, the first one is required to ensure positivity when both indices are between $d_i+1$ and $d_f$, the second one when one of the indices is between 1 and $d_i$ and the other between $d_i+1$ and $d_f$, and the last one when both indices are between 1 and $d_i$.
Requiring that the last two inequalities given in Eq.~\eqref{eqn:three-constraints} are saturated implies
\begin{equation}
  \alpha = \frac{\lambda}{2}\beta,\quad\text{and}\quad\gamma=\left(1-\frac{\lambda}{2}\right)\beta.
\end{equation}
Plugging these back into Eq.~\eqref{eqn:ird_up_scalar} and requiring that the inequality is saturated allows us to deduce that
\begin{equation}
  \frac{1}{\beta d_{i}}=2\left[1-\frac{1}{d_f}-(\lambda-1)\left(1-\frac{d_i}{d_f}\right)\right].
\end{equation}
To see that this corresponds to a non-negative value of $\beta$ note that $\lambda \leq 2$ implies that
\begin{equation}
  2\left[1-\frac{1}{d_f}-(\lambda-1)\left(1-\frac{d_i}{d_f}\right)\right] \geq 2\left[1-\frac{1}{d_f}-\left(1-\frac{d_i}{d_f}\right)\right] = \frac{ 2 ( d_{i} - 1 ) }{d_{f}} \geq 0.
\end{equation}
We immediately see that $\gamma \geq 0$, which means that the assignment given above is a feasible point for the dual given in Eq.~\eqref{eqn:ird_sdp}.
The resulting upper bound reads
\begin{equation}
  \label{eqn:frightening}
  \ird_{A,B}\leq\frac{\lambda-\frac{2}{d_f}-2(\lambda-1)\left(1-\frac{d_i}{d_f}\right)}{2\left[1-\frac{1}{d_f}-(\lambda-1)\left(1-\frac{d_i}{d_f}\right)\right]} =
  \frac{1}{2} \bigg[ 1 + \frac{ (\lambda - 1) d_{i} - 1 }{ (2 - \lambda) d_{f} + ( \lambda - 1 ) d_{i} - 1 } \bigg].
\end{equation}
It is immediate that whenever $d_i=d_f$ we recover exactly the upper bound given in Eq.~\eqref{eqn:ird_up}.

As an example we can compute the upper bound for the embedding of qubit MUBs into higher dimensions introduced in Section~\ref{sec:higher_dim}.
For this example, $d_i=2$, $d_f=d$, $f=2$, and $\lambda=1+1/\sqrt2$ so that we get
\begin{equation}
  \label{eqn:qubitMUB_up}
  \ird_\mathrm{qMUB}(d)\leq\frac12\left(1+\frac{\sqrt2}{d+\sqrt2}\right),
\end{equation}
which turns out to be the correct value, see Eq.~\eqref{eqn:qubitMUB_low} for a matching lower bound.

Note that this procedure can also be applied to sets of more than two measurements.
Although we do not go into the details in this case, Table~\ref{tab:irdlowmore} contains the values obtained by embedding a complete set of MUBs in higher dimensions by adding rank-one projective measurements onto the computational basis of the remaining $(d_f-d_i)$-dimensional space, e.g., $M_a=\ketbra{a}$ in Eq.~\eqref{eqn:ird_low_embedding}.

\begin{table}[hb!]
  \centering
  \begin{tabular}{|c|c|c|c|c|c|}
    \hline
    \backslashbox{$d_i$}{$d_f$}
    {}&    2   &    3   &    4   &    5   &    6   \\\hline
    2 & 0.5774 & 0.5273 & 0.4975 & 0.4778 & 0.4605 \\\hline
    3 &        & 0.4818 & 0.4514 & 0.4314 & 0.4114 \\\hline
    4 &        &        & 0.4309 & 0.4128 &  0.4   \\\hline
    5 &        &        &        & 0.6863 & 0.3620 \\\hline
  \end{tabular}
  \caption{Values obtained for $\ird$ using the embedding procedure described in Appendix~\ref{app:ird_up}.
    Specifically, the values correspond to the embedding of a complete set of MUBs in dimension $d_i$ into dimension $d_f$.
    For example, the value $4/10$ in the last column comes from the embedding of 5 MUBs from dimension $d_i=4$ to dimension $d_f=6$.
    Although we present numerical values all values are analytical.
    Note also that the upper bound obtained via the construction explained in Appendix~\ref{app:ird_up} only gives an upper bound on $\ird$.
    In all cases shown in this table, this bound is tight as there exists a parent POVM reaching exactly the same value.
    Such a parent is not given in this paper.
  As they provide upper bounds on the lowest value achievable by $\ird$, they can be compared to Table IV in Ref.~\cite{BQG+17}.}
  \label{tab:irdlowmore}
\end{table}

\tocless\subsection{Incompatibility random robustness}

\tocless\subsubsection{Lower bound}
\label{app:irr_low}

For a pair of rank-one measurements, it is possible to refine the ansatz defined in Eq.~\eqref{eqn:irr_low_ansatz} by tuning the relative weight of the anticommutator, but this does not lead to any general bound on $\irr$ as this measure is not monotonic under post-processings.

\tocless\subsubsection{Upper bound with addition of zero outcomes}
\label{app:irr_up}

Here we show how to tighten the upper bound introduced in Section~\ref{sec:irr_up} in the presence of zero POVM elements, which we then use in Section~\ref{sec:irr_special}.
We consider a pair $(A,B)$ of measurements that contain zero POVM elements.
Without loss of generality we can assume that the first POVM elements are non-zero.
Then, for simplicity, we assume that $n_A=n_B=n_f$ and that the number of non-zero elements of $A$ and $B$ is the same and we denote it by $n_i$.
The other cases, namely, $n_A\neq n_B$ or the number of non-zero elements of $A$ and $B$ being different, can be treated in a similar manner. Therefore we are left with two POVMs with the same number $n_f$ of outcomes such that
\begin{equation}
  \left\{\begin{array}{lcl}A_a\neq0&&\text{if }1\leq a\leq n_i\\A_a=0&&\text{if }n_i+1\leq a\leq n_f\end{array}\right.\quad\text{and}\quad
  \left\{\begin{array}{lcl}B_b\neq0&&\text{if }1\leq b\leq n_i\\B_b=0&&\text{if }n_i+1\leq b\leq n_f\end{array}\right..
\end{equation}
Then we introduce the following ansatz for the dual in Eq.~\eqref{eqn:irr_sdp}:
\begin{equation}
  \left\{\begin{array}{lcl}X_a=\alpha\id-\beta A_a&&\text{if }1\leq a\leq n_i\\X_a=\gamma\id&&\text{if }n_i+1\leq a\leq n_f\end{array}\right.\quad\text{and}\quad
  \left\{\begin{array}{lcl}Y_b=\alpha\id-\beta B_b&&\text{if }1\leq b\leq n_i\\Y_b=\gamma\id&&\text{if }n_i+1\leq b\leq n_f\end{array}\right..
\end{equation}
Note that the only difference from Eq.~\eqref{eqn:irr_up_ansatz} is that the coefficient of the identity in the dual variable depends on whether the outcome corresponds to a zero or non-zero POVM elements.
The scalar constraint of the dual in Eq.~\eqref{eqn:irr_sdp} reads
\begin{equation}
  1+\sum_{a=1}^{n_i}\left(\alpha\tr A_a-\beta\tr A_a^2\right)+\sum_{b=1}^{n_i}\left(\alpha\tr B_b-\beta\tr B_b^2\right)\geq\sum_{a=1}^{n_i}\frac{\alpha d-\beta\tr A_a}{n_f}+\sum_{a=n_i+1}^{n_f}\frac{\gamma d}{n_f}+\sum_{b=1}^{n_i}\frac{\alpha d-\beta\tr B_b}{n_f}+\sum_{b=n_i+1}^{n_f}\frac{\gamma d}{n_f},
\end{equation}
which can be further simplified by introducing $f$ defined in Eq.~\eqref{eqn:f_lambda} of the main text:
\begin{equation}
  \label{eqn:irr_up_scalar}
  1+2\alpha d\left(1-\frac{n_i}{n_f}\right)\geq\beta d\left(f-\frac{2}{n_f}\right)+2\gamma d\left(1-\frac{n_i}{n_f}\right).
\end{equation}
Assume that $\beta > 0$ and let $\lambda$ be the quantity defined in Eq.~\eqref{eqn:f_lambda} of the main text computed for the measurement pair $(A, B)$. It is easy to see that the constraints
\begin{equation}
  \label{eqn:three-constraints-2}
  \gamma\geq0,\quad\alpha+\gamma\geq\beta,\quad\text{and}\quad2\alpha\geq\beta\lambda,
\end{equation}
ensure that $X_{a} + Y_{b} \geq 0$.
The first one is required to ensure positivity when both indices are between $n_i+1$ and $n_f$, the second one when one is between 1 and $n_i$ and the other between $n_i+1$ and $n_f$, and the last one when both are between 1 and $n_i$.

Requiring that the last two inequalities given in Eq.~\eqref{eqn:three-constraints-2} are saturated implies
\begin{equation}
  \alpha = \frac{\lambda}{2}\beta,\quad\text{and}\quad\gamma=\left(1-\frac{\lambda}{2}\right)\beta.
\end{equation}
Plugging these back into Eq.~\eqref{eqn:irr_up_scalar} and requiring that the inequality is saturated allows us to deduce that
\begin{equation}
  \frac{1}{\beta d}=f-\frac{2}{n_f}-2(\lambda-1)\left(1-\frac{n_i}{n_f}\right),
\end{equation}
It is easy to check that $f > 2 (\lambda - 1)$ (which is only possible if $\lambda < 2$) guarantees that this assignment leads to strictly positive $\beta$. Then, this constitutes a feasible point for the dual given in Eq.~\eqref{eqn:irr_sdp} and we obtain
\begin{equation}
  \irr_{A,B}\leq\frac{\lambda-\frac{2}{n_f}-2(\lambda-1)\left(1-\frac{n_i}{n_f}\right)}{f-\frac{2}{n_f}-2(\lambda-1)\left(1-\frac{n_i}{n_f}\right)}.
\end{equation}
It is easy to check that if $f = 2$, the right-hand side tends to $\frac{1}{2}$ as $n_{f} \to \infty$.

\tocless\subsection{Incompatibility probabilistic robustness}

\tocless\subsubsection{Lower bound}
\label{app:irp_low}

For this measure, a natural idea would be to mix the terms $\tr(B_b)A_a + \tr(A_a)B_b$ used for $\ird$ with the terms $\sqrt{n_A/n_B}A_a + \sqrt{n_B/n_A}B_b$ used for $\irr$.
Unfortunately, our efforts in this direction did not lead to any universal lower bound.
Nevertheless, this procedure can be used for any fixed pair of measurements to obtain improved lower bounds.

\tocless\subsubsection{Upper bound}
\label{app:irp_up}

In the main text, we mention in Section~\ref{sec:higher} that the value of $\ird$ given by the qubit MUBs construction is also reachable by $\irp$ when the dimension is even.
Here we show this fact by adapting the procedure explained in Section~\ref{app:ird_up} to the measure $\irp$.

Recall that we consider pairs of rank-one projective measurements $(A,B)$ whose $d_i$ first outcomes live in the first $d_i$ dimensions of the total $d_f$-dimensional space, and whose $d_f-d_i$ remaining outcomes live in the remaining space.
For this structure, an ansatz for the dual for $\ird$ given in Eq.~\eqref{eqn:ird_sdp} has been presented in Eq.~\eqref{eqn:ird_low_embedding}.
However, this ansatz does not satisfy the additional constraints present in the dual for $\irp$ given in Eq.~\eqref{eqn:irp_sdp}, namely, $\tr X_a\leq\xi$ and $\tr Y_b\leq\upsilon$.

Assume now that $d_f=m d_i$, where $m$ is a positive integer, and that the structure of the pair of rank-one projective measurements $(A,B)$ is the following
\begin{equation}
  \label{eqn:block}
  A_a=\left\{\begin{array}{ccl}\begin{pmatrix}\tA_a&\ldots&0\\\vdots&\ddots&\vdots\\0&\ldots&0\end{pmatrix}&&\text{if }1\leq a\leq d_i\\\vdots&&\vdots\\\begin{pmatrix}0&\ldots&0\\\vdots&\ddots&\vdots\\0&\ldots&\tA_{a-(m-1)d_i}\end{pmatrix}&&\text{if }(m-1)d_i+1\leq a\leq md_i\end{array}\right.
\end{equation}
and similarly for $B_b$ with respect to $\tB_b$, where there are $m$ blocks in the matrices we write and where $\tA$ and $\tB$ are rank-one projective measurements acting on a $d_i$-dimensional space.
We can apply the procedure from Section~\ref{app:ird_up} to each $d_i$-dimensional subspace of the total $d_f$-dimensional space to get a pair of dual variables $(X^{(l)},Y^{(l)})$ for each $l\in\{1,2,\ldots,m\}$.
Then, if we define
\begin{equation}
  X_a=\frac1m\sum_{l=1}^mX_{a-(l-1)d_i}^{(l)}\quad\text{and}\quad Y_b=\frac1m\sum_{l=1}^mY_{b-(l-1)d_i}^{(l)},
\end{equation}
it clearly satisfies all constraints of the dual for $\irp$ given in Eq.~\eqref{eqn:irp_sdp}, including the trace constraint by symmetry.
This implies that for the specific block structure of Eq.~\eqref{eqn:block}, the upper bound obtained in Eq.~\eqref{eqn:frightening} for $\ird$ remains valid for $\irp$.

As an example, consider the measurement pair defined in Eq.~\eqref{eqn:qMUB4}.
For this instance, we have $d_i=2$ and $d_f=4$.
The above procedure gives the same bound as for $\ird$, which is given in Eq.~\eqref{eqn:qubitMUB_up} by setting $d=4$.
\\~\\
\tocless\subsection{Incompatibility jointly measurable robustness}
\label{app:irjm_low}

For this measure, we combine the results of Section~\ref{app:ird_low} with the relation between $\ird$ and $\irjm$ obtained in Eq.~\eqref{eqn:irdvsirjm}.
Specifically, in the primal in Eq.~\eqref{eqn:irjm_sdp}, the parent POVM $G_{ab}$ will be exactly the one we used for $\ird$ in Eq.~\eqref{eqn:ird_low_ansatz} of the main text, that is, Eq.~\eqref{eqn:ird_low_app} with $x$ and $y$ given in Eq.~\eqref{eqn:ird_low_opt}, while the parent POVM $H_{ab}$ will be of the form given in \eqref{eqn:ansatz_low_app} with $\tilde{\alpha}_{ab}=-x\tr B_b$, $\tilde{\beta}_{ab}=-x\tr A_a$, and $\gamma_{ab}=y\tr A_a\tr B_b$, so that
\begin{equation}
  \label{eqn:irjm_low_app}
  H_{ab}=\frac{1}{2(1-dx)+d^2y}\Big(\{A_a,B_b\} - x(A_a\tr B_b + B_b\tr A_a) + y\tr A_a \tr B_b\id\Big).
\end{equation}
Note that such a choice gives rise to a valid parent POVM for the noise considered in Eq.~\eqref{eqn:irdvsirjm3}, namely, $(\{[\tr(A_a)\id-\epsilon A_a]/(d-\epsilon)\}_a,\{[\tr(B_b)\id-\epsilon B_b]/(d-\epsilon)\}_b)$, where
\begin{equation}
  \label{eqn:epsilon}
  \epsilon=\frac{dx - 2}{dy-x}.
\end{equation}
Then we aim at maximising $\epsilon$ under the constraint that the operators $H_{ab}$ of Eq.~\eqref{eqn:irjm_low_app} are positive.
Since the only difference between Eq.~\eqref{eqn:irjm_low_app} and Eq.~\eqref{eqn:ird_low_app} is the sign of the middle term, the allowed region corresponds to the reflection about $x = 0$ of the allowed region given in Eq.~\eqref{eqn:ird_envelop}.
An analysis very similar to the one detailed in Section~\ref{app:ird_low} can be done in order to show that the optimal point is reached for
\begin{equation}
  x=\frac{2+\sqrt{d^2+4d-4}}{d}\quad\text{and}\quad y=\left(\frac{d+2+\sqrt{d^2+4d-4}}{2d}\right)^2,
\end{equation}
which corresponds to the feasible point presented in Eq.~\eqref{eqn:irjm_low} of the main text.
Note that, similarly to the case of $\ird$, these values of $x$ and $y$ correspond to a critical overlap:
\begin{equation}
  c_\mathrm{crit}^\mjm=\frac{-d+2+\sqrt{d^2+4d-4}}{2d}\leq\frac{1}{\sqrt{d}}.
\end{equation}
Note that this coincides with the MUB overlap only in dimension $d=2$.

To obtain a measurement-dependent refinement of the universal bound given in Eq.~\eqref{eqn:irjm_low}, we follow the approach described in Section~\ref{app:ird_low}, i.e., we maximise $\epsilon$ over a larger region determined by the values of $c_{ab}$ present in the specific measurement pair we consider.
In an analogous manner we introduce $c_-^\mjm$ and $c_+^\mjm$, where the former is taken to be 0 if no overlap is smaller that the critical one.
Finally we obtain the following measurement-dependent bound:
\begin{equation}
  \label{eqn:irjm_low_refinement}
  \irjm_{A,B}\geq\eta^\mathrm{d,low}_{A,B}+\frac{1-\eta^\mathrm{d,low}_{A,B}}{d}\cdot\frac{(1+c_-^\mjm+c_+^\mjm)d-2}{(1+c_-^\mjm+c_+^\mjm)(d-1)+c_-^\mjm c_+^\mjm d},
\end{equation}
where $\eta^\mathrm{d,low}_{A,B}$ was defined in Eq.~\eqref{eqn:ird_low_refinement}.
Note that the optimisations of the two parent POVMs appearing in the primal given in Eq.~\eqref{eqn:irjm_sdp} were performed separately. A better bound could in principle be obtained by optimising over both POVMs at the same time, but we leave this task open for future work.
\\~\\
\tocless\subsection{Incompatibility generalised robustness}
\label{app:irg_low}

For a pair $(A,B)$ of rank-one measurements, an ansatz of the form \eqref{eqn:ansatz_low_app} that is easy to analyse is defined by $\tilde{\alpha}_{ab}=(x+yc_{ab}^2)\tr B_b$, $\tilde{\beta}_{ab}=(x+yc_{ab}^2)\tr A_a$, and $\gamma_{ab}=0$, where
\begin{equation}
  c_{ab} = \sqrt{ \frac{\tr(A_{a} B_{b})}{\tr A_a\tr B_b} }
\end{equation}
if $\tr A_a\tr B_b > 0$ and $c_{ab} = 0$ otherwise.
Then
\begin{equation}
  \label{eqn:irg_low_app}
  G_{ab}=\frac{1}{2(1+dx+y)}\Big(\{A_a,B_b\} + \left(x+yc_{ab}^2\right)(A_a\tr B_b + B_b\tr A_a)\Big).
\end{equation}
If $\tr A_a\tr B_b = 0$ we immediately see that $G_{ab} = 0$, so we only need to check positivity in the case $\tr A_a\tr B_b > 0$.
Under the assumption that $x, y \geq 0$ we deduce from Eq.~\eqref{eqn:eig} that in order to have $G_{ab}\geq0$, we should have
\begin{equation}
  x + y c_{ab}^{2} + c_{ab}^{2} \pm ( x + y c_{ab}^{2} + 1 ) c_{ab} \geq 0.
\end{equation}
As shown in Section~\ref{sec:irg_low} of the main text the corresponding visibility reads
\begin{equation}
  \eta=\frac{2+\left(1+\frac1d\right)(dx+y)}{2(1+dx+y)},
\end{equation}
The goal is to maximise this $\eta$ in the positivity region of all $G_{ab}$.
Then a similar analysis to that of $\ird$ leads to the maximum $\eta=(1+1/\sqrt{d})/2$ achieved by the point $x=1/(2\sqrt{d})$ and $y=\sqrt{d}/2$ presented in Eq.~\eqref{eqn:irg_low_ansatz}.

As before to obtain a measurement-dependent refinement we define the critical overlap $c_\mathrm{crit}^\mg=1/\sqrt{d}$ (note that this coincides with the MUB overlap in every dimension).
If one of the overlaps equals $c_\mathrm{crit}^\mg$ no improvement can be obtained, so from now we assume all the overlaps to be different from $c_\mathrm{crit}^\mg$.
Let $c_-^\mg$ be the biggest overlap smaller than $c_\mathrm{crit}^\mg$ and $c_+^\mg$ the smallest bigger than $c_\mathrm{crit}^\mg$.
The optimal point corresponds to
\begin{equation}
  x = \frac{ c_-^\mg c_+^\mg }{ c_+^\mg + c_-^\mg } \quad\text{and}\quad y = \frac{1}{ c_+^\mg + c_-^\mg }
\end{equation}
and gives the following measurement-dependent refinement:
\begin{equation}
  \label{eqn:irg_low_refinement}
  \irg_{A,B}\geq\frac{2(c_-^\mg+c_+^\mg)d+(1+c_-^\mg c_+^\mg d)(d+1)}{2d(1+c_-^\mg+c_+^\mg+c_-^\mg c_+^\mg d)}.
\end{equation}
Contrary to the measurement-dependent bounds on $\ird$ and $\irjm$, namely, Eqs~\eqref{eqn:ird_low_refinement} and \eqref{eqn:irjm_low_refinement}, whenever $c_{-}^{\mg}$ tends to 0 and $c_{+}^{\mg}$ tends to 1, this bound tends to $(3d+1)/(4d)\neq1$.
This is due to the fact that the ansatz given in Eq.~\eqref{eqn:irg_low_app} does not contain the identity term, as including such a term makes the optimisation procedure difficult.
Therefore in some cases a better measurement-dependent lower bound on $\irg$ is obtained by plugging Eq.~\eqref{eqn:ird_low_refinement} into \eqref{eqn:irdvsirg}, which gives
\begin{equation}
  \label{eqn:irg_low_refinement2}
  \irg_{A,B}\geq\frac{1+c_-^\md+c_+^\md+c_-^\md c_+^\md d}{2+2(c_-^\md+c_+^\md-1)d+(1-c_-^\md)(1-c_+^\md)d^2}.
\end{equation}

\section{Details of the path used in Fig.~\ref{fig:devil}}
\label{app:path}

In Fig.~\ref{fig:devil} of the main text, we plot the value of the studied incompatibility measures on a continuous path.
Recall that we fix the first measurement to correspond to the computational basis, so the path is determined by the second measurement and it leads from $B^\mathrm{dev}$ through $B^\mathrm{qMUB}$ to $B^\mathrm{MUB}$.
In this section we provide an explicit description of this path.

The trajectory from $B^\mathrm{dev}$ to $B^\mathrm{qMUB}$ corresponds to the interval $\theta\in[\pi/4,\pi/2]$ for
\begin{equation}
  B_b(\theta)=U(\theta)\ketbra{b}U(\theta)^\dagger,\quad\text{where}\quad
  U(\theta)=\begin{pmatrix}\frac{1}{\sqrt{2}}&\frac{\sin\theta}{\sqrt2}&\frac{\cos\theta}{\sqrt2}\\\frac{1}{\sqrt{2}}&-\frac{\sin\theta}{\sqrt2}&-\frac{\cos\theta}{\sqrt2}\\0&-\cos\theta&\sin\theta\end{pmatrix}.
\end{equation}
It is easy to check that $\theta=\pi/4$ corresponds to $B^\mathrm{dev}$ defined in Eq.~\eqref{eqn:abp}, while $\theta=\pi/2$ corresponds to $B^\mathrm{qMUB}$ defined in Eq.~\eqref{eqn:abd}.

For the second part of the path, let us first explicitly state our choice of the basis $B$ unbiased to $A$ in dimension $d=3$:
\begin{equation}
  B_b^\mathrm{MUB}=U\ketbra{b}U^\dagger,\quad\text{where}\quad
  U=\frac{1}{\sqrt{3}}\begin{pmatrix}1&1&1\\1&\me^{\frac{4\mi\pi}{3}}&\me^{\frac{2\mi\pi}{3}}\\1&\me^{\frac{2\mi\pi}{3}}&\me^{\frac{4\mi\pi}{3}}\end{pmatrix}.
\end{equation}
We now choose a particular unitary $V$ that maps $B^\mathrm{qMUB}$ to $B^\mathrm{MUB}$:
\begin{equation}
  V=\begin{pmatrix}
    \frac{\sqrt2}{\sqrt3}&\frac{\sqrt3+3\mi}{6\sqrt2}&\frac{\sqrt3-3\mi}{6\sqrt2}\\
    0&\frac{\sqrt3-\mi}{2\sqrt2}&\frac{\sqrt3+\mi}{2\sqrt2}\\
    \frac{1}{\sqrt3}&\frac{-\sqrt3-3\mi}{6}&\frac{-\sqrt3+3\mi}{6}
  \end{pmatrix}.
\end{equation}
To generate a continous path we compute the principal matrix logarithm of $V$, i.e., we find a Hermitian matrix $H$ that satisfies $V = \me^{\mi H}$ and whose spectrum is contained in $(-\pi,\pi]$.
The path is given by $\me^{\mi tH} B^{\mathrm{qMUB}} \me^{-\mi tH}$ for $t\in[0,1]$,  which clearly gives $B^\mathrm{qMUB}$ for $t=0$ and $B^\mathrm{MUB}$ for $t=1$.

\section{Larger sets of measurements}
\label{app:more}

In this Appendix, we generalise some notions and techniques introduced in the main text to larger sets of measurements.
The notation of pairs used through the main text, namely, $A_a$ and $B_b$, was useful for clarity.
However, for more measurements we opt for another notation taken from nonlocality: $A_{a|x}$, where $x=1\ldots k$ labels the measurement performed and $a=1\ldots n_x$ is its outcome.
In the following, we will refer to the set of measurements $\{\{A_{a|x}\}_a\}_x$ simply as $\{A_{a|x}\}$, dropping the indices, and we will use $\sum_{a,x}$ as a shorthand for $\sum_{x=1}^k\sum_{a=1}^{n_x}$.
Similarly to Definition~\ref{def:jm} in the main text, we say that a set of POVMs $\{A_{a|x}\}$ is compatible if there exists a parent POVM $G_\vj$, where $\vj =j_1j_2\ldots j_k$ and $j_x\in\{1,\ldots,n_x\}$, such that $\sum_\vj \delta_{j_x,a} G_\vj = A_{a|x}$, that is, we obtain the original POVM elements as marginals of the parent POVM.

Similarly to Section~\ref{subsec:ir}, we can define noise models through the maps ${\bf N}:\POVM_d^{n_1,\ldots,n_k}\to\mathbb{P}(\POVM_d^{n_1,\ldots,n_k})$
such that ${\bf N}:\{A_{a|x}\} \mapsto {\bf N}_{\{A_{a|x}\}} \subseteq \POVM_d^{n_1,\ldots,n_k}$.
Given a noise model such that each noise set contains at least one jointly measurable set of measurements, we can define the corresponding incompatibility robustness measure, similarly to Definition~\ref{def:robustness},
\begin{equation}\nonumber\label{eq:robustness_app}
  \eta^\ast_{\{A_{a|x}\}} = \sup_{\hb{\eta\in[0,1]}{\{N_{a|x}\}\in {\bf N}_{\{A_{a|x}\}}}}\Big\{\eta~\Big|~\eta\cdot \{A_{a|x}\} + (1-\eta)\cdot \{N_{a|x}\}\in\JM\Big\}.
\end{equation}
For these measures, the properties discussed in Sections~\ref{subsec:ir}~and~\ref{sec:monotonicity} can also be naturally generalised to larger sets of measurements, together with the corresponding properties of the noise models ${\bf N}$.
Then it is straightforward to see that the general measures satisfy the same properties as the ones discussed for pairs in Section~\ref{sec:measures}.
These general versions can also be formulated as SDPs, and in the remainder of this Appendix we present these SDP formulations and provide lower and upper bounds on the measures.

\tocless\subsection{SDP}
\label{app:more_sdp}

Here we write the formulations of all the measures introduced in the main text as SDPs.
\begin{equation}
  \nonumber
  \begin{array}{rll}
    \ird_{\{A_{a|x}\}}=&
    \left\{
      \begin{array}{cl}
        \max\limits_{\eta,\{G_\vj\}_\vj} &\eta \\
        \text{s.t.}
        &G_\vj \geq 0,\quad \eta \leq 1 \vphantom{\sum\limits_\vj}\\
        &\sum\limits_\vj \delta_{j_x,a} G_\vj = \eta A_{a|x}+(1-\eta)\tr A_{a|x}\frac{\id}{d} \\
      \end{array}
    \right.\quad&=\left\{
      \begin{array}{cl}
        \min\limits_{\{X_{a|x}\}_{a,x}} &1 +\sum\limits_{a,x}\tr(X_{a|x}A_{a|x})\\
        \text{s.t.}
        &X_{a|x}=X_{a|x}^\dagger,\quad\sum\limits_{a,x}\delta_{j_x,a}X_{a|x} \geq 0 \\
        &1 +\sum\limits_{a,x}\tr (X_{a|x}A_{a|x})\geq \sum\limits_{a,x}\frac{\tr A_{a|x}}{d}\tr X_{a|x} \\
      \end{array}
    \right.\\\\\\
    \irr_{\{A_{a|x}\}}=&
    \left\{
      \begin{array}{cl}
        \max\limits_{\eta,\{G_\vj\}_\vj} &\eta \\
        \text{s.t.}
        &G_\vj \geq 0,\quad \eta \leq 1 \vphantom{\sum\limits_\vj}\\
        &\sum\limits_\vj \delta_{j_x,a} G_\vj = \eta A_{a|x}+(1-\eta)\frac{\id}{n_x} \\
      \end{array}
    \right.\quad&=\left\{
      \begin{array}{cl}
        \min\limits_{\{X_{a|x}\}_{a,x}} &1 +\sum\limits_{a,x}\tr(X_{a|x}A_{a|x})\\
        \text{s.t.}
        &X_{a|x}=X_{a|x}^\dagger,\quad\sum\limits_{a,x}\delta_{j_x,a}X_{a|x} \geq 0 \\
        &1 +\sum\limits_{a,x}\tr (X_{a|x}A_{a|x})\geq \sum\limits_{a,x}\frac{1}{n_x}\tr X_{a|x} \\
      \end{array}
    \right.\\\\\\
    \irp_{\{A_{a|x}\}}=&
    \left\{
      \begin{array}{cl}
        \max\limits_{\hb{\eta,\{G_\vj\}_\vj}{\{\tilde p_{a|x}\}_{a,x}}} &\eta \\
        \text{s.t.}
        &G_\vj \geq 0,\quad \tilde p_{a|x}\geq0 \vphantom{\sum\limits_\vj},\quad\sum\limits_a\tilde p_{a|x}=1-\eta\\
        &\sum\limits_\vj \delta_{j_x,a} G_\vj = \eta A_{a|x}+\tilde p_{a|x}\id \\
      \end{array}
    \right.\quad&=\left\{
      \begin{array}{cl}
        \min\limits_{\hb{\{X_{a|x}\}_{a,x}}{\{\xi_x\}_x}} &1 +\sum\limits_{a,x}\tr(X_{a|x}A_{a|x})\\
        \text{s.t.}
        &X_{a|x}=X_{a|x}^\dagger,\quad\sum\limits_{a,x}\delta_{j_x,a}X_{a|x} \geq 0 \\
        &1 +\sum\limits_{a,x}\tr (X_{a|x}A_{a|x}) \geq \sum\limits_x\xi_x \\
        &\xi_x\geq \tr X_{a|x}\vphantom{\sum\limits_{a,x}}\\
      \end{array}
    \right.\\\\\\
  \end{array}
\end{equation}
\begin{equation}
  \nonumber
  \begin{array}{rll}
    \irjm_{\{A_{a|x}\}}=&
    \left\{
      \begin{array}{cl}
        \max\limits_{\eta,\hb{\{G_\vj\}_\vj}{\{\tilde{H}_\vj\}_\vj}} &\eta \\
        \text{s.t.}
        &G_\vj \geq 0,\quad\sum\limits_\vj G_\vj = \id,\quad\tilde{H}_\vj \geq 0\\
        &\sum\limits_\vj \delta_{j_x,a} (G_\vj-\tilde{H}_\vj) = \eta A_{a|x} \\
      \end{array}
    \right.\quad&=\left\{
      \begin{array}{cl}
        \min\limits_{N,\{X_{a|x}\}_{a,x}} &\tr N\\
        \text{s.t.}
        &N=N^\dagger,\quad X_{a|x}=X_{a|x}^\dagger\vphantom{\sum\limits_{a,x}}\\
        &N\geq\sum\limits_{a,x}\delta_{j_x,a}X_{a|x} \geq 0 \\
        &\sum\limits_{a,x}\tr (X_{a|x}A_{a|x})\geq1\\
      \end{array}
    \right.\\\\\\
    \irg_{\{A_{a|x}\}}=&
    \left\{
      \begin{array}{cl}
        \max\limits_{\eta,\{G_\vj\}_\vj} &\eta \\
        \text{s.t.}
        &G_\vj \geq 0,\quad\sum\limits_\vj G_\vj = \id \\
        &\sum\limits_\vj \delta_{j_x,a} G_\vj \geq \eta A_{a|x} \\
      \end{array}
    \right.\quad&=\left\{
      \begin{array}{cl}
        \min\limits_{N,\{X_{a|x}\}_{a,x}} &\tr N\\
        \text{s.t.}
        &N=N^\dagger,\quad X_{a|x}=X_{a|x}^\dagger\vphantom{\sum\limits_{a,x}}\\
        &N\geq \sum\limits_{a,x}\delta_{j_x,a}X_{a|x},\quad X_{a|x}\geq0\\
        &\sum\limits_{a,x}\tr (X_{a|x}A_{a|x})\geq1 \\
      \end{array}
    \right.
  \end{array}
\end{equation}

\tocless\subsection{Lower bounds}
\label{app:more_low}

Here we derive lower bounds on some of the above measures in this general setting.

For $\ird$, the following bound is presented in Ref.~\cite[Eq.~(11)]{HMZ16}
\begin{equation}
  \ird_{\{A_{a|x}\}} \ge \frac1k\left(1+\frac{k-1}{d+1}\right),
  \label{eqn:more_cloning}
\end{equation}
and from \eqref{eqn:order_measures} this same bound holds for $\irjm$ and $\irg$ as well.
and from \eqref{eqn:order_measures} this same bound holds for $\irjm$ and $\irg$ as well.
Here we outline a few ways to improve on this bound.

One option is to apply the universal lower bounds for pairs, derived in the main text, successively on subsets of pairs of measurements.
Starting from $k$ measurements, we group them into $k/2$ or $(k+1)/2$ pairs, depending on the parity of $k$, and we compute the parent POVMs for these pairs defined in Eqs~\eqref{eqn:ird_low}, \eqref{eqn:irjm_low}, and \eqref{eqn:irg_low}, corresponding to the universal lower bound. Therefore we end up with $k/2$ or $(k+1)/2$ measurements, which are the parent POVMs.
We repeat this process until we end up with only one pair of measurements.
Since we use universal lower bounds, the specific pairings do not matter, and we obtain a bound that depends only on $k$ and $d$.
When $k=2^n$ for instance, we get that the lower bounds on $\ird$ and $\irg$ are the $n$th power of the corresponding lower bound for pairs, namely, Eqs~\eqref{eqn:ird_low} and \eqref{eqn:irg_low}, respectively.
Note that whenever $k$ is odd, an asymmetry is introduced by the choice of which measurement is not paired with another one, but we can overcome this problem by symmetrisation.

Let us illustrate this procedure on a triplet of measurements denoted by $(A,B,C)$.
For any pair $(A,B)$ of POVMs, we denote by $G(A,B)$ their parent POVM used to derive universal lower bounds in Section~\ref{sec:measures}, for instance, Eq.~\eqref{eqn:ird_low} for $\ird$.
Then the following POVM is a parent POVM for noisy versions of $A$, $B$, and $C$, with respect to the noise of $\ird$ in this case:
\begin{equation}
  \label{eqn:mix}
  \frac13\Big[G\big(G(A,B),C\big)+G\big(G(C,A),B\big)+G\big(G(B,C),A\big)\Big].
\end{equation}

For $\ird$ and any number of measurements $k\geq3$ and any dimension $d\geq2$, this procedure never improves on Eq.~\eqref{eqn:more_cloning}, except for triplets of qubit measurements for which it gives the bound $(1+1/\sqrt2)/3$.
Note that we outperform this bound by completely solving this case of three measurements in dimension two in Section~\ref{app:triplets}.

For $\irjm$, the above procedure is made more complex by the fact that two parent POVMs are necessary.
An alternative bound can be obtained by plugging Eq.~\eqref{eqn:more_cloning} into Eq.~\eqref{eqn:irdvsirjm}.
Note that this requires the equivalent of $\epsilon$ in Eq.~\eqref{eqn:irdvsirjm3} for more measurements, namely, a dimension-dependent number such that the set $\{(\tr(A_{a|x})\id-\epsilon A_{a|x})/(d-\epsilon)\}$ is jointly measurable.
Both procedures are possible and involve suitable combinations of the parent POVMs introduced in this work.
However, due to their complexity, we do not present the resulting bounds.

For $\irg$, we should compare the above procedure and the bound obtained by plugging Eq.~\eqref{eqn:more_cloning} into Eq.~\eqref{eqn:irdvsirg}.
For instance, for $k=3$ and $d=4$, the former gives $5/8$ and the latter $3/5$.
\\~\\
\tocless\subsection{Upper bounds}
\label{app:more_up}

The various upper bounds presented throughout the main text naturally generalise to more measurements.
We introduce the generalised quantities corresponding to Eq.~\eqref{eqn:f_lambda}
\begin{equation}
  \label{eqn:more_f_lambda}
  f=\sum\limits_{a,x}\frac{\tr A_{a|x}^2}{d}\quad\text{and}\quad\lambda=\max\limits_\vj\left\{\max\Sp\left(\sum\limits_{a,x}\delta_{j_x,a}A_{a|x}\right)\right\},
\end{equation}
and also those corresponding to Eq.~\eqref{eqn:g}
\begin{equation}
  \label{eqn:more_g}
  g^\md=\sum\limits_{a,x}\left(\frac{\tr A_{a|x}}{d}\right)^2,\quad g^\mr=\sum\limits_x\frac{1}{n_x},\quad g^\pp=\sum\limits_x\min\limits_a\frac{\tr A_{a|x}}{d},\quad\text{and}\quad g^\mjm=\min\limits_\vj\left\{\min\Sp\left(\sum\limits_{a,x}\delta_{j_x,a}A_{a|x}\right)\right\}.
\end{equation}
Using these definitions, the feasible points for the duals in Section~\ref{app:more_sdp} are
\begin{equation}
  \label{eqn:more_up_ansatz}
  \begin{gathered}
    X_{a|x}=\frac{\frac{\lambda}{k}\id-A_{a|x}}{(f-g^\md)d},\quad X_{a|x}=\frac{\frac{\lambda}{k}\id-A_{a|x}}{(f-g^\mr)d},\quad\text{and}\quad X_{a|x}=\frac{\frac{\lambda}{k}\id-A_{a|x}}{(f-g^\pp)d},\quad\text{for $\ird$, $\irr$, and $\irp$, respectively,}\\
    X_{a|x}=\frac{A_{a|x}-\frac{g^\mjm}{k}\id}{(f- g^\mjm)d}\quad\text{and}\quad N=\frac{\lambda- g^\mjm}{f- g^\mjm}\cdot\frac{\id}{d}\quad\text{for $\irjm$,}\\
    X_{a|x}=\frac{A_{a|x}}{fd}\quad\text{and}\quad N=\frac{\lambda}{f}\cdot\frac{\id}{d}\quad\text{for $\irg$.}
  \end{gathered}
\end{equation}
Note that we have implicitly assumed that $f\neq g^\ast$ for all the measures.
From the discussion below it turns out that the equality holds only when all measurement elements are proportional to the identity operator, in which case the set is trivially compatible.
These feasible points give rise to the following bounds:
\begin{equation}
  \label{eqn:more_up}
  \begin{gathered}
    \ird_{\{A_{a|x}\}}\leq\frac{\lambda-g^\md}{f-g^\md}=\eta^\mathrm{d,up}_{\{A_{a|x}\}},\quad\irr_{\{A_{a|x}\}}\leq\frac{\lambda-g^\mr}{f-g^\mr}=\eta^\mathrm{r,up}_{\{A_{a|x}\}},\quad\irp_{\{A_{a|x}\}}\leq\frac{\lambda-g^\pp}{f-g^\pp}=\eta^\mathrm{p,up}_{\{A_{a|x}\}},\\
    \irjm_{\{A_{a|x}\}}\leq\frac{\lambda- g^\mjm}{f- g^\mjm}=\eta^\mathrm{jm,up}_{\{A_{a|x}\}},\quad\text{and}\quad\irg_{\{A_{a|x}\}}\leq\frac{\lambda}{f}=\eta^\mathrm{g,up}_{\{A_{a|x}\}}.
  \end{gathered}
\end{equation}

Note that, from the inequalities in Eq.~\eqref{eqn:order_g} below and under the assumption $f>\lambda$, we have
\begin{equation}
  \label{eqn:order_up}
  \max\left\{\eta^\mathrm{d,up}_{\{A_{a|x}\}},\eta^\mathrm{r,up}_{\{A_{a|x}\}}\right\}\leq\eta^\mathrm{p,up}_{\{A_{a|x}\}}\leq\eta^\mathrm{jm,up}_{\{A_{a|x}\}}\leq\eta^\mathrm{g,up}_{\{A_{a|x}\}}.
\end{equation}

\tocless\subsubsection{Discussion of the feasible points}
\label{app:footnote13AappendixT2}

Here we first show that for all sets of measurements, the inequalities
\begin{equation}
  \label{eqn:zero_denom}
  f\geq g^\md\quad\text{and}\quad f\geq g^\mr
\end{equation}
hold, with equality if and only if all POVM elements involved are proportional to the identity.
Then we also derive the hierarchy used to derive Eq.~\eqref{eqn:order_up}, namely,
\begin{equation}
  \label{eqn:order_g}
  \min\{g^\md,g^\mr\}\geq g^\pp\geq g^\mjm\geq0.
\end{equation}
These two inequalities imply that unless all POVM elements are proportional to the identity we have $f>g^\ast$ and the bounds given in Eq.~\eqref{eqn:more_up} hold (which are generalisations to larger sets of measurements of the upper bounds given in Eqs~\eqref{eqn:ird_up}, \eqref{eqn:irr_up}, \eqref{eqn:irp_up}, \eqref{eqn:irjm_up}, and \eqref{eqn:irg_up} of the main text).

In order to prove the inequalities in Eq.~\eqref{eqn:zero_denom}, we use the Cauchy--Schwarz inequality:
\begin{equation}
  \label{eqn:Cauchy-Schwarz}
  (\tr A_{a|x})^2=[\tr(\id\cdot A_{a|x})]^2\leq\tr(\id^2)\tr(A_{a|x}^2)=d\tr(A_{a|x}^2).
\end{equation}
For $g^\md$, this implies that
\begin{equation}
  f=\sum_{a,x}\frac{\tr(A_{a|x}^2)}{d}\geq\sum_{a,x}\left(\frac{\tr A_{a|x}}{d}\right)^2 = g^\md.
\end{equation}
For $g^\mr$, we also use the concavity of the square-root, which implies that
\begin{equation}
  \sqrt{\sum_{a=1}^{n_x}\tr(A_{a|x}^2)}=\sqrt{n_x\sum_{a=1}^{n_x}\frac{1}{n_x}\tr(A_{a|x}^2)}\geq\sqrt{n_x}\sum_{a=1}^{n_x}\frac{1}{n_x}\sqrt{\tr(A_{a|x}^2)}\geq\frac{1}{\sqrt{n_x}}\sum_{a=1}^{n_x}\sqrt{\frac{(\tr A_{a|x})^2}{d}}=\sqrt{\frac{d}{n_x}},
\end{equation}
where we have used Eq.~\eqref{eqn:Cauchy-Schwarz} to get the second inequality.
This gives
\begin{equation}
  \label{eqn:f-gr-inequality}
  f=\sum_{a,x}\frac{\tr(A_{a|x}^2)}{d}=\frac1d\sum_{x=1}^k\left[\sum_{a=1}^{n_x}\tr(A_{a|x}^2)\right]\geq\sum_{x}\frac{1}{n_x} = g^\mr.
\end{equation}
Note that to have equality in the inequality in Eq.~\eqref{eqn:Cauchy-Schwarz}, the eigenvalues of $A_a$ should all be equal, that is, $A_a\propto\id$.
This shows that in order to have equality in the inequalities of Eq.~\eqref{eqn:zero_denom}, all measurement operators need to be proportional to the identity.

Regarding Eq.~\eqref{eqn:order_g}, the inequality $g^\md\geq g^\pp$ comes from
\begin{equation}
  g^\md=\sum_{x=1}^k\sum_{a=1}^{n_x}\left(\frac{\tr A_{a|x}}{d}\right)^2\geq\sum_{x=1}^k\left(\min_a\frac{\tr A_{a|x}}{d}\right)\left(\sum_{a=1}^{n_x}\frac{\tr A_{a|x}}{d}\right)=\sum_x\min_a\frac{\tr A_{a|x}}{d}=g^\pp.
\end{equation}
The inequality $g^\mr\geq g^\pp$ comes from
\begin{equation}
  g^\mr=\sum_x\frac{1}{n_x}=\sum_{x=1}^k\frac{1}{n_x}\left(\sum_{a=1}^{n_x}\frac{\tr A_{a|x}}{d}\right)\geq\sum_{x=1}^k\min_a\frac{\tr A_{a|x}}{d}=g^\pp.
\end{equation}
The inequality $g^\pp\geq g^\mjm$ comes from $\tr(M)\geq d\min\Sp(M)$ for every $d\times d$ Hermitian matrix $M$, so that
\begin{equation}
  g^\pp=\sum\limits_x\min_a\frac{\tr A_{a|x}}{d}=\min\limits_\vj\left\{\sum\limits_{a,x}\delta_{j_x,a}\frac{\tr A_{a|x}}{d}\right\}=\min\limits_\vj\left\{\frac1d\tr\sum\limits_{a,x}\delta_{j_x,a}A_{a|x}\right\}\geq\min\limits_\vj\left\{\min\Sp\left(\sum\limits_{a,x}\delta_{j_x,a}A_{a|x}\right)\right\}=g^\mjm.
\end{equation}
Lastly, the inequality $g^\mjm\geq0$ comes from the positivity of the POVM elements involved in its definition, which concludes the proof of Eq.~\eqref{eqn:order_g}.

\tocless\subsubsection{Alternative upper bounds}
\label{app:other_up}

Here we provide alternative feasible points for the duals in Section~\eqref{app:more_sdp} that give rise to upper bounds that are in some cases tighter than the ones discussed above.
Let us consider sets of POVMs $\{A_{a|x}\}$ such that no POVM element is zero.
We can define new quantities very similar to the ones of Eqs~\eqref{eqn:more_f_lambda} and \eqref{eqn:more_g}, namely,
\begin{equation}
  \label{eqn:more_f_lambda_mu_tr}
  \begin{gathered}
    f_{\tr{}}=\sum\limits_{a,x}\frac{\tr A_{a|x}^2}{d\tr A_{a|x}},\quad\lambda_{\tr{}}=\max\limits_\vj\left\{\max\Sp\left(\sum\limits_{a,x}\delta_{j_x,a}\frac{A_{a|x}}{\tr A_{a|x}}\right)\right\},\\
    g^\md_{\tr{}}=g^\mr_{\tr{}}=g^\pp_{\tr{}}=\frac{k}{d},\quad\text{and}\quad g^\mjm_{\tr{}}=\min\limits_\vj\left\{\min\Sp\left(\sum\limits_{a,x}\delta_{j_x,a}\frac{A_{a|x}}{\tr A_{a|x}}\right)\right\}.
  \end{gathered}
\end{equation}
Using these we can derive bounds similar to those in Eq.~\eqref{eqn:more_up}:
\begin{equation}
  \label{eqn:more_up_drp_tr}
  \text{for $\ird$, $\irr$, and $\irp$,}\quad X_{a|x}=\frac{\frac{\lambda_{\tr{}}}{k}\id-\frac{A_{a|x}}{\tr A_{a|x}}}{(f_{\tr{}}-g^\pp_{\tr{}})d}\quad\text{so that}\quad\max\left\{ \ird_{\{A_{a|x}\}},\irr_{\{A_{a|x}\}} \right\}\leq\irp_{\{A_{a|x}\}}\leq\frac{\lambda_{\tr{}}-g^\pp_{\tr{}}}{f_{\tr{}}-g^\pp_{\tr{}}},
\end{equation}
\begin{equation}
  \label{eqn:more_up_jm_tr}
  \text{for $\irjm$,}\quad X_{a|x}=\frac{\frac{A_{a|x}}{\tr A_{a|x}}-\frac{ g^\mjm_{\tr{}}}{k}\id}{(f_{\tr{}}- g^\mjm_{\tr{}})d}\quad\text{and}\quad N=\frac{\lambda_{\tr{}}- g^\mjm_{\tr{}}}{f_{\tr{}}- g^\mjm_{\tr{}}}\cdot\frac{\id}{d}\quad\text{so that}\quad\irjm_{\{A_{a|x}\}}\leq\frac{\lambda_{\tr{}}- g^\mjm_{\tr{}}}{f_{\tr{}}- g^\mjm_{\tr{}}},
\end{equation}
\begin{equation}
  \label{eqn:more_up_g_tr}
  \text{for $\irg$,}\quad X_{a|x}=\frac{A_{a|x}}{f_{\tr{}}d\tr A_{a|x}}\quad\text{and}\quad N=\frac{\lambda_{\tr{}}}{f_{\tr{}}}\cdot\frac{\id}{d}\quad\text{so that}\quad\irg_{\{A_{a|x}\}}\leq\frac{\lambda_{\tr{}}}{f_{\tr{}}}.
\end{equation}
Similarly to as in Section~\ref{app:footnote13AappendixT2}, the inequalities $f\geq g^\md_{\tr{}}=g^\mr_{\tr{}}=g^\pp_{\tr{}}\geq g^\mjm_{\tr{}}\geq0$ hold and give natural relations between the bounds.

For the qubit measurements mentioned in Section~\ref{sec:qubit}, namely, any rank-one POVM pair such that $A_a=\ketbra{a}$ and the Bloch vectors of $B$ lie on the $xy$-plane of the Bloch sphere, the parameters in Eq.~\eqref{eqn:more_f_lambda_mu_tr} are $f_{\tr{}}=2$ as such a pair is rank-one, $\lambda_{\tr{}}=1+1/\sqrt2$, and $ g^\mjm_{\tr{}}=1-1/\sqrt2$ due to the orthogonality of the Bloch vectors of the POVM elements of $A$ and $B$.
Therefore the upper bounds in Eqs~\eqref{eqn:more_up_drp_tr}, \eqref{eqn:more_up_jm_tr}, and \eqref{eqn:more_up_g_tr} coincide with the MUB values given in Eq.~\eqref{eqn:ir_mub2}.

For rank-one projective pairs of measurements the bounds in Eqs~\eqref{eqn:more_up_drp_tr}, \eqref{eqn:more_up_jm_tr}, and \eqref{eqn:more_up_g_tr} coincide with their counterparts in Eq.~\eqref{eqn:more_up}, but in general they are incomparable,  that is, for different measurement pairs one or the other might give the lower value.
For the pair $(A^\Lambda,B^\Lambda)$ used in Counterexample~\ref{ctrex:ird_pre}, the bound on $\ird$ in Eq.~\eqref{eqn:more_up_drp_tr} gives $3(\sqrt{13}+1)/10\approx1.3817$, whereas the one in Eq.~\eqref{eqn:more_up} gives $(9\sqrt2-1)/14\approx0.8377$.
On the other hand, for the pair $(A^\beta,B)$ used in Counterexample~\ref{ctrex:irr_post}, the bound on $\ird$ in Eq.~\eqref{eqn:more_up_drp_tr} gives $1/\sqrt2\approx0.7071$, whereas the one in Eq.~\eqref{eqn:more_up} gives $(4\sqrt2+1)/7\approx0.9510$.
This incomparability suggests that there may exist a more general way to construct such upper bounds, e.g., involving polynomials in $A_{a|x}$ in the definition of $X_{a|x}$.
We leave this question open for further work.

\tocless\subsubsection{Tightness of the upper bound on \texorpdfstring{$\irg$}{the incompatibility generalised robustness} for MUBs}

We investigate the tightness of the upper bound on $\irg$ in Eq.~\eqref{eqn:more_up} for various MUB constructions. The relation~\eqref{eqn:irdvsirg} between $\ird$ and $\irg$ is obviously also valid for more than two measurements.
Therefore, the cases in which the bounds on $\ird$ in \cite{DSFB19} are tight, that is, $\ird = (\lambda - k/d)/(k-k/d)$, give rise to tight upper bounds on $\irg$ as well.
This is because in this case Eq.~\eqref{eqn:irdvsirg} reads $\lambda / k \le \irg$, which saturates the upper bound for $\irg$ in Eq.~\eqref{eqn:more_up}.
In particular, for the standard construction of MUBs in prime power dimensions \cite{DEBZ10} the bound on $\irg$ in Eq.~\eqref{eqn:more_up} is tight when $k=d$ and $k=d+1$.

The methods developed in Ref.~\cite{DSFB19} can also be applied to show the tightness of the upper bound on $\irg$ in Eq.~\eqref{eqn:more_up} in some additional cases.
Specifically, applying the ansatz \cite[Eq.~(11)]{DSFB19} to the incompatibility generalised robustness primal leads to optimal constructions in some cases.
In particular, when the dimension is $d=2^r$, all subsets of size $k\in\{2,3,\ldots,d+1\}$ of the standard construction of complete sets of MUBs saturate the upper bound on $\irg$ in Eq.~\eqref{eqn:more_up}.

To show this, we use the notation of Ref.~\cite[Appendix~D]{DSFB19}.
In this work the authors show that for the standard MUB construction the marginals along $j_x$ of the operator $G_\vj$ defined in \cite[Eq.~(11)]{DSFB19} are diagonal in the basis $\{\mub{x}{a}\}_a$.
Thus, the corresponding value of $\eta$ in the incompatibility generalised robustness primal is
\begin{equation}
  \label{eqn:irg2r}
  \eta=\min_{a,x}\,\bmub{x}{a}\left(\sum_\vj\delta_{j_x,a}G_\vj\right)\mub{x}{a}.
\end{equation}
Moreover, by definition \cite[Eq.~(11)]{DSFB19} we have
\begin{equation}
  \sum_{a,x}\bmub{x}{a}\left(\sum_\vj\delta_{j_x,a}G_\vj\right)\mub{x}{a}=\sum_\vj\tr\left(G_\vj\sum_{a,x}\delta_{j_x,a}\pmub{x}{a}\right)=\sum_\vj\tr\left(G_\vj S_\vj\right)=\sum_\vj\tr\left(\lambda G_\vj\right)=\lambda d.
\end{equation}
Therefore, if all $\bmub{x}{a}\sum_\vj\delta_{j_x,a}G_\vj\mub{x}{a}$ are equal, regardless of $a$ and $x$, we can replace the minimum in Eq.~\eqref{eqn:irg2r} by the total sum divided by the number of terms:
\begin{equation}
  \eta=\frac{1}{kd}\sum_{a,x}\bmub{x}{a}\left(\sum_\vj\delta_{j_x,a}G_\vj\right)\mub{x}{a}=\frac{\lambda}k,
\end{equation}
and $\lambda/k$ is a lower bound for $\irg_{\{\pmub{x}{a}\}}$.
As it coincides with the upper bound for $\irg$ in Eq.~\eqref{eqn:more_up} (recall that $f=k$ for rank-one measurements), the tightness of this upper bound follows.

When $d=2^r$, one can see from \cite[Appendix~D~3]{DSFB19} that $\bmub{x}{a}\sum_\vj\delta_{j_x,a}G_\vj\mub{x}{a}$ is indeed independent of $a$ and $x$.
This shows that if $d=2^r$, then for the standard construction of MUBs, we have $\irg=\lambda/k$ for all sets of $k\in\{2,3,\ldots,d+1\}$ projective measurements onto $k$ MUBs.

Another interesting example is given by triplets of MUBs in dimension $d=4$.
From Ref.~\cite{BWB10}, we know that all possible triplets can be parametrised by three (real) parameters.
For $\ird$, depending on the choice of these parameters, we get a different robustness, whereas for $\irg$, they all give the same value, namely, $2/3$.

Note however that the bound on $\irg$ in Eq.~\eqref{eqn:more_up} is not always tight for MUBs.
For $k=4$ MUBs in dimension $d=5$, we get $\irg\approx0.5692<0.5693\approx\lambda/3$.

\tocless\subsubsection{Upper bound on \texorpdfstring{$\irjm$}{the incompatibility jointly measurable robustness} for MUBs}

Below we show that for the standard construction of MUBs in odd prime power dimensions \cite{KR04}, the bounds on $\irjm$ and $\irg$ in  Eq.~\eqref{eqn:more_up} coincide.
In order to show this, we need to prove that in this case
$ g^\mjm=0$ for all $k\in\{2,3,\ldots,d+1\}$.
When $k<d$, this is clear.
For $k=d$ and $k=d+1$, this minimum eigenvalue is reached, for instance, when we pick the first POVM element of each measurement.
We first use the notations of Ref.~\cite[Appendix~D~2]{DSFB19} to prove the general case and then give a simplified proof in the case of prime dimensions.

Here we give the proof when $k=d=p^r$, with $p$ prime and $x\in\ff$, the Galois field with $d$ elements, which singles out a particular choice of $d$ MUBs that does not include the computational basis.
The other cases, namely, $k=d+1$ and $k=d$ with one of the bases being the computational basis, can be treated similarly.
Recall that $g^\mathrm{jm}$ concerns the spectra of the operators $\sum_{a,x}\delta_{j_x,a}A_{a|x}$ for every $\vj$, see Eq.~\eqref{eqn:more_g}. If we choose $\vj = \vec{0}$, we get
\begin{equation}
  \sum_{x\in\ff}\pmub{x}{0}
  = \frac1d\sum_{l,l'\in\ff}\sum_{x\in\ff}\me^{\frac{2\mi\pi}{p}\Tr[x(l^2-l'^2)]}\ket{l}\bra{l'}
  = \id+\sum_{l\in\ff^*}\ket{l}\bra{-l},
  \label{eqn:cryptic}
\end{equation}
where the trace over the Galois field $\ff$ is defined by $\Tr a=a+a^2+\cdots+a^{p-1}$ so that it belongs to $\{0,1,\ldots,p-1\}$.
Note that the convention used here to label the POVM elements is different than the one in the main text, as it starts from 0 instead of 1.
For the operator in Eq.~\eqref{eqn:cryptic}, the vector $(\ket{l}-\ket{-l})$ is an eigenvector with eigenvalue 0 for $l\in\ff\setminus\{0\}$, which concludes the proof.

As an easier illustration, we consider the case when $d$ is an odd prime.
In this case, a complete set of MUBs is given by the computational basis $\{\ket{l}\}_{l=0}^{d-1}$ and
\begin{equation}
  \mub{x}{a}=\frac{1}{\sqrt{d}}\sum_{l=0}^{d-1}\me^{\frac{2\mi\pi}{d}(xl^2+al)}\ket{l},
\end{equation}
where $x$ labels the bases and $a$ the vectors.
Eq.~\eqref{eqn:cryptic} then takes the form
\begin{equation}
  \sum_{x=0}^{d-1}\pmub{x}{0}
  = \frac1d\sum_{l,l'=0}^{d-1}\sum_{x=0}^{d-1}\me^{\frac{2\mi\pi}{d}[x(l^2-l'^2)]}\ket{l}\bra{l'}
  = \id+\sum_{l=1}^{d-1}\ket{l}\bra{-l},
\end{equation}
for which $\ket{l}-\ket{-l}$ is an eigenvector with eigenvalue 0 for $l\in\{1,2,\ldots,d-1\}$.
\\~\\
\tocless\subsection{Most incompatible triplets of qubit measurements}
\label{app:triplets}

Below we analyse the incompatibility robustness of a triplet of qubit MUBs, and show that they are among the most incompatible triplets in dimension 2 under $\ird$, $\irp$, $\irjm$, and $\irg$.
For a triplet of projective measurements onto three qubit MUBs $(A^\mathrm{MUB},B^\mathrm{MUB},C^\mathrm{MUB})$, the quantities defined in Eqs~\eqref{eqn:more_f_lambda} and \eqref{eqn:more_g} are $f=3$, $\lambda=(3+\sqrt3)/2$, $g^\md=g^\pp=3/2$, and $ g^\mjm=(3-\sqrt3)/2$, so that the bounds of Eq.~\eqref{eqn:more_up} read
\begin{equation}
  \label{eqn:up_mub3}
  \ird_{\mathrm{3MUB}}\leq\irp_{\mathrm{3MUB}}\leq\frac{1}{\sqrt3},\quad\irjm_{\mathrm{3MUB}}\leq\sqrt3-1,\quad\text{and}\quad\irg_{\mathrm{3MUB}}\leq\frac12\left(1+\frac{1}{\sqrt3}\right),
\end{equation}
where we write $\eta^\ast_\mathrm{3MUB}$ to denote the incompatibility robustness of $(A^\mathrm{MUB},B^\mathrm{MUB},C^\mathrm{MUB})$.

Now we derive universal lower bounds for the above measures for triplets of qubit measurements, and show that a triplet of MUBs saturates these.
We start with $\ird$, which is post-processing monotonic, and therefore it is enough to derive bounds on it for rank-one triplets $(A,B,C)$, for which we introduce
\begin{equation}
  \begin{aligned}
    &G_{abc}=\frac{1}{2(9-\sqrt3)}\bigg\{\Big[A_aB_bC_c+A_aC_cB_b+B_bC_cA_a+B_bA_aC_c+C_cA_aB_b+C_cB_bA_a\Big]\\
    &+\frac{3\sqrt3-4}{2}\Big[\tr(B_b)\tr(C_c)A_a+\tr(A_a)\tr(C_c)B_b+\tr(A_a)\tr(B_b)C_c\Big]+\frac{9-5\sqrt{3}}{2}\tr(A_a)\tr(B_b)\tr(C_c)\id\bigg\}.
  \end{aligned}
\end{equation}
We show that this is a valid feasible point for the primal for $\ird$ in Section~\ref{app:more_sdp} together with $\eta=1/\sqrt3$.
The correctness of the marginals is immediate.
The positivity follows from a tedious but straightforward computation in which we express the eigenvalues of $G_{abc}$ as functions of the overlaps between $A_a$ and $B_b$, $B_b$ and $C_c$, and $C_c$ and $A_a$ (which is possible, because we are dealing with $2 \times 2$ matrices).
This shows that
\begin{equation}
  \ird_{A,B,C}\geq\frac{1}{\sqrt3},
\end{equation}
which also holds for non-rank-one triplets by post-processing monotonicity of this measure.

Regarding the other measures, the above inequality immediately holds for $\irp$ due to the obvious generalisation of Eq.~\eqref{eqn:order_measures} to triplets of measurements.

For $\irjm$, the method described in Eq.~\eqref{eqn:irdvsirjm2} can be used for triplets as well to get $\ird_{A,B,C}+(1-\ird_{A,B,C})\epsilon/2\leq\irg_{A,B,C}$, where $\epsilon=\sqrt{3}-1$ because
\begin{equation}
  \frac{\tr(A_a)\id-\left(\sqrt3-1\right)A_a}{3-\sqrt3}=\frac{1}{\sqrt3}\Big[\tr(A_a)\id-A_a\Big]+\left(1-\frac{1}{\sqrt3}\right)\tr\Big[\tr(A_a)\id-A_a\Big]\frac{\id}{2},
\end{equation}
and similarly for $B_b$ and $C_c$.
The validity of $\epsilon$ is then guaranteed by applying the bound obtained just above on $\ird$ to the measurements $(\{\tr(A_a)\id-A_a\}_a,\{\tr(B_b)\id-B_b\}_b,\{\tr(C_c)\id-C_c\}_c)$.

For $\irg$, the method described in Eq.~\eqref{eqn:irdvsirg} can be used for triplets as well to get $\ird_{A,B,C}+(1-\ird_{A,B,C})/2\leq\irg_{A,B,C}$.

Therefore, we have proven that
\begin{equation}
  \frac{1}{\sqrt{3}}\leq\ird_{A,B,C}\leq\irp_{A,B,C},\quad\sqrt3-1\leq\irjm_{A,B,C},\quad\text{and}\quad\frac12\left(1+\frac{1}{\sqrt3}\right)\leq\irg_{A,B,C}.
\end{equation}
As a triplet of projective measurements onto three qubit MUBs reaches these lower bounds from Eq.~\eqref{eqn:up_mub3}, they are among the most incompatible triplets of qubit measurements with respect to $\ird$, $\irp$, $\irjm$, and $\irg$.

\end{document}